# Asteroid thermal modeling in the presence of reflected sunlight with an application to WISE/NEOWISE observational data

Nathan Myhrvold

*Intellectual Ventures, Bellevue, WA 98052 U.S.A. Nathan@nathanmyhrvold.com*

June 1, 2016

## Abstract

This study addresses thermal modeling of asteroids with a new derivation of the Near Earth Asteroid Thermal (NEATM) model which correctly accounts for the presence of reflected sunlight in short wave IR bands. Kirchhoff's law of thermal radiation applies to this case and has important implications. New insight is provided into the $\eta$ parameter in the NEATM model and it is extended to thermal models besides NEATM. The role of surface material properties on $\eta$ is examined using laboratory spectra of meteorites and other asteroid compositional proxies; the common assumption that emissivity $\epsilon = 0.9$ in asteroid thermal models may not be justified and can lead to misestimating physical parameters. In addition, indeterminacy in thermal modeling can limit its ability to uniquely determine temperature and other physical properties. A new curve fitting approach allows thermal modeling to be done independent of visible band observational parameters such as the absolute magnitude $H$. These new thermal modeling techniques are applied to observational data for selected asteroids from the WISE/NEOWISE mission. The previous NEOWISE analysis assumes Kirchhoff's law does not apply. It also deviates strongly from established statistical practice and systematically underestimates the sampling error inherent in observing potentially irregular asteroids from a finite sample of observations. As a result, the new analysis finds asteroid diameter and other physical properties that have large differences from published NEOWISE results, with greatly increased error estimates. NEOWISE results have a claimed $\pm10\%$ accuracy for diameter estimates, but this is unsupported. NEOWISE results for 102 asteroids appear to have had the diameter from prior radar and occultation studies simply copied rather than being due to NEOWISE thermal modeling. Diameter estimates from bootstrap calculations appears to be no better than $\pm31.5\%$ accurate when compared to diameters from radar, stellar occultations and spacecraft. NEOWISE errors for parameters like visible band albedo $p_v$ and near-IR albedo $p_{IR1}$ and $p_{IR2}$ are even higher, calling into question their utility.

Keywords: asteroids, near-Earth objects, NEATM, WISE, NEOWISE

Highlights:

- A new derivation of the NEATM with reflected sunlight is presented, with a new fitting methodology.
- Kirchhoff's law affects thermal modeling when IR bands contain reflected sunlight.
- Asteroid diameter estimates with WISE/NEOWISE data are far more inaccurate than previously reported.

- IR and visible band albedos estimated by NEOWISE may have limited utility.
- Examination of the NEOWISE studies covering ~157,000 asteroids finds systematic areas for improvement in the data analysis.

## 1. Introduction

Space-telescope observations in the infrared (IR) wavelength band have long generated information about asteroids that is unique or hard to obtain via other means, including estimates of their diameters *D* and, when coupled with optical observations, of their visible albedos $p_v$ as recently reviewed by Mainzer, Trilling and May (Mainzer et al., 2015b).

The WISE space telescope and the associated NEOWISE project have used asteroid thermal modeling to create such estimates on about 157,000 asteroids, far more than all previous observations combined. WISE was a NASA MIDEX (medium-class explorer) space-telescope mission. It observed in 4 bands: W1, W2, W3, and W4, which were centered on wavelengths of 3.4, 4.6, 12, and 22 $\mu m$ respectively (Wright et al., 2010). NEOWISE is a NASA mission that adds to WISE the post-processing capability necessary to identify and observe asteroids and other small solar system bodies (Mainzer et al., 2011a). WISE and NEOWISE data are stored in the Infrared Science Archive (IRSA)/WISE image archive and are available for download.

The relatively short wavelength W1 and W2 bands collect both emitted thermal infrared light as well as reflected sunlight. Other IR space telescopes, such as IRAS (Delbó, 2004; Hestroffer, 2011; Tedesco et al., 2002a, 2002b; Usui et al., 2014) and AKARI (Usui et al., 2014, 2012, 2011a, 2011b) did not have this issue. IRAS had bands at 12, 25, 60 and 100 $\mu m$, so its lowest wavelength band is roughly equivalent to W3. Asteroid observations with AKARI were carried out in two bands S9W (6.7-11.6 $\mu m$) and L18W (13.9-25.6 $\mu m$), which are again longer wavelength than the WISE W1, W2 bands. As a result these previous space telescopes did not have to contend with reflected sunlight. Others like the Midcourse Space Experiment and Spitzer Space Telescope do have short wavelength IR bands.

The sheer size of the WISE/NEOWISE data set makes it critically important to understand its modeling issues, including the impact of reflected sunlight. Most prior studies which did contend with reflected sunlight in the IR, including NEOWISE, assume that Kirchhoff's law does not apply to asteroid thermal modeling when combined with reflected solar light.

This belief, though widespread, is mistaken. Kirchhoff's law does apply, and it has important implications, particularly to determination of IR albedo. I also examine the role of the "beaming" parameter $\eta$ in the Near Earth Asteroid Thermal Model (NEATM) and the relationship between this parameter and emissivity, thermal inertia, rotation, and surface features. New thermal models are proposed which take extend the variable $\eta$ concept from NEATM to other cases.

NEATM and most other thermal models are based in part on the assumption that thermal IR emissivity can be assumed to be $\epsilon = 0.9$. This assumption is assessed by using laboratory spectra of meteorites and minerals as a proxy for asteroids. The lab spectra reveal that $\epsilon \neq 0.9$ for many materials, and this has a direct impact on thermal modeling results. In addition, I show that there is substantial indeterminacy in asteroid thermal modeling due to both assumptions about emissivity and the existence of multiple competing solutions with different physical properties.

The NEOWISE group, led by principal investigator Amy Mainzer, has published numerous thermal-model fits (see Table 1), and resulting asteroid physical properties—including diameter $D$, visible bad albedo $p_v$ and infrared albedo covering the W1 and W2 bands. Asteroid physical properties from the associated studies are available from the electronic archives of the journals in which the papers were published. I refer to the NEOWISE studies listed in Table 1 collectively as the "NEOWISE papers" because they were published by collaborations that include members of the NEOWISE group, even though some collaborators and some aspects of the work may have been performed outside the official scope of the NASA/JPL NEOWISE project.

The suite of NEOWISE papers together conclude that their results allow relatively precise measurement of asteroid diameter in the majority of cases. While some of the NEOWISE papers do include caveats that the errors can be higher in some cases and that the results must be interpreted with caution, the authors nevertheless assert widely that their results are accurate estimates of diameter, visible albedo, and infrared albedo. As an example, Masiero reports that "Using a NEATM thermal model fitting routine we compute diameters for over 100,000 main belt asteroids from their IR thermal flux, with errors better than 10%"(Masiero et al., 2011).

The validity of this statement is important because numerous researchers have used the NEOWISE results to draw conclusions about many important topics in solar-system science (Bauer et al., 2013; Faherty et al., 2015; Mainzer et al., 2012b, 2012c, 2011e; Masiero et al., 2015a, 2015b, 2013; Nugent et al., 2012; Sonnett et al., 2015).

The NEOWISE results also have strong implications for the distribution of NEO size and albedo, which in turn strongly influences discussion about a search for potentially hazardous earth-impacting asteroids (Grav et al., 2015; Mainzer et al., 2015a, 2015b; Myhrvold, 2016). This underscores the importance of having asteroid diameter and albedo estimates which are as accurate as possible. It is also important to have good understanding of the inevitable error in such estimates.

Understanding the limitations and techniques necessary to analyze the WISE/NEOWISE observational data set is also important in its own right. The data set covers enormously more asteroids than all previous IR studies combined. It is likely to be the largest such archive available to astronomers for many years, and has potential for the application of more detailed and sophisticated thermophysical modeling than the simple thermal models considered here(Alí-Lagoa et al., 2013; Hanuš et al., 2015; Koren et al., 2015). The data from WISE/NEOWISE must therefore be understood both for the enormous opportunity that it represents as well as the challenges in using it given some of its intrinsic limitations.

The present study critically examines the NEOWISE results and the broad conclusion that they are accurate estimates of asteroid physical parameters by performing an independent analysis of the WISE/NEOWISE data. I find systematic flaws in the data analysis, error analysis, and physics of the thermal model. As a consequence, the NEOWISE results are severely compromised in accuracy, and may be of limited utility. Some of this is due to fundamental attributes of the WISE observations, so even when reanalyzed the results have substantial error.

This study is organized into 5 principal sections. Section 2 derives the NEATM in a new way, explaining the role of Kirchhoff's law in asteroid thermal modeling in the presence of reflected sunlight. It includes derivation of additional models based on the FRM model, and a simple blackbody model. It also offers a new approach to fitting asteroid thermal models based on using a temperature parameter rather than using the $\eta$ parameter for fitting, which improves both numerical performance, reduces the likelihood of false minima and allows thermal modeling in cases where absolute magnitude $H$ is unknown, thus decoupling thermal modeling from optical band observations.

The derivation also offers new insight into the role of the $\eta$ parameter and its interpretation. The widespread assumption that emissivity $\epsilon = 0.9$ across all IR bands is examined by using laboratory spectra for meteorites and minerals thought to be possible asteroid constituents. Since $\epsilon \neq 0.9$ for many of these materials within the WISE bands, this effects $\eta$ and the resulting temperature and diameter. A surprising finding is that fitting the W3, W4 bands can have multiple solutions at very different subsolar temperatures and diameters. Each effect is a source of indeterminacy in asteroid thermal models.

Section 3 covers detailed data reduction from WISE/NEOWISE observations, both as done by the NEOWISE team, and alternative approaches. It is relevant to any potential user of WISE/NEOWISE observational data because the analysis methods laid out in the NEOWISE papers are incompletely described and thus are irreproducible. In additions the NEOWISE methods deviate from established statistical practice. The NEOWISE error estimates ignore population sampling error.

Section 4 performs an empirical analysis of the ~157,000 asteroid model fits published in the NEOWISE papers. The checks reveal certain important inconsistencies and show that the results have a sensitive dependence on the values of absolute visible-band magnitude $H$. A surprising finding is that the purported "best fit" NEOWISE models do not go near the data points in most cases, missing data from entire bands for the majority of asteroids. Partly this seems to be due to an incorrect calculation of incident solar flux by NEOWISE, but even with that corrected there is substantial discrepancy.

The key method of determining accuracy of NEOWISE thermal modeling is to compare parameters like diameter $D$ with estimates from other means, such as radar, stellar occultations and spacecraft flybys (referred to here by the acronym ROS). Comparison between the NEOWISE diameters and an independent data set of such measurements reveals troubling numerical coincidences. In the 123 cases for 102 asteroids, the NEOWISE diameter estimate matches the prior ROS value from the literature exactly, to the nearest meter. The odds of this occurring by chance appear to be infinitesimal, suggesting that there may be a software bug or some other issue that compromises the NEOWISE results.

Unfortunately this also means that it is impossible to assess the accuracy of NEOWISE estimates because the asteroids one would want to use for comparisons have already had the estimates set equal to the ROS values they would be compared to.

Section 5 presents the results of bootstrap based analysis on a subset of the asteroids in the NEOWISE results. The new results are strikingly different than those obtained by NEOWISE. The resulting curves do fit the data without missing entire bands. However the estimated errors are far larger and the scatter between the new analysis and NEOWISE is considerable. The ratio of diameters estimated with the bootstrap to NEOWISE, $D_{\text{bootstrap}}/D_{\text{NEOWISE}}$ has a median value of 1.036 and a 95% confidence interval of $0.775 \leq D_{\text{bootstrap}}/D_{\text{NEOWISE}} \leq 3.74$ for randomly chosen main belt asteroids, when restricted to the best cases (eliminating low temperature cases with even greater scatter). While the median result is only 3.6% larger, the asymmetry in the distribution shown by the 95% confidence interval alters distribution toward larger diameters, especially for diameters $D \leq 131{,}334$m.

The visible band albedo $p_v$ and near IR albedos $p_{IR1}$ and $p_{IR2}$ (for the W1 and W2 bands respectively) are off by even larger factors. The NEOWISE results for these parameters may not have much utility as a result.

The proximate cause of the high error estimates is simply scatter in the data points, likely caused by irregular shape or other deviations from the model which assumes a uniform spherical asteroid. In some cases there are a sufficient number of points relative to the variability to sample the irregularity and achieve low error, but for most asteroids the combination of scatter and insufficient observational data leads to large estimated errors.

Thermal modeling indeterminacy leads to cases where straightforward application of the NEATM and other models have anomalously low temperature and very large diameters. While it is possible to prune these solutions it appears that this problem may have a physical origin and bears further investigation.

Error analysis deals with the internal precision of estimated parameters. Accuracy is the comparison of the thermal modeling results to asteroids with diameters measured by radar, stellar occultation and other methods. Although NEOWISE results frequently claim an accuracy of $\pm 10\%$, as in the Maisero et al quote above, it appears that no quantitative calculation of accuracy was done in prior NEOWISE studies, and as noted above the NEOWISE results appear to have the ROS derived diameters simply copied to their result, making it impossible to make a fair comparison.

The bootstrap methods revise the error limits that are possible to $\pm 31.5\%$ for the asteroids where we have ROS diameters. It is likely that the errors are larger in the general case because those asteroids have much higher error estimates. Data tables in the appendix give physical parameter estimates for the 2488 asteroids studied here.

## 2. Asteroid Thermal Models with Reflected Sunlight

A previous study (Myhrvold, 2016) found that the proposed NEOCam mission (Mainzer et al., 2015a) made the assumption that one could choose an IR band albedo $p_{IR}$ independently of emissivity $\epsilon$ in situations where a thermal emission is combined with reflected light. The same approach is taken in the NEOWISE papers listed in Table 1, which assume that the IR albedo $p_{IR}$, which is defined for bands W1 and W2 together, or $p_{IR1}$ and $p_{IR2}$ for the bands individually, can take on arbitrary values, while the emissivity within those same bands is assumed to be fixed at $\epsilon = 0.9$. Similar assumptions have been used in other studies that use NEATM modeling with thermal infrared observations in bands that have substantial reflected sunlight, including (Emery et al., 2006; Hargrove et al., 2012; Lim et al., 2011; Mueller et al., 2011) and Trilling et al., 2010.

Unfortunately, this assumption is flawed by a basic physics error. Maxwell's equations are invariant under time reversal, so emission and absorption are fundamentally the same physics. Kirchhoff's law of thermal radiation is an expression of this fact: it holds that emissivity $\epsilon$ and albedo $p$ are simply related, $\epsilon + p = 1$. This condition is also required to obtain thermal equilibrium and the Planck spectrum.

A possible reason for the confusion on the applicability of Kirchhoff's law is that there are many different expressions of both emissivity and albedo. Some are integrated over wavelength $\lambda$, broadly or within a band, and some are integrated over a geometric surface. Kirchhoff's law may not apply across all of them without taking care to use the relevant emissivity and albedo. Although it is certainly important to keep these considerations in mind when applying Kirchhoff's law, they do not mean that the law can be ignored. As will be shown below, when reflected sunlight is added to the NEATM model (or other asteroid thermal models), it is possible to keep track of these complications and still apply Kirchhoff's law.

The effect of this violation of Kirchhoff's law is to add an extra and unwarranted degree of freedom to the W1, W2 bands. Kirchhoff's law requires that their in-band emissivity, which gates thermal emission, is related to their in-band albedo, which gates reflected sunlight, so higher reflected sunlight comes at the cost of lower thermal emission. The non-Kirchhoff approach allows thermal emission at $\epsilon = 0.9$ regardless of the amount of reflected sunlight in the same band.

In principle Kirchhoff's law applies to any asteroid thermal model, but several factors limit its impact on most previous studies. The first is the IR passbands used for observation – in the case of IRAS and Akari, for example, the observational bands are long enough wavelength that there is comparatively little reflected sunlight. As a result these studies ignored reflected sunlight altogether. The error in flux this would introduce is likely small compared to other errors, and thus would likely not impact results very much.

In the case of NEOWISE the situation is reversed. There are only four WISE bands of which two are short enough wavelength to have a significant amount of reflected sunlight. Kirchhoff can thus impact 50% of the observational data. In addition NEOWISE does not simply model the diameter, it also explicitly solves for IR albedo in the W1, W2. The albedo determination in those bands is very strongly dependent on Kirchhoff's law – which is all about absorption and emission in those bands. Thus it is not surprising that previous studies did not deal with Kirchhoff's law, but it is very important to NEOWISE.

## 2.1 Kirchhoff's Law Applied to a Flat-Plate Model

I will start by considering a toy model: a flat, circular, Lambertian plate of diameter $D$ that both emits and reflects light in space. The plate is perpendicular to both the viewing direction and incident solar light, as illustrated in Figure 1. We can assume that the plate is held in thermal equilibrium at a constant temperature $T = T_\mathrm{p}$.

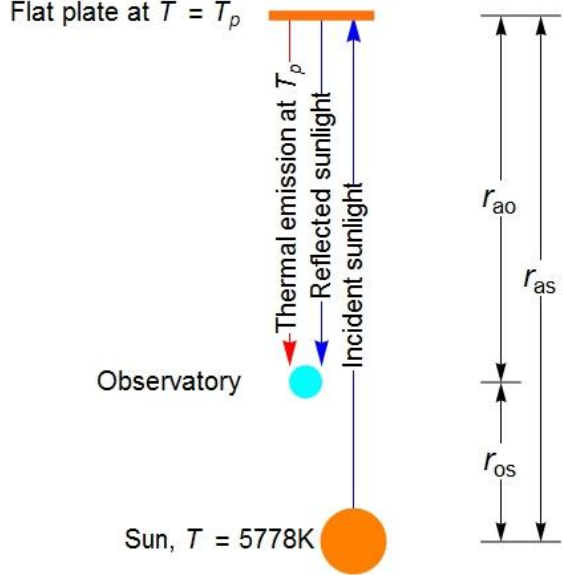

**Figure 1. Flat plate model of an asteroid with both reflection and thermal emission.** See text for details.

The plate has directional-hemispherical reflectivity $r_{\text{dh}}(\lambda)$ and spectral emissivity $\epsilon(\lambda)$, in the notation of Hapke (Hapke, 2012).

An observer will see both reflected solar radiation and emitted thermal IR. Assuming the asteroid is a perfect blackbody, the emitted thermal flux $F_{\text{thermal}}(\lambda)$ and incident solar flux $F_{\text{sun}}(\lambda)$ at a distance of 1AU from the sun are given by

$$F_{\text{thermal}}(\lambda) = \pi\, B_\nu(T_p, c/\lambda)$$
$$F_{\text{sun}}(\lambda) = \frac{\pi\, R_{\text{sun}}^2}{\text{AU}^2}\, B_\nu(5778\text{ K}, c/\lambda)$$
$$B_\nu(T, \nu) = \frac{2h\nu^3}{c^2 \left(e^{\frac{h\nu}{k_B T}} - 1\right)} \quad ,$$

(1)

where $B_\nu(T, \nu)$ is the Planck distribution, $R_{\text{sun}}$ is the radius of the sun, AU is the earth-sun distance and the effective blackbody temperature of the sun is assumed to be 5778 K. Note that fluxes are typically measured in Jy units, so the frequency form of the Planck distribution is used. Kirchhoff's law clearly holds in this case, so for a non-blackbody asteroid we have

$$r_{\text{dh}}(\lambda) = p(\lambda) = 1 - \epsilon(\lambda) \quad ,$$

(2)

and the total flux as observed can be then written as

$$F_{\text{obs}}(T_p, \lambda) = \frac{D^2}{4\,\text{AU}^2\, r_{\text{ao}}^2} \left( \epsilon(\lambda) F_{\text{thermal}}(T_p, \lambda) + \frac{(1 - \epsilon(\lambda))}{r_{\text{as}}^2} F_{\text{sun}}(\lambda) \right).$$

(3)

The emissivity is thus

$$\epsilon(\lambda) = \frac{D^2 F_{\text{sun}}(\lambda) - F_{\text{obs}}(\lambda)\, 4\, \text{AU}^2\, r_{\text{ao}}^2\, r_{\text{as}}^2}{D^2(F_{\text{sun}}(\lambda) - F_{\text{thermal}}(\lambda)\, r_{\text{as}}^2)} \quad . \tag{4}$$

Equation (4) develops a singularity when the reflected and emitted flux are equal and $F_{\text{sun}}(\lambda) - F_{\text{thermal}}(\lambda)\, r_{\text{as}}^2 = 0$. The reason, as can be seen in Equation (3), is that the total flux at a given wavelength $\lambda$ is effectively a linear combination of reflected and emitted light, with a linear parameter of $\epsilon(\lambda)$. If the two sources are equal in flux, then any value of $\epsilon(\lambda)$ gives the same answer, so one cannot solve for $\epsilon(\lambda)$ in Equation (4). Whether such equality of flux occurs within the range of $\lambda$ of interest depends on temperature $T_{\text{p}}$.

The flat-plate model removes the complexity of the various kinds of albedo and emissivity making it quite obvious that Kirchhoff's law must hold in this case.

NEATM is more complex; a Lambertian sphere in pointwise radiative equilibrium. The spherical geometry introduces some geometric factors, and we can use the empirical phase law for asteroids to obtain phase-angle factor and phase integral. Modulo these additional factors, however, the fundamental physics of the NEATM model is the same as that of the flat-plate model. Indeed, it is common to use a faceted model in which each facet is a flat plate, albeit one oriented at an angle to the sun and the observer.

## 2.2 NEATM Model

Instead of a flat plate, now consider an asteroid that is assumed to be spherical; $\theta$ and $\varphi$ are in conventional spherical coordinates. Figure 2 shows the observational geometry.

Assume that the absorption of radiation at a point is Lambertian and thus given by

$$\text{Absorbed}(\theta, \varphi) = \cos\Phi \int_0^\infty (1 - r_{\text{dh}}(\lambda)) F_{\text{sun}}(\lambda)\, d\lambda \; . \tag{5}$$

The quantity $r_{\text{dh}}(\lambda)$ is the directional-hemispherical reflectance in the notation of Hapke (Hapke, 2012); in general, it is a function of wavelength $\lambda$.

Define the weighted average of an arbitrary function $h(\lambda)$ over a spectral distribution $F(T, \lambda)$ as

$$\langle g(\lambda) | F(T, \lambda) \rangle = \int_0^\infty g(\lambda) F(T, \lambda)\, d\lambda \Big/ \int_0^\infty F(T, \lambda)\, d\lambda \quad . \tag{6}$$

I thank Bruce Hapke for suggesting this notation to me. The absorption Equation (5) can then be rewritten as

$$\text{Absorbed}(\theta, \varphi) = \frac{S}{r_{as}^2} (1 - \langle r_{dh}(\lambda) | F_{sun}(\lambda) \rangle) \cos \Phi \quad , \tag{7}$$

where the weighted average is over the Planck distribution $F(T, \lambda) = B(T, \lambda)$ and $S$ is the solar constant. The thermal radiation emitted at a point is then

$$\begin{aligned}\text{Emitted}(\theta, \varphi) &= \eta \int_0^\infty \epsilon(\lambda) B(T(\theta, \varphi), \lambda) \, d\lambda \\ &= \eta \left( \sigma T^4(\theta, \varphi) - \langle r_{dh}(\lambda) | B(T(\theta, \varphi), \lambda) \rangle \right) \quad ,\end{aligned} \tag{8}$$

where $\epsilon(\lambda)$ is the spectral emissivity and $\eta$ is a model-dependent parameter called the beaming parameter, discussed below.

The fundamental idea in NEATM (Harris, 1998) and before it, in the standard thermal model (Lebofsky and Spencer, 1989; Lebofsky et al., 1986; Veeder et al., 1989) is that radiative equilibrium occurs at each illuminated point on the asteroid. As a result, the relevant albedo (reflectivity) is $r_{dh}$; it is *not* the geometric albedo $p$. The relevant emissivity is the spectral emissivity at a point (again, using the terminology of Hapke—see discussion below), *not* the "bolometric emissivity". These distinctions are crucial because they imply that Kirchhoff's law must hold. At each point on the surface,

$$\text{Absorbed}(\theta, \varphi) = \text{Emitted}(\theta, \varphi) \quad , \tag{9}$$

which implies that

$$\frac{S}{r_{as}^2} (1 - \langle r_{dh}(\lambda) | B(5778K, \lambda) \rangle) \cos \Phi = \eta \left( \sigma T^4(\theta, \varphi) - \langle r_{dh}(\lambda) | B(T(\theta, \varphi), \lambda) \rangle \right) \quad . \tag{10}$$

Equation (10) is the fundamental expression of thermal equilibrium in the NEATM model. Note that the treatment above differs from the usual derivation because it is careful to use the correct terms for albedo and emissivity.

Equation (10) holds at every point $\theta, \varphi$ on the surface of the asteroid. However pointwise properties are not observed for most asteroids because they typically are not disk resolved. Instead, most derivations of the NEATM model use the following, rather poor, approximation:

$$A \approx A_v = p_v q \quad , \tag{11}$$

where $A$ is the Bond albedo

$$A = \langle r_{dh}(\lambda) | B(5778K, \lambda) \rangle \quad , \tag{12}$$

$A_v$ is the Bond albedo in the visible band, $p_v$ is the visible band geometric albedo, and $q$ is the empirically derived visible-band phase integral, typically obtained by using the HG asteroid phase function, parameterized by slope parameter $G$.

The reasoning behind the approximation of Equation (11) is:

> Since the peak of the Sun's spectral energy distribution occurs at visible wavelengths, the Bond albedo is customarily assumed to be equal to the total Bond albedo at V band (~0.56 µm), $A_v$. (Mainzer et al., 2015b),

While it is true that the peak occurs in the visible band, that by itself is insufficient to justify Equation (11) because the visible band contains a very small portion of the total incident solar energy (Figure 3).

The visible band, defined as 0.435 µm $\leq \lambda \leq$ 0.635 µm, contains only 25.4% of the incident solar energy. Even this overestimates the amount of solar energy returned from asteroids in the form of visible albedo, however. Visible-light geometric albedo is typically measured in a more realistic astronomical-filter passband, such as the Johnson-V band $s_{\mathrm{jv}}(\lambda)$. That filter passband captures a mere $\langle s_{\mathrm{jv}}(\lambda)|B(5778\mathrm{K},\lambda)\rangle = 0.113$, or 11.3%, of the total solar energy—such a small fraction of the total that Equation (11) is not guaranteed to be a good approximation of the true Bond albedo.

A second reason to avoid the simple approximation is that visible-band albedo does not measure energy; in modern observations, $A_v$ is based on an observational photon count within a given observational passband (i.e., a filter). Photon counts do not follow the distribution $B(T,\lambda)$ but instead the Planck photon count distribution

$$B(T,\lambda) = \frac{2hc^2}{\left(e^{\frac{hc}{k_\mathrm{B} T \lambda}} - 1\right)\lambda^5}, \quad N(T,\lambda) = \frac{2c^2}{\left(e^{\frac{hc}{k_\mathrm{B} T \lambda}} - 1\right)\lambda^4} \quad . \tag{13}$$

The distributions $B$ and $N$ are, in general, distinct and peak at different wavelengths. In this terminology, the visible geometric albedo $p_\mathrm{v}$ is effectively given by the weighted average over $F(T,\lambda) = s_{\mathrm{jv}}(\lambda)N(T,\lambda)$:

$$p_\mathrm{v} = \frac{1}{q}\langle r_{\mathrm{dh}}(\lambda)|s_{\mathrm{jv}}(\lambda)N(5778\mathrm{K},\lambda)\rangle \quad . \tag{14}$$

If asteroids were perfectly gray bodies, then $r_{\mathrm{dh}}(\lambda)$ would be a constant independent of wavelength $\lambda$, and the approximation (11) would be correct. We know this is not the case, as evidenced by a long line of research culminating in asteroid classification systems, the most recent of which is the Bus-DeMeo classification (DeMeo and Carry, 2013; DeMeo et al., 2009), which has shown that asteroids can have quite different spectral characteristics in the visible and near-IR. The near-IR variations are particularly important because it contains the majority of the incident solar flux (Figure 3).

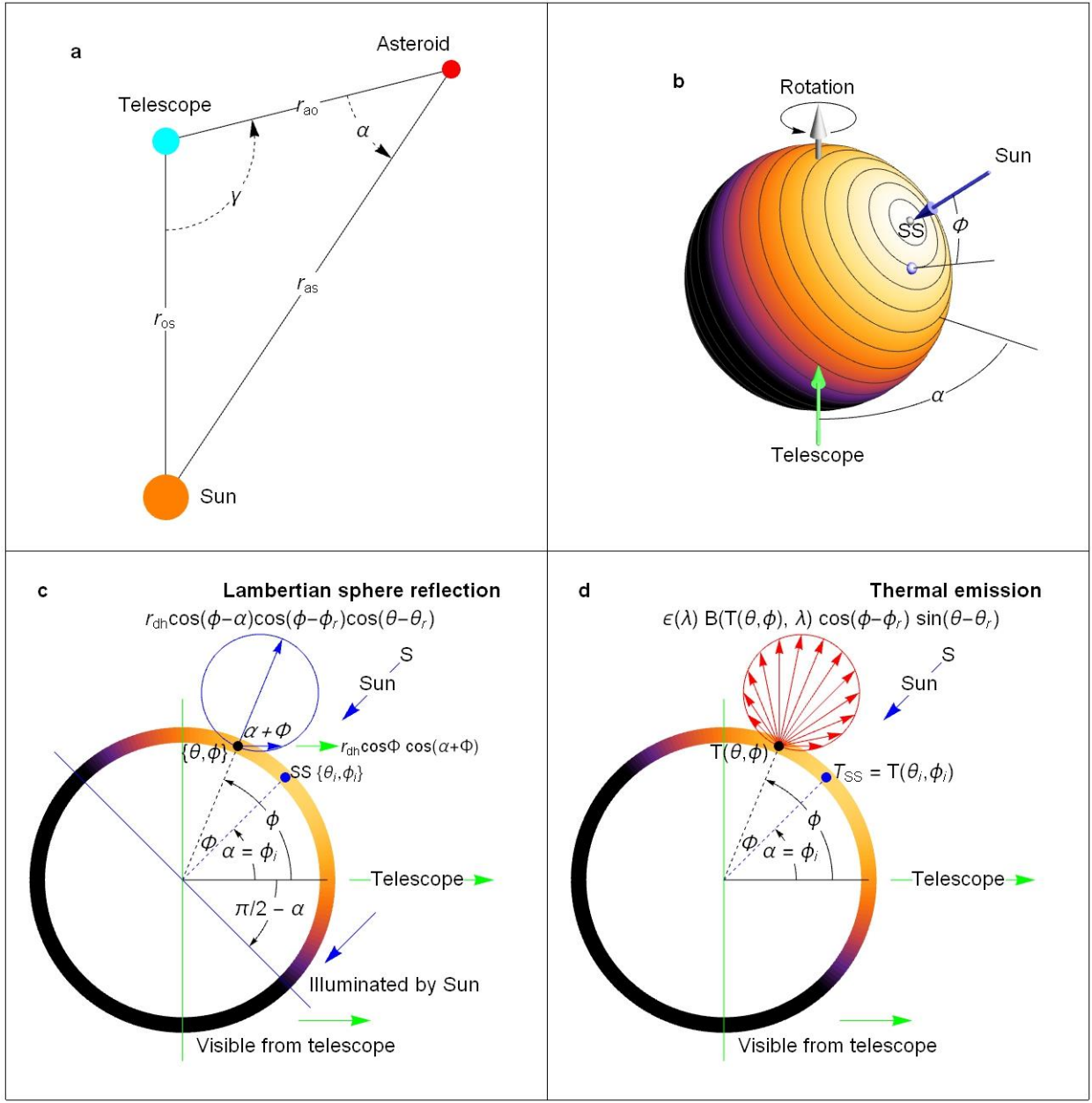

**Figure 2. Reflection and emission from a spherical asteroid in the NEATM model.** Expressions here assume an asteroid positioned as shown in (a) and having physical properties, including directional-hemispherical reflectivity $r_{dh}(\lambda)$ and emissivity $\epsilon(\lambda)$, that are constant across the surface of the asteroid. Panel (b) shows the geometry of the subsolar point SS, observed at phase angle $\alpha$. All points (like the black example point) that are at the same central angle $\Phi$ from the subsolar point have the same absorbed solar energy and thus the same temperature in the NEATM model, as indicated by shading and contour lines. Panels (c) and (d) show a cross section through the asteroid and the expressions used to estimate Lambertian reflection/absorption and thermal emission.

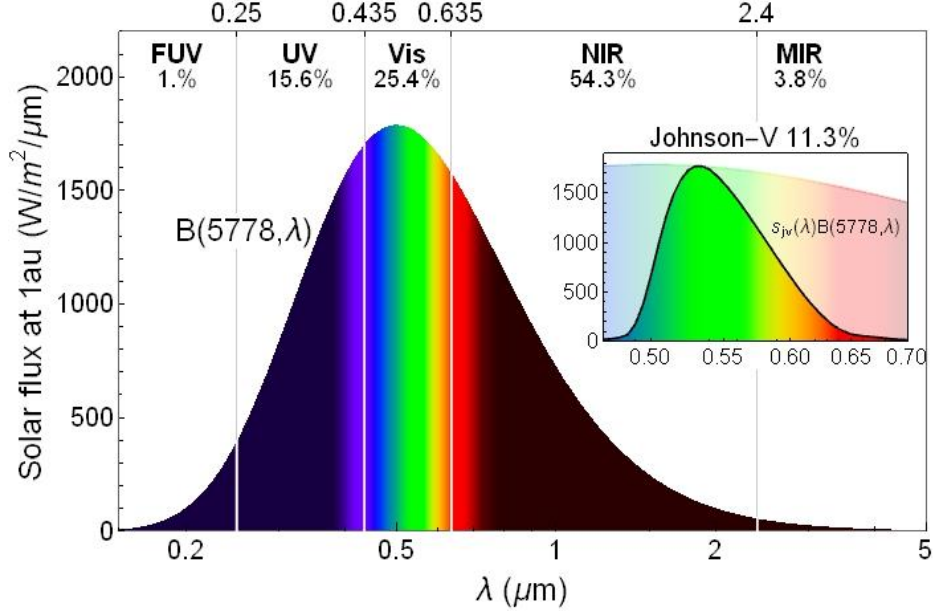

**Figure 3. Distribution of incident solar energy in different wavelength passbands.** The vertical lines define simple, idealized passbands for far-ultraviolet, ultraviolet, visible, near-infrared, and mid-infrared radiation. The solar spectrum is modeled as a Planck distribution at $T = 5778$ K, and the percentages indicate the integrated fraction of the flux in each passband. The inset chart shows an example of a more realistic observational passband, the Johnson-V band, superimposed on the same Planck distribution. Only 11.3% of the total incident solar energy falls within the Johnson-V band.

The right side of Equation (10) includes a weighted average of $r_{\mathrm{dh}}(\lambda)$ over the Planck distribution at the temperature of the point on the asteroid $T(\theta, \varphi)$. This temperature will generally be a factor of 10 or more lower than the effective blackbody temperature of the sun. Most NEATM derivations approximate the right side of Equation (10) inadequately, by using

$$0.9 \, \sigma T^4 \approx \epsilon_{\mathrm{B}} 0.9 \, \sigma T^4 \approx \epsilon_{\mathrm{B}} \sigma T^4 \approx \sigma \, T^4(\theta, \varphi) \epsilon_{\mathrm{NEATM}} \quad , \tag{15}$$

Where $\epsilon_B$ is sometimes called the "bolometric emissivity" (Emery et al., 2014, 2006; Hanuš et al., 2015; Lagoa et al., 2011; Lim et al., 2011; Lindsay et al., 2015; Marchis et al., 2012; Mueller, 2012).

That terminology is misleading because the term "bolometric" is typically used in astronomy to mean a value taken across all wavelengths. While $1 - r_{\mathrm{dh}}(\lambda)$ could be properly called "bolometric emissivity," its weighted average by the Planck distribution at a specific temperature cannot be. The latter effectively measures $r_{\mathrm{dh}}(\lambda)$ over a relatively narrow range of wavelengths around the peak of the Planck distribution. This misleading terminology encourages the misconception that $\epsilon_\mathrm{B}$ is somehow a different type of emissivity. This choice of terms may be one origin of the mistaken idea that the emissivity does not conform to Kirchhoff's law.

The correct calculation is the weighted average of $r_{\mathrm{dh}}(\lambda)$ over the Planck distribution of the surface temperatures $T(\theta, \varphi)$

$$\epsilon_{\text{NEATM}} = 1 - \left\langle r_{\text{dh}}(\lambda) \middle| \int_{-\pi/2}^{\pi/2} \int_0^{\pi} B(T(\theta,\varphi),\lambda) \sin^2\theta \cos\varphi \, d\theta d\varphi \right\rangle \quad . \tag{16}$$

Although virtually all treatments of NEATM make use of the approximation in Equation (15), few empirical studies have attempted to verify its accuracy, in part because doing so requires knowledge of $r_{\text{dh}}(\lambda)$. One way to test the approximation is to analyze laboratory spectra from meteorite specimens or from lunar or terrestrial minerals that have been widely used as proxies for asteroids.

I gathered 325 such spectra ranging from 0.4 to 50 $\mu$m (although not all spectra cover the full range) from (Baldridge et al., 2009; Byrnes et al., 2007; Cahill et al., 2010; Cloutis and Gaffey, 1993; Cloutis et al., 2012c, 2012d, 2012e, 2012f, 2011a, 2011b, 2010; Cloutis, 1997; Cloutis et al., 2015, 2013a, 2013b, 2012a, 2012b; Dunn et al., 2013; Emery et al., 2006; Gritsevich et al., 2012; Hanna and Sprague, 2009; Hiroi et al., 2010; Izawa et al., 2010; Krause et al., 2011; Lane et al., 2011; Lim et al., 2011; Morlok et al., 2010; Moskovitz et al., 2010; Ostrowski et al., 2011; Paul et al., 2014; Ramsey and Christensen, 1998; Reddy et al., 2012a; Rice et al., 2013; Sanchez et al., 2014; Thomson and Salisbury, 1993; Trigo-Rodríguez et al., 2013; Vernazza et al., 2010)

Using this data set, I calculated the approximation as well as the exact value (within the context of the model) by using Equations (11) and (15). The results are shown in Figure 4.

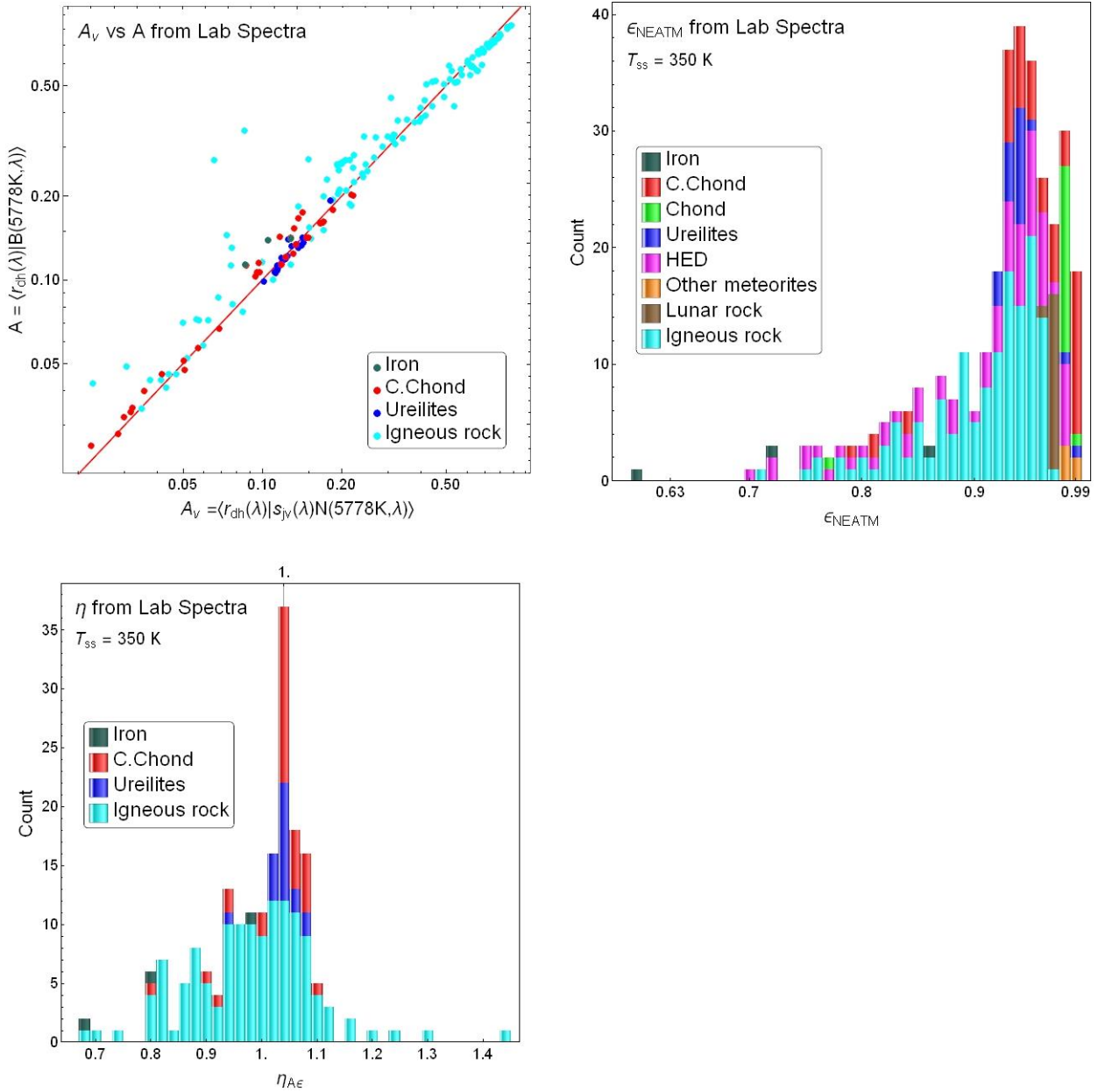

**Figure 4. Comparison of typical NEATM approximations to $A$, $\epsilon$, and $\eta$ from laboratory spectra.** The quantities and their approximations were calculated for hypothetical asteroids by using laboratory spectra for meteorites and lunar and terrestrial rock samples thought to be possible asteroid compositional proxies. *Top left:* The approximated albedo $A_v$ from Equation (11) versus the Bond albedo $A$ from Equation (12). *Top right:* A histogram of the weighted average emissivity $\epsilon_{\text{NEATM}}$ obtained from Equation (16), assuming subsolar temperature $T_{ss} = 350$ K. The NEATM approximation of $\epsilon = 0.9$ is shown as a vertical line. The distribution of the beaming parameter $\eta_{A\epsilon}$ resulting from material properties alone altering the estimates in equations (11) and (15) is shown *bottom left*.

Figure 4 demonstrates that while the approximation of Equation (11) works reasonably well for iron, carbonaceous chondrites, and ureilites, it deviates substantially from the exact Bond albedo for some igneous minerals. For this particular sample of materials, $0.64 \leq \epsilon_{\text{NEATM}} \leq 1$. The approximations are clearly imperfect.

Figure 4 is only illustrative of how the approximation may fail to account for the properties of real materials. Surface properties such as porosity and space weathering may modify the source material spectra (Gaffey, 2010; Izawa et al., 2015; Vernazza et al., 2012, 2010). The set of laboratory spectra studied here do not comprise a full portfolio of asteroid compositional proxies—it contains, for example, no ices, which may be present in many asteroids of the outer solar system—and the samples in the proxy set may not be representative of all relevant asteroid families.

The usual derivation of NEATM (Harris and Lagerros, 2002; Harris, 1998; Kim et al., 2003) nevertheless uses approximations (11) and (15) to obtain a temperature distribution across the asteroid:

$$T_{\text{NEATM}}(\theta, \varphi, r_{\text{as}}) = T_{\text{ss}}(r_{\text{as}})(\sin \theta \cos \phi)^{0.25} \quad , \tag{17}$$

where the maximum temperature across the surface is the subsolar temperature given by

$$T_{\text{ss}}(r_{\text{as}}) = \left(\frac{S(1 - p_v q)}{0.9 \sigma \eta \, r_{\text{as}}^2}\right)^{0.25}. \tag{18}$$

The parameter $\eta$ was originally introduced to account for infrared radiation that is "beamed" into space, and is called the "beaming parameter" as a result. This is often connected in accounts of the NEATM model as being due to surface features such as craters or surface roughness that cause a strong opposition effect, as in this quote:

> This model uses a beaming parameter, η, to account for the tendency of surface features such as craters to "beam" radiation back to the observer at low phase angles. (Mainzer et al., 2012a)

Other accounts emphasize the role of $\eta$ in thermal inertia.

> The NEATM uses the so-called beaming parameter η to account for cases intermediate between zero thermal inertia (the Standard Thermal Model, or STM; Lebofsky et al. 1978) and high thermal inertia (the Fast Rotating Model, or FRM; Veeder et al. 1989; Lebofsky & Spencer 1989). (Grav et al., 2011b)

In the case of $\eta > 1$, either the surface "beaming" effect or the thermal inertia effect are valid interpretations, each of which emphasizes a different physical effect. In reality it is the combination of those effects that contribute to $\eta$. In both cases, when $\eta > 1$, the subsolar temperature $T_{\text{ss}}$ is lower than it would be with perfect energy conservation, because some of the energy is lost—either beamed into space or carried to an unobserved portion of the asteroid by conduction, rotation, and thermal inertia effects.

However, many studies—including the original account by Lebofsky for the predecessor STM model—find that $\eta < 1$ for some asteroids (Lebofsky and Spencer, 1989; Lebofsky et al., 1986). A value of $\eta$ lower than unity means that the asteroid has a *higher* temperature and emits *more* energy than it absorbs. In such cases, neither beaming nor thermal inertia are valid physical interpretations.

The explanation is that approximations (11) and (15) can underestimate the true absorbed energy, in which case $\eta < 1$ restores it to the correct value. It is also possible to have $\eta > 1$ without any thermal inertia or beaming effects, if the approximations overestimate the true absorbed energy. The NEATM model has achieved its great success precisely because $\eta$ acts to "calibrate out" the bad estimates, whether high or low. The role of $\eta$ is to enforce conservation of energy and quantitatively account for unobserved energy due to errors in several areas: estimates of absorption and reflection, surface effects and thermal inertia.

It is thus useful to think of $\eta$ as being composed of three factors: $\eta_{A\epsilon}$, which accounts for bringing the $A$ and $\epsilon$ approximations into line, $\eta_{\text{surf}}$ to account for the effects of surface porosity and topography and any "beaming" they may produce, and $\eta_{\text{in}}$ to account for effects due to rotation and thermal inertia. Thus

$$\eta = \eta_{A\epsilon}\, \eta_{\text{surf}}\, \eta_{\text{in}} \qquad (19)$$

Figure 4c illustrates that hypothetical asteroids with the laboratory spectra can have $0.64 \leq \eta_{A\epsilon} \leq 1.42$ on a smooth asteroid even without considering either surface roughness or thermal inertia.

Equation (19) and Figure 4 have strong implications for the practice of assigning values of $\eta$ based on the overall asteroid group. This is used by the NEOWISE papers in cases where there is insufficient data for thermal modeling e.g.: $\eta = 1.0$ for main belt asteroids (Masiero et al., 2011), $\eta = 0.77$ for Hildas (Grav et al., 2011a), $\eta = 1.4$ for NEO (Mainzer et al., 2011b). Equation 19 and the examples in Figure 4 show that $\eta_{A\epsilon}$ can have a strong material dependency. As a result we would expect that $\eta_{A\epsilon}$ or $\eta_{\text{in}}$ would not be a single value across large scale asteroid groups such as the main belt which are known to contain multiple asteroid families on either a dynamical, or spectral class.

Note that there are no fundamental limits on the value of $\eta$. On the low side, the limit depends on how an estimate comes from approximations (11) and (15). On the high side, the Fast Rotating Model (FRM) derives $\eta_{\text{in}} = \pi$ purely on the grounds of thermal inertia. That is not an upper bound on $\eta$, despite claims to the contrary in the NEOWISE papers (Tables 1, 3), or even on $\eta_{\text{in}}$. As shown in Equation (19), the thermal inertia effect is just one multiplicative factor alongside others. An asteroid with high thermal inertia which tumbles sufficiently fast on more than one axis has $\eta_{\text{in}} = 4$ (Myhrvold, 2016).

As a multipurpose calibration constant, $\eta$ is conceptually very useful. But using $\eta$ as a true free parameter in least-squares regression can be problematic if another parameter within Equation (18), such as $p_v$ (or expression for $p_v$ based on $D$ and $H$ via Equation (5)) is simultaneously being fit. In effect, the parameters act to cancel one another out and increase the likelihood that numerical algorithms for finding the best fit become trapped in a local minimum or fail due to other numerical problems.

To avoid such problems, one can reparametrize Equation (18) by introducing a temperature $T_1$, which has the physical interpretation of $T_{ss}$ at $r_{as} = 1$, thus

$$T_{\text{ss}}(r_{\text{as}}) = \frac{T_1}{\sqrt{r_{\text{as}}}}, T_1 = \left(\frac{S(1-p_{\text{v}}q)}{0.9\,\sigma\,\eta}\right)^{0.25}. \tag{20}$$

Fitting on $T_1$ eliminates $\eta$ from the calculation. The flux of emitted light at the surface of the asteroid then becomes

$$F_{\text{NEATM}}(T_1,\alpha,\lambda) = \int_{\alpha-\pi/2}^{\alpha+\pi/2}\int_0^{\pi} B_\nu\left(\frac{T_1}{\sqrt{r_{\text{as}}}}\max(0,\sin\theta\cos\phi)^{0.25},\frac{c}{\lambda}\right)\sin^2\theta\cos(\alpha-\varphi)d\theta\,d\varphi \quad . \tag{21}$$

The integrals are a sum of Planck distributions and can be easily evaluated using standard numerical integration techniques. Many NEATM implementations, including NEOWISE papers, use faceted, high face count polyhedra to approximate the spherical asteroid as a series of flat polygons. This can work well if the face count is high enough. However it is both simpler, and more accurate to use automatic adaptive numerical integration routines, as are typically provided in most numerical software libraries. These algorithms automatically create sampling intervals for the function and assures the maximum numerical accuracy. Other implementation details are discussed below.

Although insufficient details are presented to know for sure, the NEOWISE papers appear to use Equation (18) with $p_{\text{v}}$ expressed in terms of $D$ and $H$ from (5). This is unnecessary—the effect of $\eta$ compensates for any change in $p_{\text{v}}$.

Note that one implication of (20) and (21) is that the absolute magnitude $H$ is not involved in the regression. This is important because in some contexts we do not have good optical observations of an asteroid but do have WISE/NEOWISE and/or other IR observations (for example for asteroids that were newly discovered by WISE/NEOWISE and may not have optical follow-up). It is also important because the values of $H$ in standard databases such as the Minor Planet Center, tend to change whenever new survey data is deposited, as discussed in (Pravec et al., 2012). Numerically speaking, $H$ is problematic because it occurs in the exponent of expressions for $p_{\text{v}}$, while other variables do not. This is exacerbated if $H$ is used as fitting parameter, as done in the NEOWISE work.

Recasting the NEATM equations using $T_1$ reveals an important point – neither $H$ nor $p_{\text{v}}$ nor $\eta$ are fundamental to the thermal modeling of asteroids. They are each important parameters in their own right, but the essence of the thermal model is fitting the observed asteroid spectrum (via discrete band observations) to the model spectrum, and that only requires $T_1$.

Conversely, given $T_1$ and $H$ one can solve for $p_{\text{v}}$ via Equation (5) and $\eta$ from Equation (20). In previous thermal modeling papers $\eta$ has been used primarily a diagnostic parameter, to allow comparison of different approaches. However it also has a physical interpretation in terms of Equation (19), and this has been proposed to help identify metallic asteroids (Harris and Drube, 2014), for example. Use of $T_1$ as the primary fitting parameter in no way diminishes the use of $\eta$ for such purposes. Indeed exploitation of the multi-factorial origin of $\eta$ as expressed in Equation (19) and Figure 4 may provide further insights into physical interpretation of $\eta$.

## 2.3 An Extended FRM Model

Although the NEATM model has been very popular and has great utility, other asteroid thermal models have been used as well. The Fast Rotating Model, also known as the Isothermal Latitude Model (Harris and Lagerros, 2002; Kim et al., 2003; Lebofsky et al., 1986, 1978; Usui et al., 2011a, 2011b; Veeder et al., 1989) assumes that an asteroid has a high enough thermal inertia and a fast enough rate of rotation that the thermal equilibrium condition of Equation (9) holds on average value along each line of constant latitude. The temperature distribution then simplifies to

$$T_{\text{FRM}}(\theta, \varphi, r_{\text{as}}) = \frac{T_1}{\sqrt{r_{\text{as}}}} (\sin \theta)^{0.25} \quad . \tag{22}$$

The distribution of flux with wavelength $\lambda$ is different than that predicted by NEATM in Equation (18) and does not depend on the phase angle $\alpha$.

FRM was derived around the same time as the STM, which had a fixed value of $\eta$. Averaging around a constant latitude in effect similarly fixes $\eta = \pi$. Subsequently, Harris introduced the concept of treating $\eta$ as a free variable (Harris, 1998). Although FRM has not been updated to accommodate a variable $\eta$, Equation (19) makes it clear that it could be, simply by fixing $\eta_{\text{in}} = \pi$, but leaving $\eta_{A\epsilon}$ and $\eta_{\text{surf}}$ free as fitting parameters. In practice, this is readily achieved by using $T_1$ as a fitting parameter, as in Equation (20). To distinguish the variable-$\eta$ form of FRM from the conventional, fixed-$\eta$ FRM, I refer to the former as the Extended Fast-Rotating Model (EFRM).

Like conventional FRM, EFRM assumes that the asteroid's axis of rotation is always perpendicular to the plane between the Sun and the observer. Although it is possible to generalize FRM (and EFRM) to arbitrary axes as GFRM (Myhrvold, 2016) I have not done so in this study.

The NESTM model (Wolters and Green, 2009) combines NEATM and FRM in an attempt to better model the unilluminated side of the asteroid, which may improve model fitting at high phase angles. It is not considered here because it requires an extra parameter, which blends the effect of NEATM and FRM. This study explores diverse models compared to NEOWISE, but they all have the same number of free parameters.

## 2.4 Isothermal Blackbody Model

The simplest temperature distribution an asteroid could attain is that of a blackbody having constant temperature. Such an assumption could hold for an asteroid that tumbles on multiple axes, as a multi-axis limit of the FRM as shown in (Myhrvold, 2016). An asteroid for which $\eta = 1$ when held still would tend toward $\eta = 4$ in the limit of high thermal inertia and fast tumbling on multiple axes. In that case, the constant temperature across the asteroid is given by

$$T_{\text{BB}}(\theta, \varphi, r_{\text{as}}) = \frac{T_1}{\sqrt{r_{\text{as}}}} \quad . \tag{23}$$

Apart from the tumbling case, by fitting to a simple thermal spectrum at constant temperature we can construct a very simple model that serves as a point of comparison to reveal, by way of statistical model comparison, under what circumstances models such as NEATM and FRM are actually warranted.

## 2.5 HG-NEATM

The dependence of the NEATM model given by Equation (21) on phase angle is essentially that of a Lambertian sphere. It is a widely used approximation, but we know that visible light opposition effects make low phase angles much brighter and large phase angles much dimmer than a Lambertian model supports. The HG empirical phase model, and many other empirical models, were created to address this deficiency.

The astronomical community does not yet possess sufficient IR light curves to establish the IR-band equivalent of the HG model. Nor do we yet understand how that relates to the visible-light phase curve. Work on this problem is ongoing (Davidsson and Rickman, 2014; Davidsson et al., 2015; Hanuš et al., 2015; Honda et al., 2007; Reddy et al., 2012b; Rozitis and Green, 2011; Sanchez et al., 2012; Vernazza et al., 2010; Wright, 2007). For the present study, it would be useful to compare NEATM to a different model.

The EFRM and isothermal blackbody models both lack phase-angle dependence. A reasonable ansatz is to adopt the HG phase function for IR emission to obtain an HG-NEATM model:

$$F_{\text{HG-NEATM}}(T_1, \alpha, \lambda) = \frac{\psi_{\text{HG}}(\alpha, G)}{q} \int_{-\pi/2}^{\pi/2} \int_0^\pi B_\nu\left(\frac{T_1}{\sqrt{r_{\text{as}}}} \max(0, \sin\theta \cos\phi)^{0.25}, \frac{c}{\lambda}\right) \sin^2\theta \cos\varphi \, d\theta \, d\varphi. \quad (24)$$

We further assume that the same value of the slope parameter $G$ obtained from visible light observation (or assumed to the default value of $G = 0.15$) can be used for emitted thermal radiation.

One of the effects responsible for the phase-angle dependence is shadow hiding: at opposition, one sees all illuminated surfaces, but when phase angle increases, the asteroid dims as shadows cast by surface features reduce the overall intensity. Equation (24) posits, in effect, that IR wavelengths are emitted only from illuminated surfaces; shadows are treated in the same way that NEATM treats unilluminated portions of the asteroid. NEATM is the low thermal inertia limit of instantaneous equilibrium (as in Equation (9)), and thus it seems appropriate to use the same phase function.

A realistic asteroid that has finite thermal inertia will deviate from this (i.e. $\eta_{\text{in}}$ in Equation (19)), but using the same phase function and parameter for optical and IR seems like a reasonable starting point. One could go further and either adopt a different phase function, or fit a different slope parameter $G_{\text{IR}}$ for thermal emission, but that will be left for a future study.

Note, however, that this is not intended to be a full derivation— it is merely an intuitive motivation to create a point of comparison. That comparison is shown graphically in Figure 5, which illustrates the spectra of the different models and their behavior over a range of phase angles.

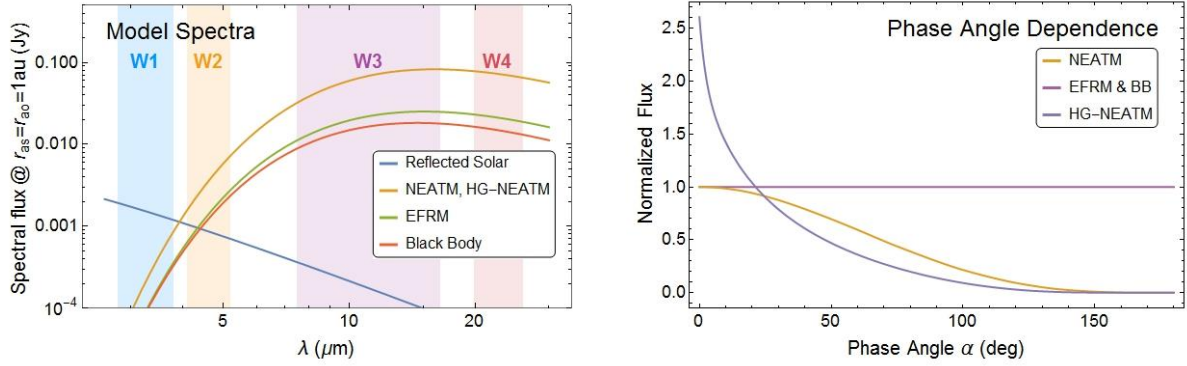

**Figure 5. Comparison of spectral flux and phase-angle dependence across several asteroid models.** The left panel shows spectra for $T_{ss} = 350$ K and $\alpha = 0$ as predicted by three models, along with the spectrum of reflected sunlight. Note that HG-NEATM has the same spectrum as NEATM. The right panel shows flux as a function of phase angle $\alpha$, in the case of HG-NEATM the curve is shown for $G = 0.15$.

## 2.6 Adding Reflected Sunlight

Asteroid thermal models were originally developed using observations that consisted almost exclusively of thermal radiation. WISE observations, however, capture a considerable amount of reflected solar radiation, which must be accounted for in the analysis.

The solar flux incident on an asteroid at 1AU from the sun is

$$F_{\text{Sun}}(\lambda) = \frac{\pi R_{\text{sun}}^2}{\text{AU}^2} B_\nu\left(5778\,\text{K}, \frac{c}{\lambda}\right) \quad . \tag{25}$$

Where AU is the earth-sun distance. Note that $S = \int_0^\infty F_{\text{Sun}}(\lambda)\,d\lambda$, where $S = 1360.8\ w/m^2$ is the solar constant. The total observed flux at a distance $r_{\text{ao}}$ from an asteroid of diameter $D$ is then

$$F_{\text{obs}}(\lambda, \alpha, r_{\text{as}}, r_{ao}) = \frac{D^2}{4\,\text{AU}^2\,r_{\text{ao}}^2}\left(\epsilon(\lambda)\,F_{\text{model}}\left(\frac{T_1}{\sqrt{r_{\text{as}}}}, \alpha, \lambda\right) + p(\lambda)\frac{\psi_{\text{HG}}(\alpha, G)}{\text{AU}^2\,r_{\text{as}}^2}F_{\text{Sun}}(\lambda)\right) \tag{26}$$

$$p(\lambda) = \frac{(1 - \epsilon(\lambda))}{q} = \frac{r_{\text{dh}}(\lambda)}{q} \tag{27}$$

$$\epsilon(\lambda) = 0.9,\ p(\lambda) = p_{\text{IR}}\,. \tag{28}$$

Where the distances $r_{\text{as}}$ and $r_{\text{ao}}$ are in units of AU. The equations above assume that Kirchhoff's law holds and incorporates the assumption used in the NEOWISE papers that the same phase function $\psi_{\text{HG}}$ can be used for IR as well as for visible light. The relationship between IR and visible phase functions is still being investigated; future work may be able to improve this flux model further.

Equation (27) is the expression of Kirchhoff's law in this context and (28) is the form used by the NEOWISE papers, which violates it.

Some applications aim to determine emissivity from a set of spectral observations $F_{obs}$, which can be done using

$$\epsilon(\lambda) = \frac{D^2 \psi_{HG}(\alpha, G) F_{Sun}(\lambda) - 4 AU^2 q\, r_{as}^2\, r_{ao}^2\, F_{obs}(\lambda)}{D^2 \left( \psi_{HG}(\alpha, G) F_{Sun}(\lambda) - q\, r_{as}^2 F_{model}\left(\frac{T_1}{\sqrt{r_{as}}}, \alpha, \lambda\right) \right)} . \tag{29}$$

Equation (29) is the equivalent of (4) for the flat-plate model and has many of the same features, including the fact that $\epsilon(\lambda)$ can become undefined in cases that drive the denominator to zero.

## 2.7 Converting Modeled Flux to WISE Magnitudes

Flux obtained from Equation (26) must be converted into WISE magnitudes in order to be used to fit a thermal model. The magnitudes are given by the WES (Cutri et al., 2011) are

$$Wi = -2.5 \log_{10} \frac{f_i}{z_i}$$
$$z = (306.682, 170.663, 31.3684, 7.9525)\text{ Jy} \quad , \tag{30}$$

where $f_i$ is the total flux in Jy in band $i$. Wright (Wright, 2013) developed a remarkably simple set of quadrature formulas for the integrated WISE flux $f_i$ for an arbitrary function of $V(\lambda)$:

$$\begin{aligned} f_1 &= 0.5117\, V(3.0974\text{ μm}) + 0.4795\, V(3.6298\text{ μm}) \\ f_2 &= 0.5811\, V(4.3371\text{ μm}) + 0.4104\, V(4.9871\text{ μm}) \\ f_3 &= 0.1785\, V(8.0145\text{ μm}) + 0.4920\, V(11.495\text{ μm}) + 0.2455\, V(15.256\text{ μm}) \\ f_4 &= 0.7156\, V(21.15\text{ μm}) + 0.2753\, V(24.69\text{ μm}) \quad . \end{aligned} \tag{31}$$

Equation (31) allows us to estimate reflected solar light from $V(\lambda) = F_{Sun}(\lambda)$,

$$f_{Sun} = (6.52 \times 10^{13}, 3.86 \times 10^{13}, 7.22 \times 10^{12}, 2.10 \times 10^{12})\text{ Jy} \quad . \tag{32}$$

Equations (30) to (32) yield the following function for adjusted WISE magnitudes:

$$\widehat{Wi} = -5ld + lz_i + 2.5\log_{10}4 - 2.5\log_{10}\left(\vec{\epsilon}_i f_{\text{model }i}\left(\frac{T_1}{\sqrt{r_{as}}}, \alpha\right) + (1-\vec{\epsilon}_i)\frac{\psi_{\text{HG}}(\alpha, G)f_{\text{Sun}_i}}{q\,\text{AU}^2 r_{as}^2}\right) \quad (33)$$

$$lz = (6.2172, 5.58035, 3.74123, 2.25126) \quad,$$

where $ld = \log_{10} D$ is used instead of $D$ as a fitting variable, and $f_{\text{model}}$ is the result of taking model fluxes such as $V(\lambda) = F_{\text{NEATM}}(T_1, \alpha, \lambda)$ and applying Equation (31). An efficient numerical implementation is to collect a grid of values of $f_{\text{model}_i}$ as a function of $T_{ss}$ and $\alpha$ (I use ~10,000 points for each band) so that $f_{\text{model}_i}(T_{ss}, \alpha)$ can be implemented as a table lookup with linear interpolation between table entries. Equation (33) can then be used for model fitting.

Five different parameterizations are possible for the vector of emissivities $\vec{\epsilon}$. Each represents a different assumption; the one that fits best, according to objective model selection criterion such as AIC$_c$, should be chosen. I discuss the choice of 0.9 in the next section.

$$\vec{\epsilon} = \begin{cases} (0.9 & 0.9 & 0.9 & 0.9) \\ (\epsilon_1 & 0.9 & 0.9 & 0.9) \\ (0.9 & \epsilon_2 & 0.9 & 0.9) \\ (\epsilon_{12} & \epsilon_{12} & 0.9 & 0.9) \\ (\epsilon_1 & \epsilon_2 & 0.9 & 0.9) \end{cases} \quad (34)$$

## 2.8 Indeterminacy and the Limitations of Asteroid Thermal Modeling

One way to determine the diameter of an asteroid is to relate its flux in the visible band to its diameter and albedo by using Equation (5). It is not enough to know the flux (and therefore $H$) because the estimated diameter $D$ depends on the albedo $p_v$. Knowledge of $H$ alone is thus indeterminate: any given value of $H$ simply describes a curve $D^2 p_v = $ constant from Equation (5). Since $p_v$ is known to span a wide range of values, this approach has poor specificity for $D$.

It is widely believed that thermal modeling of asteroids does not have the same indeterminacy problems. That is only partially true; indeterminacy is still present. Consider Equation (33) with the vector of emissivities $\vec{\epsilon} = (\epsilon_1, \epsilon_2, \epsilon_3, \epsilon_4)$. We can plot surfaces of constant WISE magnitude $\widehat{Wi}$, as shown in Figure 6.

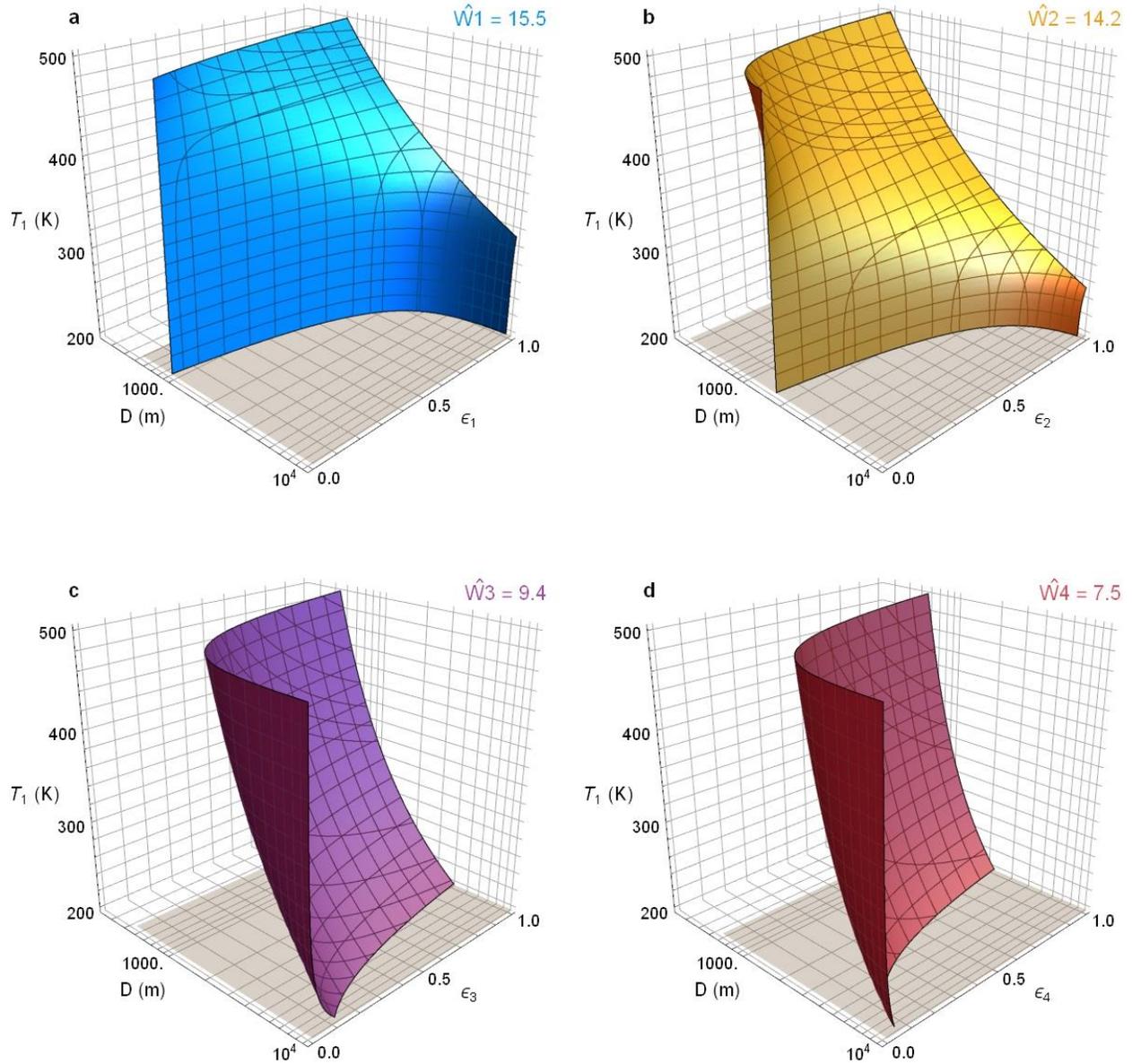

**Figure 6. Surfaces of constant WISE magnitude.** The WISE magnitude is invariant across the surface shown in each panel, which plots Equation (33). Numerical values for the $\widehat{W}\iota$ were taken from WISE observations of asteroid 152941 at $r_{as} = 1.30$, $\alpha = 50°$. Each band is indeterminate (i.e., the surface is not a single point) if emissivity is allowed to vary. The range of variation in $D$ for $0 \leq \epsilon_i \leq 1$ is shown by the shaded rectangle on the bottom of each graph. For any given temperature $T_1$, the range (shaded area at bottom of plot) is smaller for the predominately thermal bands $i = 3, 4$ (panels c and d) than it is for bands $i = 1, 2$ (panels a and b), which include a large reflected-light component.

Each band is indeterminate if it has its own value of emissivity $\epsilon_i$. As a result, it is not possible to fit emissivity in all four bands. The W3, W4 bands are dominated by thermal emission in this example, but if we were to allow separate emissivities $\epsilon_3, \epsilon_4$, the variability in the emissivity would mask the spectral signal from temperature.

As shown in Figure 6, the degrees of indeterminacy due to variation in emissivity for the thermally dominated W3 and W4 bands are almost an order of magnitude smaller (in terms of diameter) than they are for the W1 and W2 bands, which include substantial contributions from reflected sunlight. Nevertheless, some degree of indeterminacy still occurs if emissivity is treated as a free variable. Indeterminacy with diameter is resolved only if $\epsilon_3 \equiv$ constant, $\epsilon_4 \equiv$ constant. Note that these constants need not be the same.

The NEOWISE papers use $\epsilon_3 = \epsilon_4 = 0.9$, a choice that is common to most NEATM modeling efforts. It is important to recognize, however, that this choice is rather arbitrary. In particular, it has nothing to do with the choice of $\epsilon_B = 0.9$ in Equation (15). Counterintuitively, the impact of the free parameter $\eta$ ensures that the value of $\epsilon_B$ is actually irrelevant—no matter what (finite) value one uses for $\epsilon_B$ in (15), $\eta$ compensates for it. Hence, we are free to choose any value for $\epsilon_3, \epsilon_4$, including $\epsilon_3 \neq \epsilon_4$, as long as each value is constant.

In principle, we can calculate the appropriate values of $\epsilon_3, \epsilon_4$ by using Equation (16) and

$$\epsilon_3 = \langle \epsilon_{\text{NEATM}}(\lambda, T_1, r_{\text{as}}) | w_3(\lambda) \rangle \quad , \tag{35}$$

Where $w_3(\lambda)$ defines the WISE filter passband for band W3; $\epsilon_4$ is similar.

The left panel of Figure 7 shows how well $\epsilon_3 = \epsilon_4 = 0.9$ works in practice, when compared to values derived using the laboratory spectra for proxies of asteroid composition. It suggests that the choice $\epsilon_3 = \epsilon_4 = 0.9$ may be reasonable for some asteroids, but it does not work as well for others.

Like Figure 4, Figure 7 is intended merely to illustrate a general issue: the assumption that $\epsilon_3 = \epsilon_4 = 0.9$ is not necessarily the only possibility. The laboratory spectra used here do not span the full range of possibilities for asteroid surface material. In particular, they omit ices, and their coverage of rocky asteroids may be neither representative nor comprehensive. Surface properties and space weathering may also have an impact. Some laboratory spectra attempt to account for such factors to some degree, but it is unclear to what degree those adjustments are successful.

We thus face fundamental limitations in using thermal modeling to determine the diameter and other physical properties of asteroids. If we do not know the composition of the objects, we cannot know $\epsilon_3, \epsilon_4$, and that in turn limits estimation of other properties such as diameter $D$. In this study, lacking any better choice, I use $\epsilon_3 = \epsilon_4 = 0.9$, the same value used in the NEOWISE papers.

Choosing fixed values of $\epsilon_3, \epsilon_4$ still leaves us with a potential indeterminacy issue of a different kind. The center panel of Figure 7 shows that the color difference between the W3 and W4 bands can be achieved at two temperatures that differ by more than 400 K. Because the absolute flux in both bands scales with diameter (Equation (33)), observed W3 and W4 fluxes could perfectly match two different solutions $(T_1, D)$. The example shown in Figure 7 is taken from a single WISE observation for asteroid 208, but the behavior occurs generically as a consequence of Equation (33).

The two solutions occur due to the reflected sunlight component in Equation (33), as can be shown by setting the parameter $\epsilon = 1$. Although there is, in general, very little reflected light in the W3, W4 bands, there is enough that when the peak of the thermal emission spectrum is shifted far enough to one side or the other of the bands that it has an influence. Note also that this phenomenon also occurs for the blackbody, EFRM and HG-NEATM models as well as for NEATM.

When fitting curves to WISE observational data, either of the two solutions is equally valid for fitting a flux through the data points. In some cases, data in the W1 and W2 bands can rule out one of the two temperature possibilities. But when the W1 or W2 bands are temperature-insensitive (i.e., dominated by reflected light), we may get rather poor resolution in temperature and thus in diameter; the least-squares fit may move toward either the high or the low temperature value. As will be shown below, this problem occurs in the analysis of many asteroids.

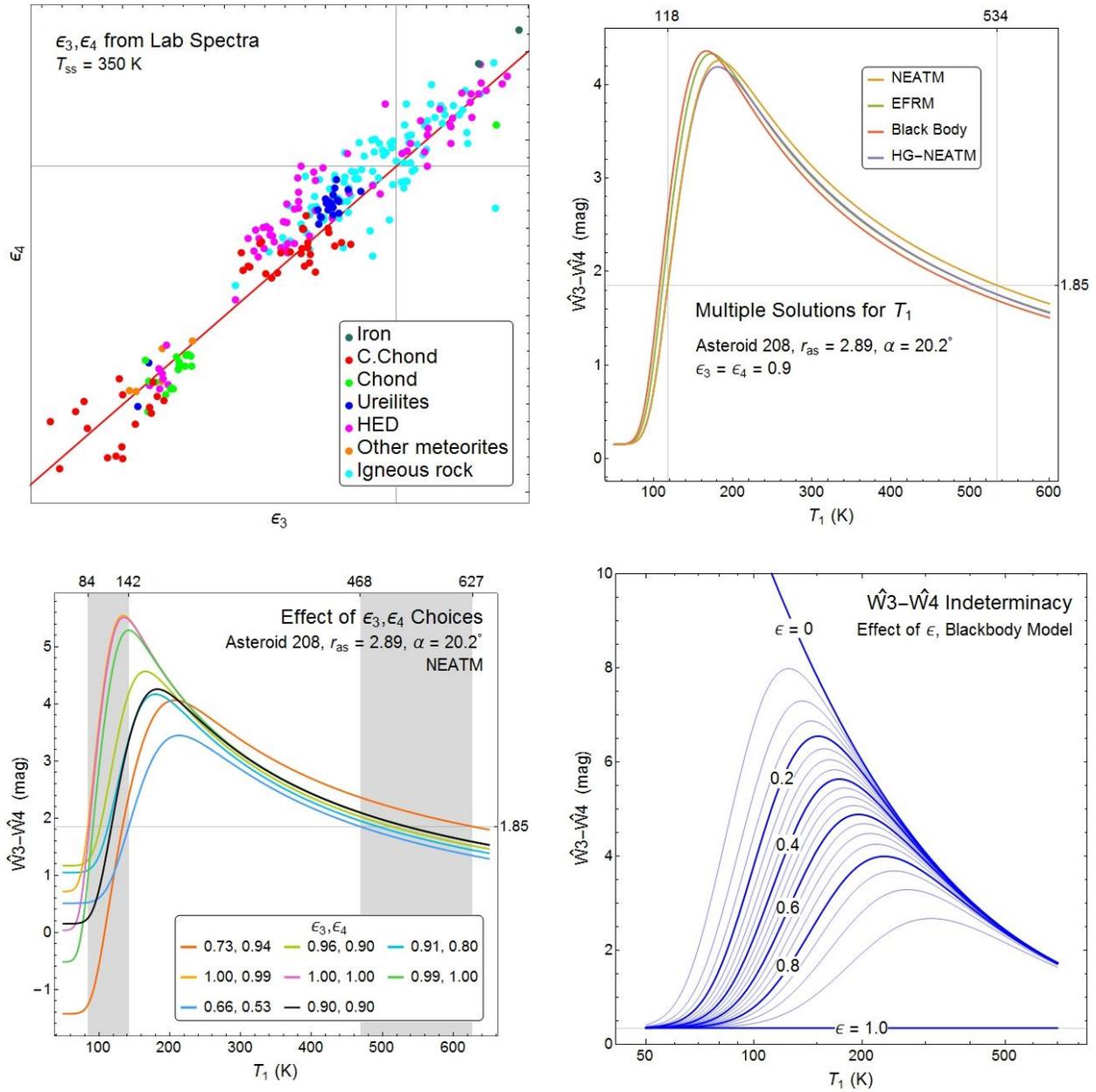

**Figure 7. Indeterminacy in bands W3, W4.** The *top left* plot shows that the point $\epsilon_3 = \epsilon_4 = 0.9$, commonly used in NEOWISE papers, does not match the values calculated from laboratory spectra for asteroid material proxies calculated with Equation (36) for $T_{ss} = 350$ K and $\alpha = 0°$. The next plot (*top right*) shows that, even when $\epsilon_3 = \epsilon_4 = 0.9$ are fixed, the $\widehat{W3} - \widehat{W4}$ band difference does not occur at a unique temperature. The NEATM model yields the same value of $\widehat{W3} - \widehat{W4} = 1.85$ for $T_1 = 118$ K and $T_1 = 534$ K, for example. Similar indeterminacy occurs for the other models as well. The *bottom left* plot shows the effect of using example values from the right plot for $\epsilon_3, \epsilon_4$ within the NEATM model. (Note that rounding makes several values appear to be 1.00.) Different assumptions for $\epsilon_3, \epsilon_4$ have a large effect on the temperature $T_1$. The effect of $\epsilon = \epsilon_3 = \epsilon_4$ on the indeterminacy is shown on the *bottom right*, using the blackbody model.

## 2.9 Methods for Fitting Models

To fit the thermal models, Equations (3) were used to adjust the WISE data, which was then fit by using the models described in Equation (33). NEATM, EFRM, Blackbody, and HG-NEATM models were fit. For each model, all 5 choices of emissivity vector (Equation (32)) were used, yielding a total of 20 combinations of models and parameterizations.

Standard nonlinear regression library functions from Mathematica 10.1 were used to fit curves. Similar functions are available for R, Matlab, and Python. The advantage of using standard libraries is that the code is more likely to be mature and debugged than code written only for asteroid fitting. Multiple optimization algorithms were used to obtain fits, starting in each case with a set of 4 initial values based on heuristics. For most cases, Levenberg-Marquardt produced the best results. $AIC_c$ was used to determine which model produced the best fit.

The fitting procedure was performed for 4 data sets, Mixed FC+3B+PC, Mixed FC, MBR FC+3B+PC, MBR FC summarized in Table 1. Following initial curve fitting, 500 bootstrap samples were drawn with replacement from each asteroid data set. During this resampling, the original observation structure was respected. A single draw thus contained 1 to 4 bands of data gathered as part of the same exposure. This is appropriate because a single draw is the complete WISE observation made at a particular rotational phase, and solar phase angle. Drawing bootstrap trials randomly with replacement thus samples the population of possible observations and allows us to get a handle on population sampling error.

Each of the bootstrap samples were fit with the same 20 model/parameterization combinations described above, using as a starting point the best fit achieved during the initial fitting run. Again, $AIC_c$ was used to select the best of these fits. The diameter $D$ was recovered from these fits by using model averaging with Akaike weights (Burnham and Anderson, 2002; Claeskens and Hjort, 2008).

## 3. WISE Data and Its Analysis

This section covers the details of WISE observational data and the data analysis process starting with the data as downloaded from IRSA to data suitable for thermal modeling. It is important both to document what was done in this study, as well as provide guidance to those using WISE/NEOWISE data for future asteroid studies.

The WISE mission operated in 4 distinct phases (Wright et al., 2010). The first, or full-cryo (FC), phase involved all of the bands. As the cryogen was depleted, that phase was followed by a three-band (3B) phase, during which only W1, W2, and W3 were available. The next, post-cryo (PC), phase collected observation from only the W1 and W2 bands. For this study, I use data from each of these first 3 mission phases. Recently, asteroid observations with the WISE satellite restarted after a hiatus as the NEOWISE reactivation mission (Mainzer et al., 2014a), but no data from that mission phase is considered here.

Observations of asteroids by WISE/NEOWISE have informed many publications by the NEOWISE team as well as by authors not affiliated with the project. This study focuses on 7 papers that reported the bulk of the primary model fits to data from the FC mission. The work here is also relevant to 3 other NEOWISE papers that present fits for data gathered during the 3B and PC missions. Table 1 summarizes the basic features of these papers.

It is worth noting that many of the papers in Table 1 have titles that include "Preliminary", as they were part of the main WISE/NEOWISE result release in 2011. In the five years since the publications of the NEOWISE group have focused primarily on application of these results, or on the 3B and PC mission. No revised version of the original studies has been published, and the astronomical community has treated the "preliminary" results as input for hundreds of follow on studies.

|  | Source Studies | | | | This Study | |
|---|---|---|---|---|---|---|
|  | Reference | Abbreviation | Topic | Count | FC + 3B + PC | FC |
| FC | (Mainzer et al., 2014b) | Mainzer/Tiny | Tiny NEO | 106 | 32 | 30 |
| | (Grav et al., 2011b) | Grav/JT:Pre | Jovian Trojans | 1742 | 124 | 80 |
| | (Grav et al., 2011a) | Grav/Hilda | Hildas | 1023 | 125 | 89 |
| | (Mainzer et al., 2011c) | Mainzer/TMC | Thermal model parameters | 49 | 46 | 46 |
| | (Mainzer et al., 2011b) | Mainzer/NEO | NEO | 428 | 144 | 135 |
| | (Masiero et al., 2011) | Masiero/MB | MBA (selected) | 129478 | 1819 | 1789 |
| | | | MBA (random) | | 235 | 161 |
| | (Masiero et al., 2014) | Masiero/MB:NIR | MBA $p_{IR1}, p_{IR2}$ | 2835 | 1240 | 1240 |
| 3B + PC | (Mainzer et al., 2012a) | Mainzer/PP:3 | MBA, NEO | 116 | 45 | 41 |
| | (Masiero et al., 2012a) | Masiero/MB:3 | MBA | 13669 | 896 | 894 |
| | (Grav et al., 2012) | Grav/JT:Tax | Jovian Trojans $p_{IR1}, p_{IR2}$ | 478 | 70 | 61 |
| Other | (Ryan and Woodward, 2010) | RW | STM, NEATM | 118 | 92 | 92 |
| | (Ďurech et al., 2010) | | | 105 | 63 | 63 |
| | (Shevchenko and Tedesco, 2006) | | | 63 | 44 | 44 |
| | (Durech et al., 2011) | | | 60 | 33 | 33 |
| | (Magri et al., 2007) | | | 55 | 43 | 43 |
| | (Magri et al., 1999) | | | 33 | 20 | 20 |
| | (Shepard et al., 2010) | ROS | Radar, Occultations, Spacecraft | 19 | 16 | 16 |
| | (Timmerson et al., 2009) | | | 3 | 3 | 3 |
| | (Benner et al., 2006) | | | 1 | 1 | 1 |
| | (Keller et al., 2010) | | | 1 | 1 | 1 |
| | (Ostro et al., 1990) | | | 1 | 1 | 1 |
| | (Veeder et al., 1989) | | | 1 | 1 | 1 |
| | (Busch, 2010) | | | 1 | 0 | 0 |
| | (Mahapatra et al., 2002) | | | 1 | 0 | 0 |
| | (Tedesco et al., 2002b) | IRAS | IRAS | 2228 | 1627 | 1616 |
| Totals | This Study | Mixed | Selected | NA | 2253 | 2129 |
| | | MBR | Random main belt | 300 | 235 | 161 |

**Table 1. Summary of published NEOWISE papers and numbers of asteroids analyzed in each.** FC: primary papers that analyze WISE observations from the full-cryo mission phase. 3B + PC: primary papers analyzing observations from the three-band (3B) and post-cryo (PC) phases. Other: non-NEOWISE papers that estimate diameters. For the FC missions, the NEOWISE group reported model fits that estimate the parameters $D, H, p_v, p_{IR}, \eta$; where noted, the studies may determine separate W1 and W2 albedos $p_{IR1}, p_{IR2}$. For the 3B and PC mission data, the shaded counts are asteroids for which more limited models were fit, which in general did not determine all of the parameters. FC + 3B + PC column: the number of asteroid that met the criteria of this study (i.e., had at least three data points in each band), after pooling data from 3 mission phases; FC column: the number that met the criteria when using only FC data. Note that many asteroids appear in multiple NEOWISE or other source studies.

## 3.1 Selected Data Subset

I used three approaches to create a data set of sufficient size for analysis. For those papers reporting 106 or fewer model fits, data for all of the asteroids noted in the paper were downloaded from IRSA. For papers having more than 106 fits, I used a sampling technique. Lists were constructed of all of the unique values for each parameter ($D, H, p_v, p_{IR}, \eta$, and, where applicable, $p_{IR1}, p_{IR2}$) from the published results. The 10 largest values, 10 smallest values, and 10 values centered on the 25th, 50th, and 75th percentiles were identified for each parameter. The asteroids having these parameters in the published fits were then selected, and the corresponding data were downloaded from IRSA. My goal was to obtain a sample of reasonable size that spans the full range of variation found in the NEOWISE paper results. Note that some asteroids appear in multiple NEOWISE papers. Masiero/MB:Pre, for example, contains all of the asteroids studied (with different model parameterization) in Masiero/MB:NIR as a proper subset.

Independent studies that collected asteroid diameter estimates from radar, stellar occultation and spacecraft visits (hereinafter "ROS") were also used as input (see Table 1). In addition, Ryan and Woodward (Ryan and Woodward, 2010) published STM and NEATM estimates for 118 asteroids, of which 92 were found to have WISE/NEOWISE data in IRSA, and had FC+3B+PC data. Prior to NEOWISE the largest thermal modeling effort was due to IRAS; of these 1728 asteroids are common to both studies.

This combined set of asteroid data from these sources will be referred to as the Mixed set. All mission phases were downloaded, yielding two distinct samples of asteroid data: one from the FC mission alone (thus directly comparable to the NEOWISE FC papers) and a second comprising data from the FC, 3B, and PC missions pooled together. In addition, a completely random selection of 300 asteroids from the largest list (Masiero/MB:Pre) was also made, which will be referred to as the MBR set. Once again both FC and FC + 3B + PC versions were gathered for these asteroids. In all cases, only asteroids having at least three data points in each band were retained for analysis. Table 1 summarizes the counts and data sources.

## 3.2 Processing of Raw Data by NEOWISE

WISE observations from IRSA must be processed for analysis. Sections 4 and 5 discuss thermal modeling and fitting; here I address the initial stage of raw data processing which prepares data for subsequent thermal modeling. Since this process is important for any independent group wishing to use the data for future studies (for example for thermophysical modeling), and because there are substantial issues with previous descriptions of the data processing, I will cover it in some detail.

Table 2 summarizes rules that are common to all NEOWISE papers. Table 3 lists aspects of methodology that differ among the papers (or, where no details are given, that potentially differ). As obtained from IRSA, WISE data comprises the WISE magnitudes $W1, W2, W3, W4$ corresponding to each of the infrared bands, along with their error estimates. Note that I use italic $W$ (e.g., $W1$) to indicate the magnitude as a variable or datum, and roman W (e.g., W1) to connote the band. By using the modified Julian date of the observation, one can obtain distance and phase angle information from ephemeris software (e.g., JPL Horizons, used here).

Ideally WISE records all 4 bands in a single observation. Some observations, however, omit one or more bands due to the quality and artifact flags, which are stored separately for each band. The top row of Table 2 lists the flag values for data used in this analysis.

WISE observations begin to reach sensor saturation at the following magnitude thresholds,

$$W1 \leq 8.1, W2 \leq 6.7, W3 \leq 3.8, W4 \leq -0.4 \quad , \tag{36}$$

and they become fully saturated by these thresholds (Cutri et al., 2011):

$$W1 < 2.0, W2 \leq 1.5, W3 \leq -3.0, W4 \leq -4.0 \quad . \tag{37}$$

The NEOWISE papers address saturation in two ways, as explained by (Mainzer et al., 2011b):

> As described in Mainzer et al. (2011b, hereafter M11B) and Cutri et al. (2011), we included observations with magnitudes close to experimentally-derived saturation limits, but when sources became brighter than W1 = 6, W2 = 6, W3 = 4 and W4 = 0, we increased the error bars on these points to 0.2 magnitudes and applied a linear correction to W3 (see the WISE Explanatory Supplement for details).

This is problematic because there are no further details in "Mainzer et al. 2011b" (referenced here as (Mainzer et al., 2011c). Cutri et al. 2011 is the WISE Explanatory Supplement (hereinafter WES), which contains no "linear correction" to W3. The NEOWISE corresponding authors did not reply to correspondence on the topic. Other authors have recently described a linear correction to the WISE bands (Lian et al., 2014), but it is very slight.

The thresholds described in the passage above differ markedly from Equation (37), in some cases by 2 magnitudes. Most of the NEOWISE papers include a similar passage, but the thresholds given in each vary (see Table 3). The NEOWISE papers are silent as to the origin or reasons behind these rules or why they differ. It may represent evolving thinking over a period of time, or it may be specific to the asteroid family studied in each of the papers.

One procedure, common to all of the papers is the decision to increase the error on selected points by 0.2 magnitude, ostensibly because they are partially saturated. The WISE pipeline as described in the WES already makes saturation adjustments. As noted by others (Alí-Lagoa et al., 2013; Hanuš et al., 2015), this value is sufficiently high that the corresponding data becomes essentially irrelevant if used in a minimum $\chi^2$ or weighted least-squares analysis. Thus, for that method of analysis at least, increasing the error to that level has essentially the same effect as simply deleting the data points.

The NEOWISE papers enforce a minimum error of 0.03 magnitude (Table 2). The reasoning given is that 0.03 magnitude is the maximum instrumental error possible with WISE. But the WISE pipeline already makes an estimate of the error under the appropriate conditions; it is unclear why it should be increased arbitrarily to the maximum possible. It appears that these choices may have been intended to compensate for flaws in the error analysis discussed below in section 4.2.

The NEOWISE papers discard any data that exceed another threshold, considering it to have been compromised by fully saturated sensors. The procedure is typically described as it is by (Masiero et al., 2011):

> Following those authors (Mainzer et al. 2011b; Cutri et al. 2011) we did not use objects brighter than W3 = −2 and W4 = −6 for thermal modeling.

Again, the referenced paper by Mainzer et al. offers no further details. The cited paper by Cutri et al. is the WES, and it lists the thresholds in Equation (37) as the full saturation limits; these differ markedly from those in the quoted passage. Other NEOWISE papers impose different limits as summarized in Table 3. Again this may be due to evolving thinking, new information, or some issue specific to the group of asteroids studied in each paper, but it is not explained.

The NEOWISE papers group WISE data into "epochs," which are periods of time when the WISE cadence and asteroid orbit produce a set of observations separated by a maximum time interval. NEATM model fitting is not performed on the overall data set, but only on a per-epoch basis.

The maximum interval between observations to count as the same epoch is 3 days in one paper (Mainzer et al., 2011d) but 10 days in another (Mainzer et al., 2011b) and not discussed in the others. An epoch typically includes 8–12 observations over 36 hours, but some asteroids were observed hundreds of times in a single epoch which could last for many days. For any given asteroid, observations can also span multiple epochs. Among the 1087 asteroids sampled from the papers in Table 1, 66.6% were observed during a single epoch, 28.6% during two epochs, 4.3% in three epochs, and 0.5% in four epochs. All of these epochs occurred during the FC mission.

| Description | Condition | Justification |
|---|---|---|
| WISE flags | ph_qual = A,B,C and cc_flags = 0,p | Avoid bad observations |
| Constraint on $\eta$ | $0 \leq \eta \leq \pi$ | "theoretical maximum" |
| Fit or assign $\eta$? | Two "thermally dominated" bands | Necessary for thermal modeling |
| Minimum error | $\sigma \geq 0.03$ | Maximum instrumental error |
| Error in partial saturation | $\sigma = 0.2$ | Partial saturation |
| Constraint on H | $H_{\text{MPC}} - 0.3 \leq H \leq H_{\text{MPC}} + 0.3$ | "error in H" |

**Table 2. Data analysis rules common to all NEOWISE FC papers in table 1.** See text for details.

Several rules govern the selection of epochs for thermal modeling as well as the selection of data to be used from each epoch. The first rule excludes any epochs having fewer than three observations in at least one WISE band. The NEOWISE papers differ on whether all observations are counted or the observation is counted only if it meets a criterion, such as having an estimated error $\sigma_i < 0.25$ mag (for band $i$) or having $SNR > 4$. The second rule excludes any band $i$ for which the number of observations $n_i < 0.4\, n_{\max}$, where the maximum number of observations in any other band is $n_{\max} = \max_{i=1,..,4} n_i$.

Grav et al. (Grav et al., 2011b) justify these rules as follows:

In order to avoid having low-level noise detections and/or cosmic rays contaminating our thermal model fits, we required each object to have at least three uncontaminated observations in a band. Any band that did not have at least 40% of the observations of the band with the most numerous detections (W3 or W4 for the Trojans), even if it has 3 observations, was discarded.

Yet the $n_i \geq 0.4\ n_{\max}$ rule (or 40% rule) would not appear to offer an effective means to achieve the described goal of protecting against the effects of low-level noise and cosmic rays. In the case of low-level noise, the WISE pipeline flags, an error estimate, or a simple check against sensitivity limits would seem better able to identify suspect data. The WES provides several measurements of sensitivity, based on comparisons to external sources, noise models, and internal repeatability.

Cosmic rays detected or inferred by the WISE pipeline trigger artifact flags that lead to exclusion of the corresponding data. Undiagnosed cosmic rays would presumably be recorded as an anomalously bright observation. In either case, the 40% rule seems an inappropriate tool for eliminating the effects of cosmic rays.

Figure 8 shows how the 40% rule affects the analysis of WISE data for 3 example asteroids. For asteroid 2021, the rule eliminates the W1 band in two different epochs even though the majority of the excluded data are brighter than the W1 external source sensitivity limit of 16.83. Moreover, for this asteroid the W2 band includes many data points that appear to be equally close to the W2 limit of 15.6.

The situation is reversed for asteroid 2048: W2 is discarded, but W1 remains. For asteroid 162843, the rule excludes none of the bands, despite the fact that the ranges of magnitudes in the W1 band are similar to those of asteroid 2021 and very close to the sensitivity limit.

| Paper | $p_{IR}$ | Epoch | MC | Saturation | Band Condition | W3 Correction |
|---|---|---|---|---|---|---|
| Grav/Hilda | $p_{IR1} = p_{IR2}$ | | 100 | $W3 < 4$ | $n_{max} \geq 3$<br>$n_i \geq 0.4\, n_{max}$ | $W3 < 4$ |
| Grav/JT:P | $p_{IR1} = p_{IR2}$ | | 50 | $W1 < 6$<br>$W2 < 6$<br>$W3 < 4$<br>$W4 < 3$ | $n_{max} \geq 3$<br>$n_i \geq 0.4\, n_{max}$ | |
| Mainzer/TMC | $p_{IR1} = p_{IR2}$ | 3 days | 25 | $W3 < 4$<br>$W4 < 3$ | $n_{max} \geq 3$<br>$n_i \geq 0.4\, n_{max}$ | $W3 < 4$ |
| Mainzer/Tiny | $p_{IR1} = p_{IR2}$ | | | | | |
| Mainzer/NEO | $p_{IR1} = p_{IR2}$ | 10 days | 50 | $W1 < 6$<br>$W2 < 6$<br>$W3 < 4$<br>$W4 < 0$ | $n_{max} \geq 3$<br>$n_i \geq 0.4\, n_{max}$ | $W3 < 4$ |
| Masiero/MB:Pre | $p_{IR1} = p_{IR2}$ | | | $W3 < 4$<br>$W4 < 3$ | $n_{max} \geq 3$<br>$\sigma_i < 0.25$<br>$n_i \geq 0.4\, n_{max}$ | $-2 < W3 < 4$ |
| Masiero/MB:NIR | $p_{IR1}, p_{IR2}$ | | | | $n_{max} \geq 3$<br>$SNR > 4$<br>$W1 \geq 0.5$ ref<br>$W2 \geq 0.1$ ref | |

**Table 3. Summary of analysis methodology for each NEOWISE study.** The $p_{IR}$ column shows the treatment of albedo for the $W1$ and $W2$ bands, either set equal or separately estimated. The MC column is the number of Monte Carlo trials used for error analysis; Epoch is the maximum gap allowed between observations in an epoch; Saturation is the onset of partial saturation threshold for setting the error $\sigma_i = 0.2$. The band condition is the set of rules for keeping or discarding an observation epoch, or a band within the epoch. The number of observations in each band is $n_i$ and the maximum across all bands is $n_{max} = \max_{i=1,..,4} n_i$. In order to retain an epoch there must be one band having at least 3 observations. Some studies impose a further condition that the 3 observations must have $SNR > 4$ or $\sigma_i < 0.25$, or that the W1, W2 band data be a fraction of reflected light. To retain data in a band, $n_i \geq 0.4\, n_{max}$. The W3 Correction column lists the range of values for which a linear correction is applied to that band. Shaded boxes indicate that the method, rule, or criterion is not mentioned in the corresponding paper.

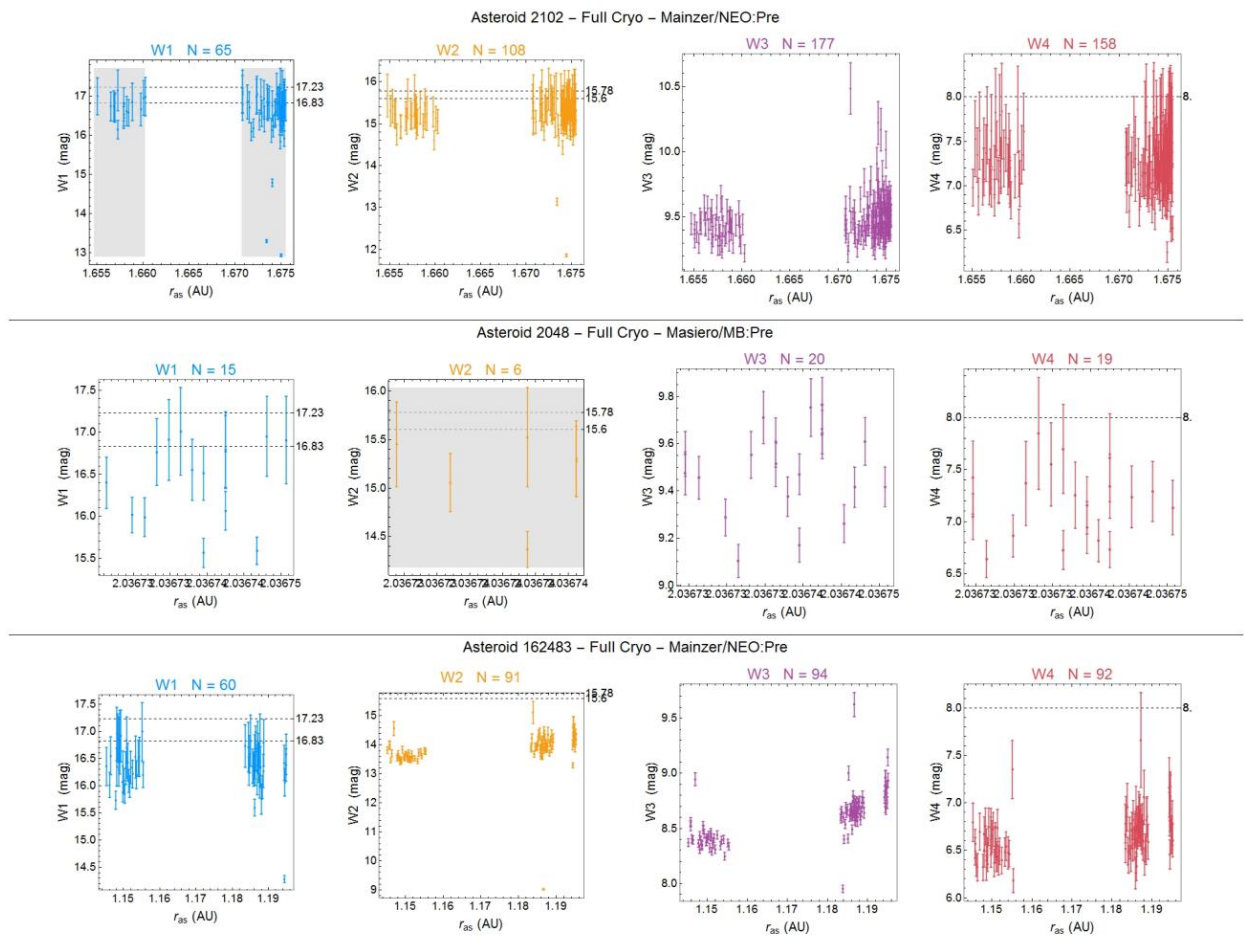

**Figure 8. Examples of the $n_i \geq 0.4\ n_{max}$ or 40% rule.** Observations omitted by the rule are shaded gray. Dashed lines show the sensitivity limits. In the case asteroid 2102 (*top*), W1 band observations in two different epochs are eliminated by the rule. For asteroid 2048 (*middle*), the entire W2 band is eliminated; in both cases, other bands appear to be just as close to the sensitivity limits. For asteroid 162483 (*bottom*) no bands are eliminated, despite the fact that some W1 band data falls very close to the sensitivity limit. See section 2.3 for description of the plotting technique.

The impact of the 40% rule varies from one NEOWISE paper to the next, as shown in Figure 9. For a random sample of 300 main belt asteroids from Masiero/MB:Pre, the rule deletes at least one band from 86% of the sample. That is a very large amount of data to discard; normally one would require an exceptionally strong reason for such a dramatic ad hoc intervention.

The NEOWISE papers require data in two "thermally dominated" bands to perform thermal modeling. There does not seem to be an a priori criterion for this – instead the NEOWISE papers appear to run the curve fitting then decide ex post facto which bands are thermally dominated by looking at the results.

The net effect of the "thermally dominated" rule is to eliminate the W1 and often W2 bands in many situations. One side effect is that this can tend to mask problems associated with the Kirchhoff's law, or with calculation of incident solar radiation (see section 4 and Figure 16 below). Indeed this may be the origin of the rule.

When two thermally dominated bands are unavailable, the papers set values for the parameter $\eta$ to an assumed value characteristic of the broad class of asteroid (NEO, main belt, etc.). Since the focus of the present study is thermal modeling, the selection criteria for all asteroids analyzed in this study is that they have data in all four bands, which effectively eliminates asteroids where the value of parameter $\eta$ was assumed by the NEOWISE group.

In this study both the data, and the error estimates from the WISE pipeline, as downloaded from the IRSA database are used without the modifications used in Table 2, except as provided for in the WES.

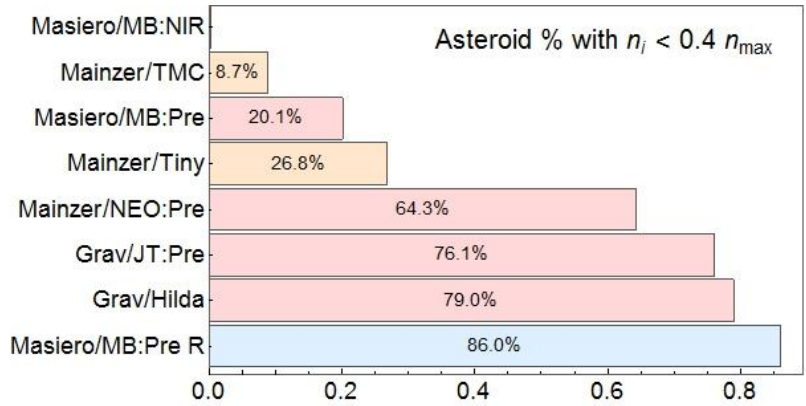

**Figure 9. Percentage of asteroids in samples from NEOWISE papers that have one or more band eliminated by the $n_i \geq 0.4\, n_{max}$ or 40% rule.** Colors indicate the asteroid samples used. Orange: all asteroids listed in the paper. Red: a selected subset, as described in the text. Blue: an unbiased random sample of 300 asteroids from Masiero/MB:Pre.

## 3.3 Plots of the WISE Data

To visualize the WISE data and the fits to it, it is convenient to plot the points and fits on 2-D charts. In the raw WISE data, the magnitudes $W1$ through $W4$ are a function of the distance from the asteroid to the sun $r_{as}$ and of the distance from the asteroid to the observer (WISE) $r_{ao}$. Similarly, any model of thermal emission or reflected light will also be a function of $r_{ao}$ and $r_{as}$ as well as model-specific parameters.

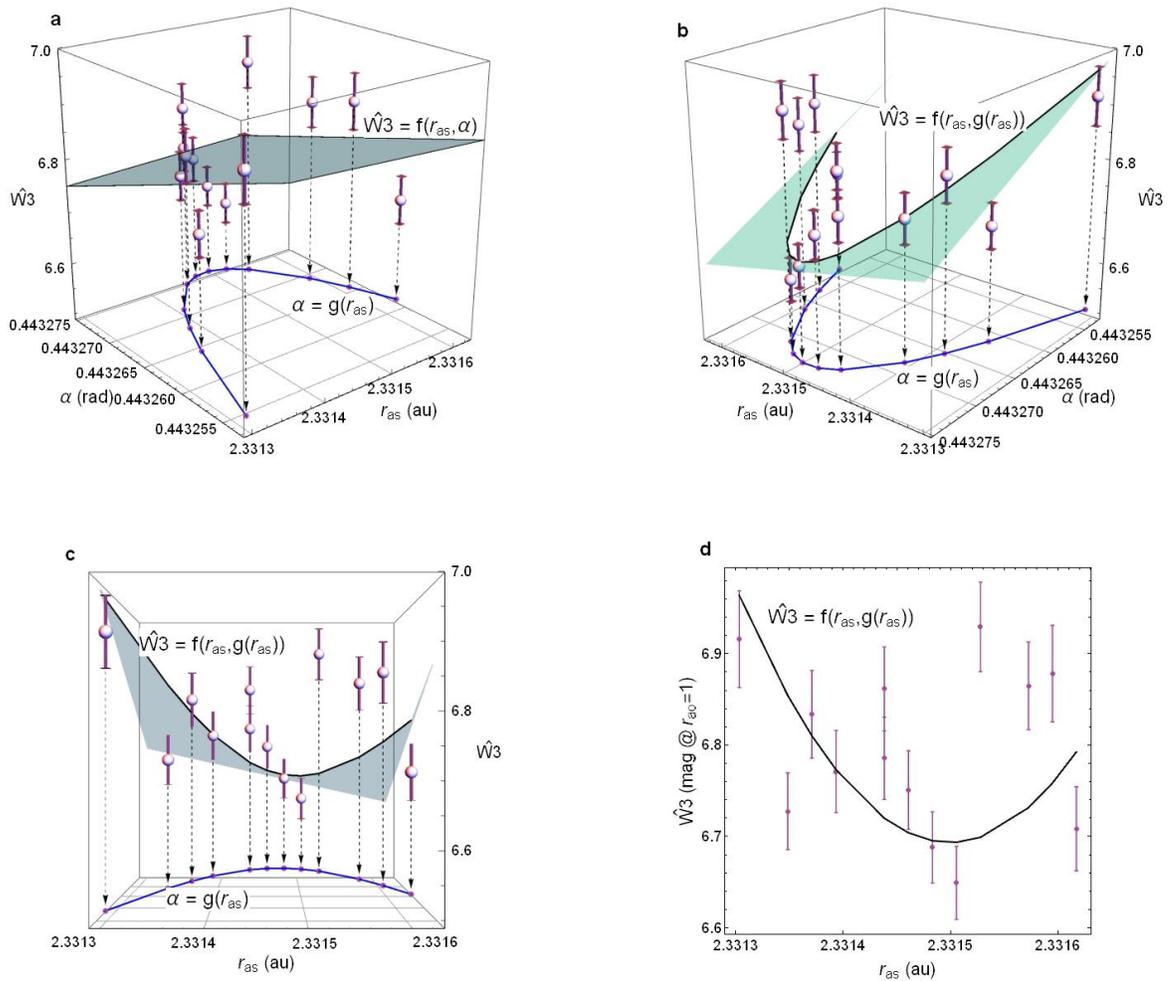

**Figure 10. Example 2-D plots of the WISE data.** (a) A 3-D plot of W3 band points as a function of $r_{as}$ and $\alpha$. The surface $\widehat{W3} = f(r_{as}, \alpha)$ is a model fit to the data. (b) There is a 1:1 relationship between $\alpha$ and $r_{as}$, viz. $\alpha = g(r_{as})$. We can cut the model plane using that function forming the line $f(r_{as}, g(r_{as}))$. (c) A perspective view of (b), oriented to look down the $\alpha$ axis. (d) The equivalent orthogonal 2-D plot of the data shown in (c).

Any light, whether emitted or reflected, is affected by the inverse-square law (i.e., the asteroids are not disk-resolved by WISE). We can thus make the adjustment

$$\widehat{W1} = W1 + 5\log_{10} r_{ao} \quad , \tag{38}$$

and so on for the other bands. The quantity $\widehat{W1}$ is analogous to absolute magnitude $H$: it is the magnitude that WISE would detect if the asteroid were at $r_{ao} = 1$ AU.

We can then replace the $r_{ao}$ axis with the phase angle $\alpha$, which is computed from $r_{ao}, r_{as}$. This still leaves us with a 3-D plot (Figure 10).

The nature of the WISE observation is such that there is a 1:1 relationship between $\alpha$ and $r_{as}$. As a consequence, we can always make a function $= g(r_{as})$. The simplest way to implement $g(r_{as})$ numerically is as a piecewise linear function. We can then project out the $\alpha$ axis as shown in Figure 10, leaving a 2-D plot.

The same transformation can be done to either thermal models, or models of reflected light, $f(r_{as}, r_{ao}, ...)$, where the ellipses refer to model parameters. Such a model can be written in terms of a function of $r_{as}$:

$$\hat{f}(r_{as}, \alpha, ...) = f(r_{as}, r_{ao}, ...) + 5 \log_{10} r_{ao} \quad . \tag{39}$$

The function $\hat{f}(r_{as}, \alpha, ...)$ is used to fit the data points. Although this is a function of two independent variables $r_{as}, \alpha$, a convenient performance optimization is to implement it as a function of 4 variables: $r_{as}, \alpha, \psi, q$, where $\psi = \psi_{HG}(\alpha, G)$ is the HG phase function at $\alpha$ and slope parameter $G$, and $q$ is the HG phase integral. These may be calculated up front, which saves considerable time in the curve-fitting process by eliminating the need to evaluate these variables many times.

Model fits to $\hat{f}(r_{as}, \alpha, ...)$, whether from a fitting process or from the NEOWISE papers, can then be plotted as $\hat{f}(r_{as}, g(r_{as}), ...)$ because we need to evaluate the model only at the data points, either for fitting or for plotting.

## 3.4 Comparing Parameter Estimate Distributions

The quantitative assessment of precision and accuracy is complicated by the fact that both the thermal modeling estimates (from either the bootstrap analysis here or from NEOWISE fits) and the diameter estimates provided by other means are known only from their statistical distributions. Diameter estimates from NEOWISE, or from the literature are given as a value with a standard error, so the diameter distribution can be modeled as a normal distribution. NEOWISE fits are similar. The bootstrap trials yield a set of diameter estimates.

The distribution of diameter estimates for asteroid 192 is shown in the top left of Figure 11, for three sources of estimates: NEOWISE, radar, occultations or spacecraft (ROS) and the bootstrap methods discussed below (bs). In order to evaluate the ratio between these we can draw 10,000 Monte Carlo samples from each distribution and take the ratio of the samples.

Mathematically, we can form the ratio of two diameter estimates $D'$ and $D$ for asteroid $i$, and draw

$$\rho_{i,j} = \frac{D'_{i,j}}{D_{i,j}} \, , i = 1, ... N_a, \qquad j = 1, ..., 10000 \, , \tag{40}$$

where we assume that there are $N_a$ asteroids in a trial, and $j$ is an index over the Monte Carlo samples. Histograms for $\rho$ from diameter estimates for asteroid 192 are shown for 10,000 samples in top right of Figure 11.

In order to assess the overall accuracy across many asteroids we can form trials which draw one sample from the ratio distribution for each asteroid. There are two approaches that are commonly used in scientific error analysis. The most common is to quote a mean and standard error. In order to take this approach one must create Monte Carlo trials that draw one sample from the ratio distribution of each asteroid, and record the mean and standard deviation. A histogram of the standard deviation across 10,000 such trials is shown in the lower left of Figure 11. The total expected error $E_\rho$ at the 1 sigma level is then

$$E_\rho = \max \begin{pmatrix} |\bar{\rho} - 1 + \text{quantile}(SD_{1\text{ trial}}(\rho), 0.6827)|, \\ |\bar{\rho} - 1 - \text{quantile}(SD_{1\text{ trial}}(\rho), 0.1827)| \end{pmatrix}, \quad (41)$$

where $SD_{1\text{ trial}}$ is the standard deviation over each trial of $N_a$ asteroids, and the means $\bar{\rho}$ and is also over all trials.

The reason to use a quantile based approach in (41) is that it is more robust when doing Monte Carlo or bootstrap evaluation. The result is that the expected error found across a trial of $N_a$ asteroids is $\leq \pm E_\rho$ at the 1 sigma level. Different confidence levels could be achieved by changing the threshold values ( 0.6827, 0.6827) in Equation (41).

A problem with quoting the estimated mean and standard error is that it assumes symmetric errors, and with an asymmetric distribution it can be misleading. A different approach is to give the estimated 95% confidence interval for the ratio $\rho$. In that case we again draw trials sampling each asteroid once, finding the quantiles of the distribution of $\rho$

$$q_{low} = \text{quantile}(\rho, 0.025), \; q_{mid} = \text{quantile}(\rho, 0.5), q_{hi} = \text{quantile}(\rho, 0.975). \quad (42)$$

Histograms for these quantiles are shown in the lower right of Figure 11. The final 95% confidence interval estimate is then.

$$95\%CI = \{\text{quantile}(q_{low}, 0.025), \quad \text{quantile}(q_{mid}, 0.5), \quad \text{quantile}(q_{hi}, 0.975)\}. \quad (43)$$

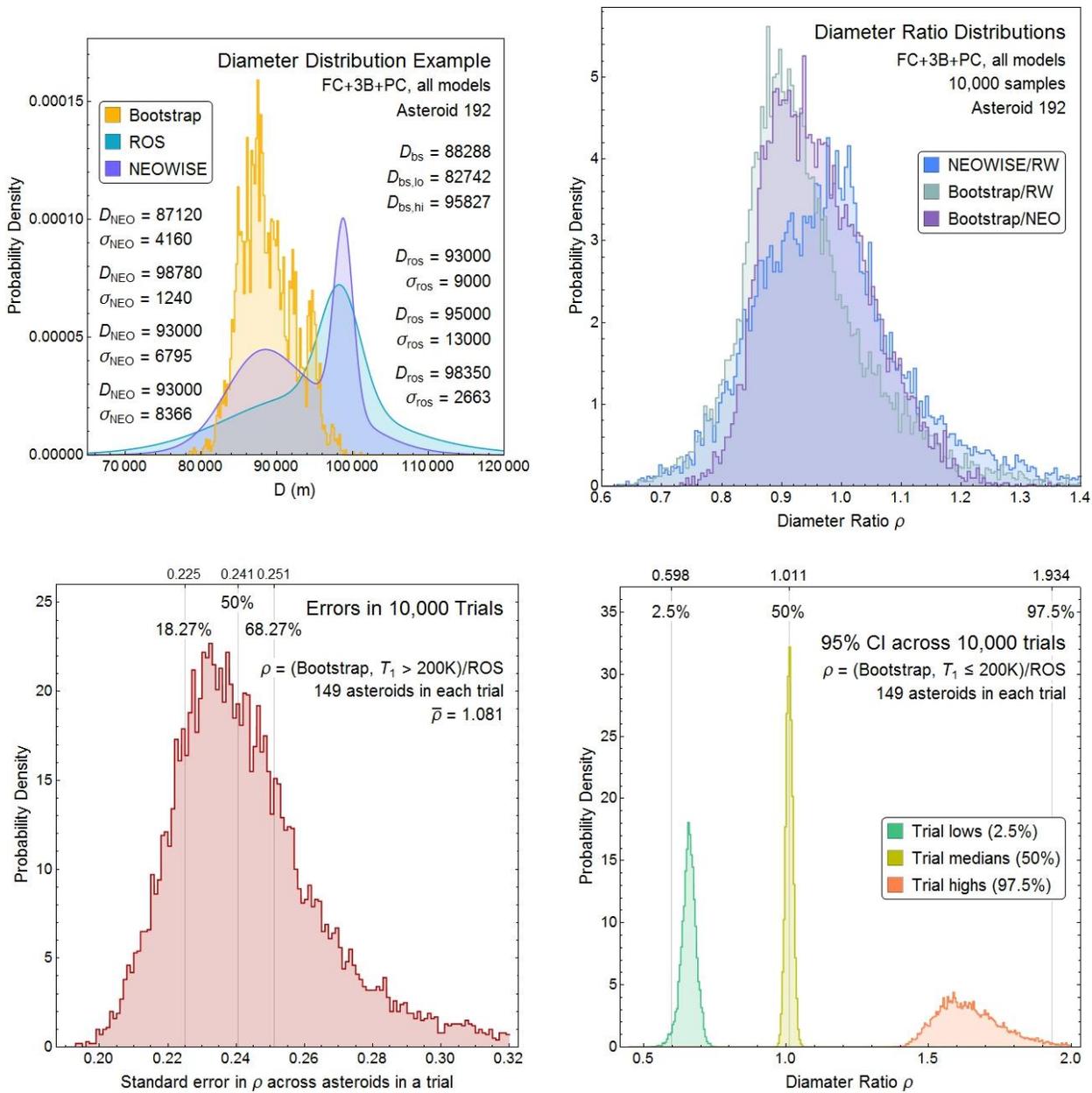

**Figure 11. Statistical evaluation of diameter estimation accuracy.** The values of diameter *D* estimated from thermal modeling and from radar and other observations are known from their statistical distributions, with an example shown (*upper left*) for asteroid 192 (Nausikaa).  Distributions for estimates from NEOWISE, the bootstrap approach in this study, and results from radar, occultations and spacecraft visits (ROS) gathered from the literature. Monte Carlo samples drawn from these distributions can be divided to get a histogram of the diameter ratio (*upper right*).  Trials drawing one sample from the ratio distribution for each asteroid allow calculation of the standard deviation across a trial, a histogram of 10,000 standard deviation trials is shown (*lower left*).  Histograms of the 95% confidence interval on 10,000 trials, along with their quantiles are shown lower right.

# 4. Empirical Study of Results Reported by NEOWISE

This section covers empirical consistency checks across all ~157,000 asteroids from the papers in Table 1. As such it offers some insight into a broader set of asteroids than covered by thermal modeling.

The NEOWISE papers, like many previous asteroid studies, adopt the convention that asteroids are spherical. There is a well-known relationship for spherical asteroids (Harris and Lagerros, 2002; Kim et al., 2003) that

$$H = 5 \log_{10}\left(\frac{1{,}329{,}000}{D\sqrt{p_\mathrm{v}}}\right) \quad , \tag{44}$$

where diameter $D$ is in meters. Because each of the parameters in Equation (45) is included in the NEOWISE results, we can check for consistency to see whether this relation holds true for each solution, as it ought to. As Figure 12 shows, this check reveals many inconsistencies.

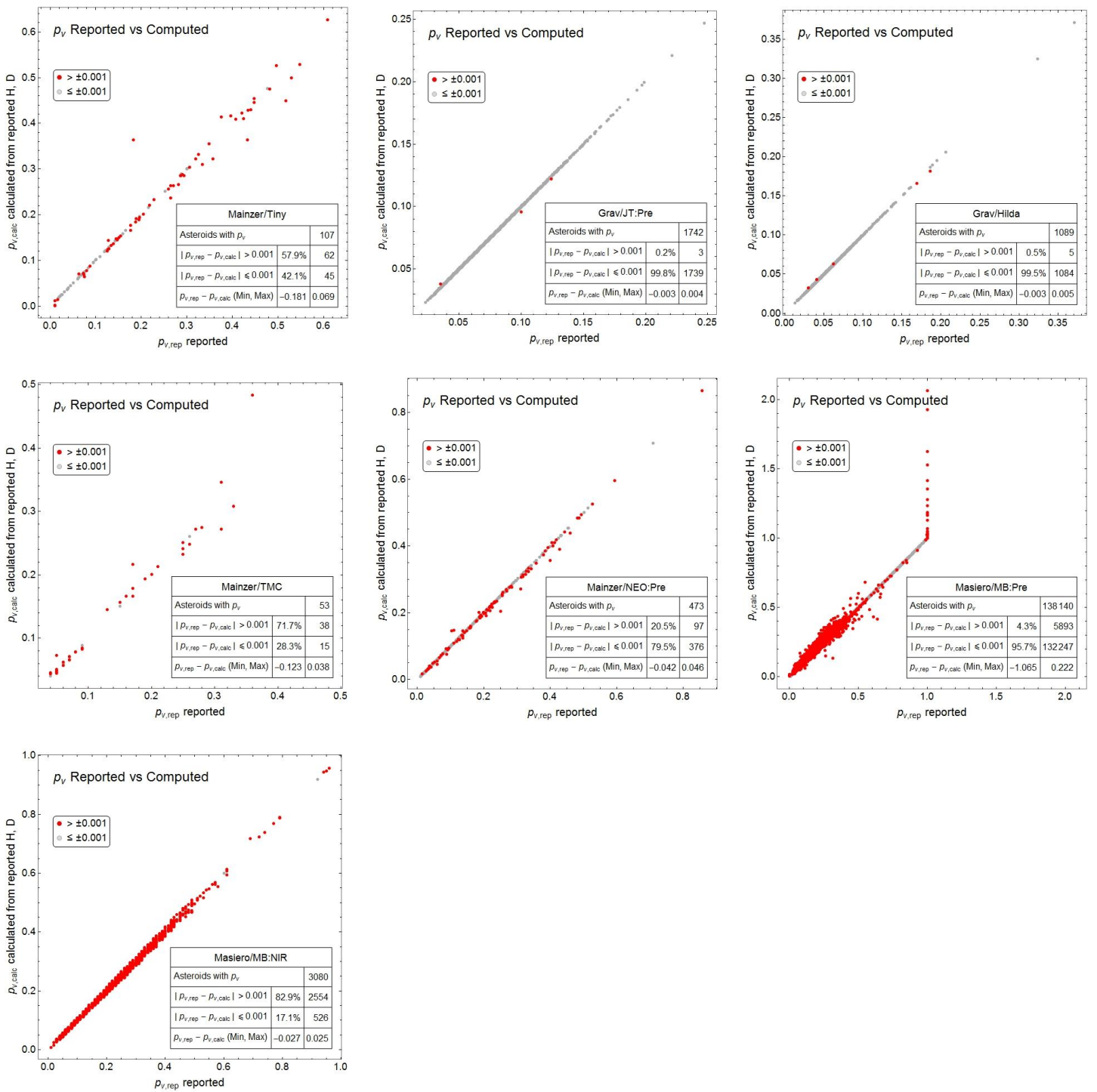

**Figure 12. Inconsistent asteroid properties in NEOWISE results.** Each panel plots the visual albedo $p_v$, as reported in NEOWISE papers, versus $p_v$ as calculated by Equation (45). The gray points match to within the reported number of digits. The red points do not match.

Note that Figure 12 is purely an internal check, which is be independent of the estimated errors. The threshold of 0.001 is to remove rounding errors in the reporting which is quoted to that many digits. However, as the min and max figures in the plot legends show the difference can be considerable. Although the NEOWISE papers report the albedo results as being derived from best fits (see discussion below), any true best fit procedure should produce values that all match those generated by Equation (45). The inconsistencies plotted in Figure 12 are troubling because they suggest either error in the fitting process, or some undocumented aspect of the analysis procedure that violates Equation (45).

As an additional check for consistency, I compared the reported values of $H$ compare to values obtained from MPCORB.DAT from the Minor Planet Center. Figure 13 shows that most values of $H$ differ from today's values. The scatter seen in Figure 13 can be explained in part by changes made in the MPC's values of $H$ since the NEOWISE papers downloaded them prior to publications, likely in 2010 or 2011, but also by the fact that the NEOWISE papers fit $H$ as well as fitting the other parameters. Because of this latter effect, mismatches would be apparent even when comparing to the same version of MPCORB.DAT used by the NEOWISE analyses. The mismatch ranges between -4.1 to +3.2 magnitudes which is a very large range of variation.

Figures 4 and 5 raise the question of how the $p_v$ results would change if today's MPCORB.DAT values were used and the fits were consistent. As shown in Figure 14, if we assume the validity of Equation (45), then very few results remain the same; the majority change by a large amount. This sensitivity analysis shows that the NEOWISE results are highly dependent on $H$ values. This is not surprising because if Equation (45) is inverted to show the dependency of $H$ on $p_v$ it is in the exponent. This unfortunate fact makes any determination of $p_v$ sensitively depend on absolute magnitude.

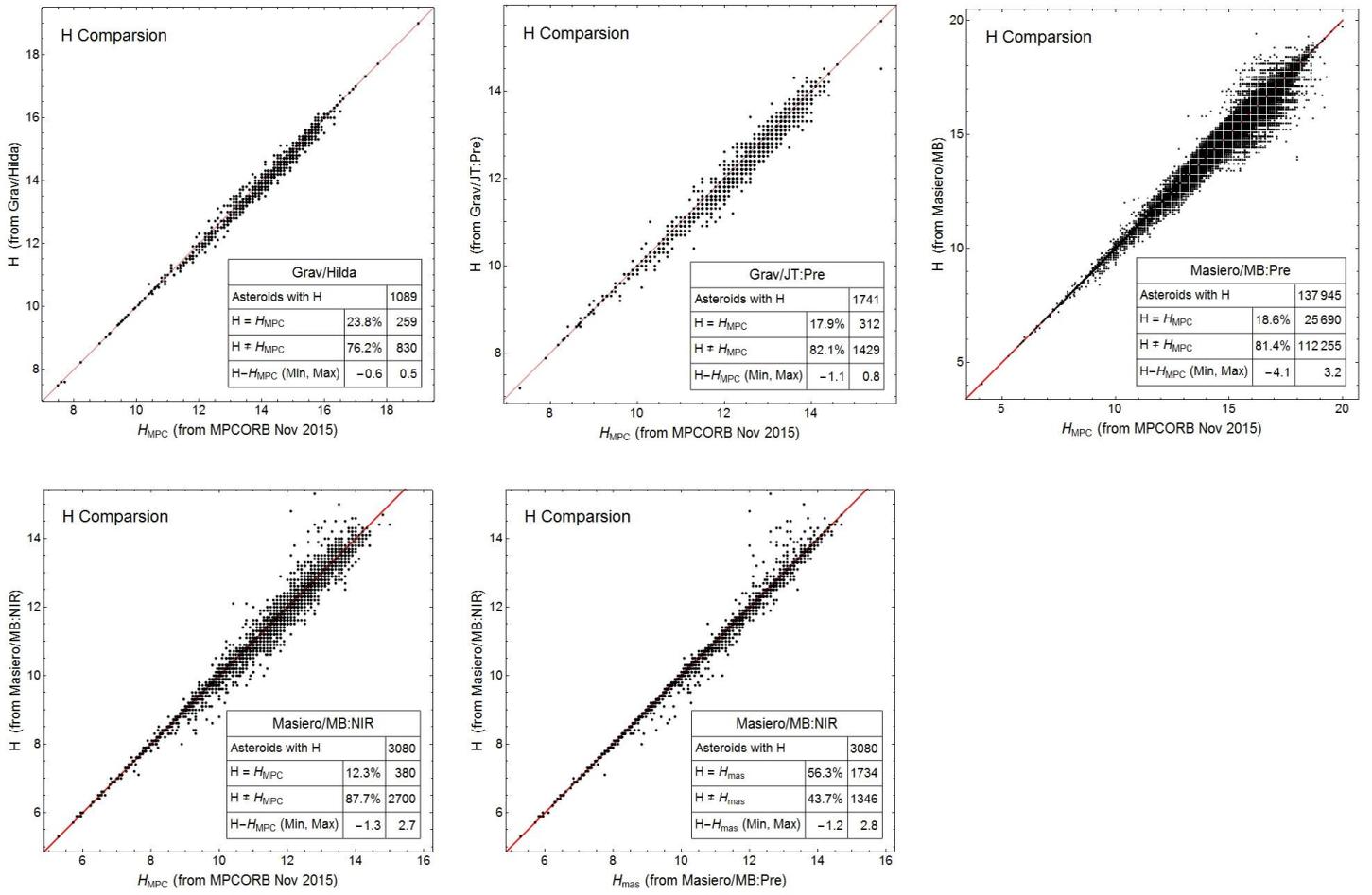

**Figure 13. Comparison of *H* values reported in NEOWISE papers to those provided by the Minor Planet Center.** Matching points lie on the red line $y = x$. Points on either side of the line differ between the two sources. The lower right plot compares values of *H* as determined by Masiero/MB:Pre to those from Masiero/MB:NIR.

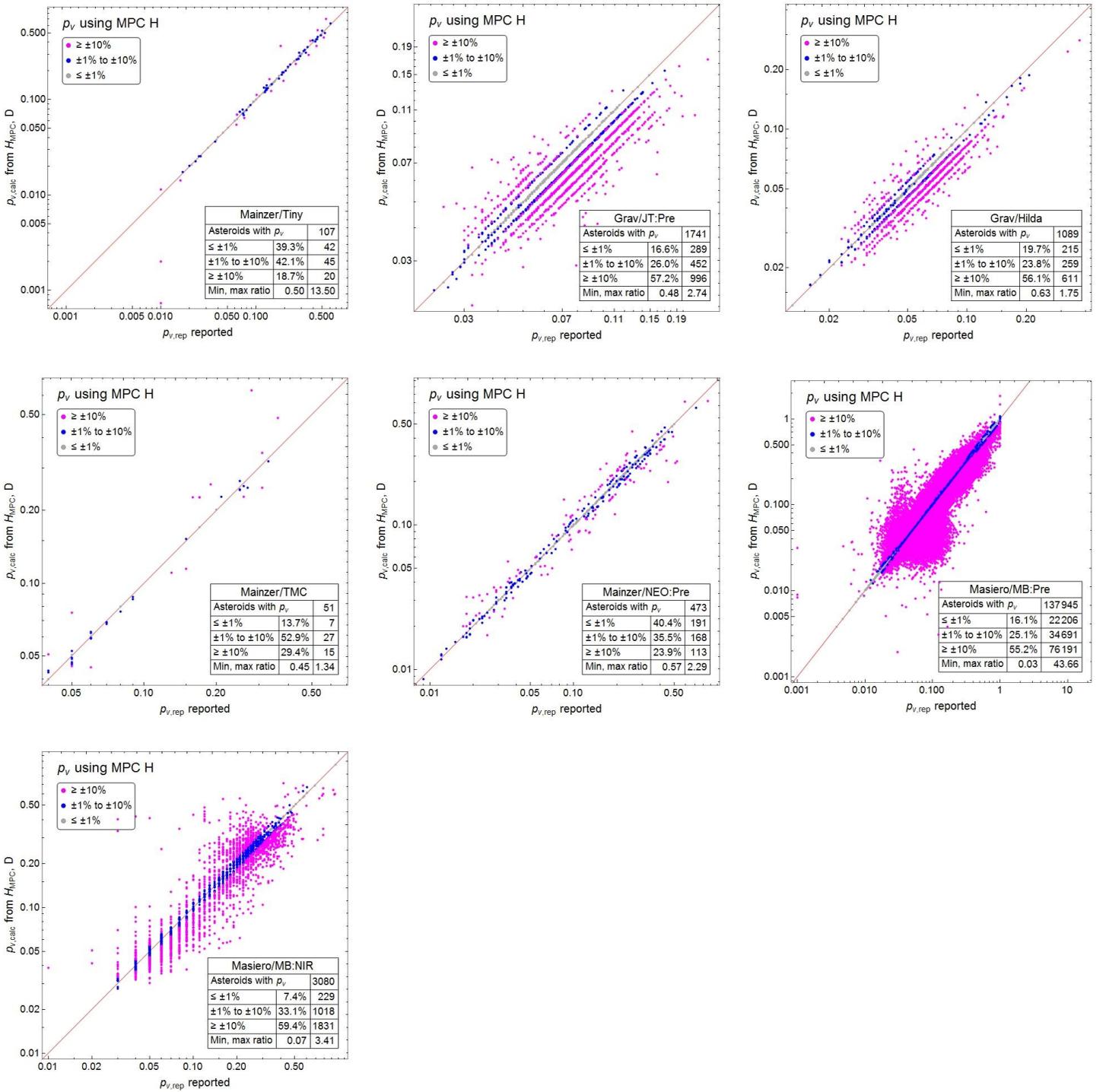

**Figure 14. Changes in $p_v$ when 2015 values of $H$ from MPC are used.** Values of $p_v$, updated to reflect values of $H$ from the MPC as of November 2015 (calculated using Equation (5)), are plotted against the values reported in the NEOWISE papers. Differences are highlighted in blue (between ±1% and ±10%) and magenta (> ±10%). In the case of main belt asteroids from Masiero/MB:Pre using 2015 values increases some $p_v$ by a factor of 43, and reduces others by a factor of 0.03, and the majority (55%) change by more than ±10%.

The most surprising empirical discovery from examination of the NEOWISE results is that the majority of the "best fits" do not pass near the data points they purport to fit. Figure 15 shows representative examples, plotted along with a best fit determined by alternative methods described in section 4 below.

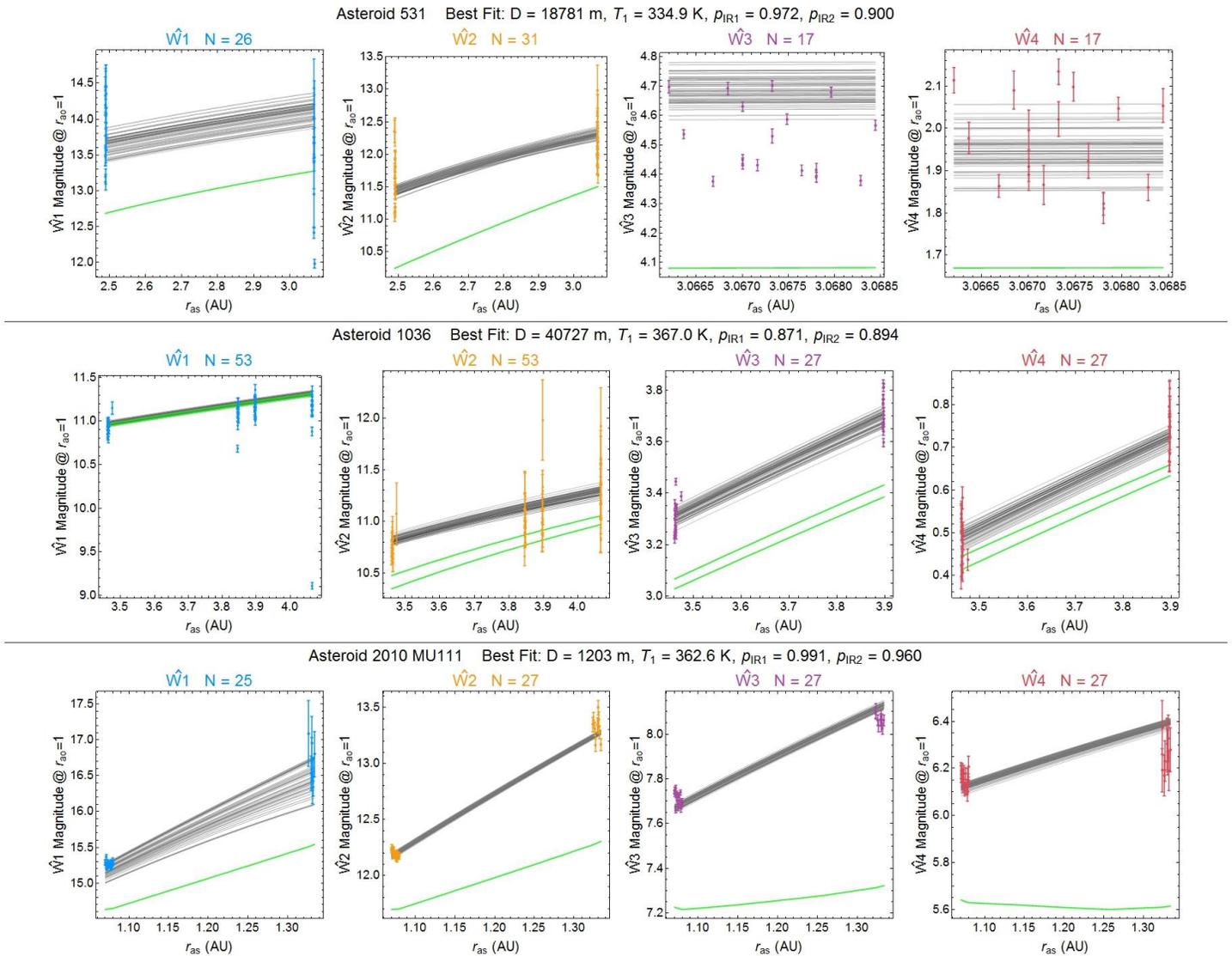

**Figure 15. Examples of NEOWISE "best fits" to WISE asteroid data compared to alternative "best" fit**. Each panel plots all 4 WISE bands for the asteroid noted, using only FC data. *Green lines*: best fits, as reported in the NEOWISE papers, multiple lines indicate fits from different epochs and/or different NEOWISE papers. *Gray lines*: best fits for bootstrap trials using the methods in this study – the Akaike averaged consensus parameters are given as the best fit.

The poor performance of the NEOWISE fits as shown in Figure 15 suggests that the fluxes calculated from NEOWISE best fit parameters (plotted as green lines in Figure 15) might not be the same as those used by the NEOWISE team in producing the fits. The NEOWISE papers do not describe their flux calculations nor do they offer sufficient test cases or exemplars to verify this directly. The corresponding authors did not reply on this topic when queried.

However, one can get an indirect assessment by looking at the histogram of residuals from the NEOWISE fits to the data. One would expect that best fits to data would have residuals that that would tend toward zero median for a sufficiently large number of data points, while a systematic error in calculating the fluxes would present itself as a residual distribution shifted away from zero. These are plotted in Figure 16, and show that there does not seem to be a systematic deviation between the fluxes calculated here and those used by the NEOWISE papers.

The fits in Figure 15 frequently miss the cloud of data points in an entire band. To tally the prevalence of this behavior, I counted cases in which all data points being fit fall entirely above or entirely below the model curve for all data points in a WISE band (as in Figure 15, for bands W1, W2 for asteroid 102). Such fits "miss" the band altogether, whereas normally a best-fit curve have some data points fall both above and below the line (i.e. have both positive and negative residuals. It should be noted that this is an extremely crude way to judge goodness of fit, but given the very poor quality fits in the NEOWISE results even this crude measure is informative.

This is shown in bottom row of Figure 16. Except for fits published in Mainzer/NEO:Tiny, the majority of NEOWISE fits miss at least two bands of data. Of the NEOWISE fits for 300 randomly selected main belt asteroids from Masiero/MB:Pre, only 8.7% pass through all four bands of data, the behavior one would expect from a best-fit model. This is particularly puzzling because the NEOWISE fits have two additional free parameters.

It appears that one reason for missing the data points is due to the series of data analysis rules, such as the 40% rule or the thermally dominated rule which have the potential to eliminate entire bands from the fit. The ad hoc modifications to the estimated observational error can also, in effect, eliminate data points by making their weighting in a WLS approach so low that they are effectively ignored by the curve fitting algorithm.

The result is that the NEOWISE fits appear to be based on a subset of the data, coupled with constraints on the parameters such as $\eta$. Unfortunately because the NEOWISE fitting is not documented enough to replicate I could not determine in more detail the origin of the fits that do not pass near the data points.

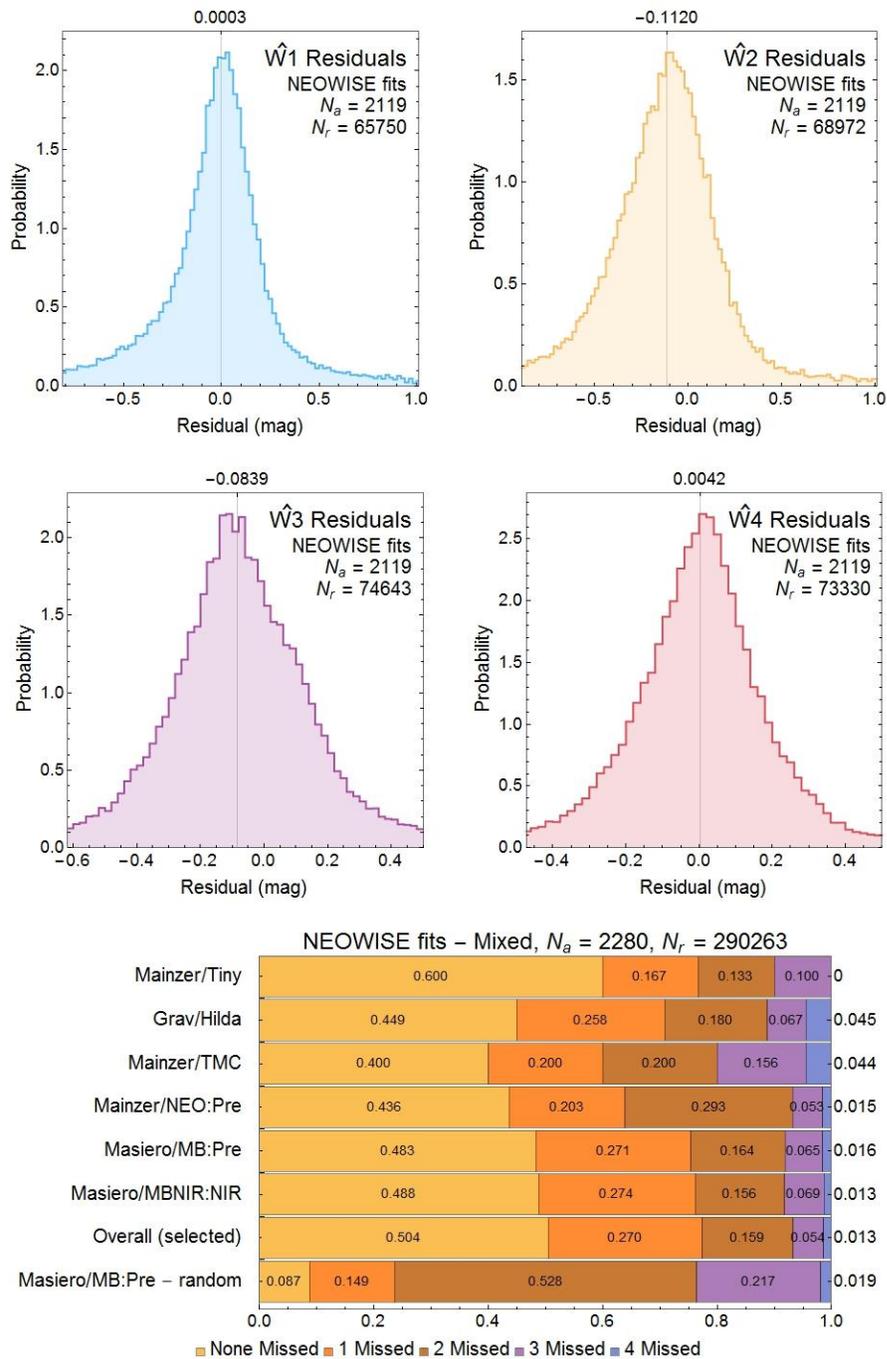

**Figure 16. Residuals and band missing for NEOWISE model fits.** Histograms of the residuals from the NEOWISE fits are shown for each WISE band. The fact that the residuals cluster with a median value near zero shows that there is no systematic difference between the NEOWISE flux calculations and those done here. The bottom row plot shows the prevalence of band misses: a model fit "misses" a band of data if all data points in that band are either below or above the fit line (see Figure 15 for examples). In the NEOWISE papers, only about half of the published model fits pass through all bands of data (shown in the chart as "none missed"), and only 8.7% for randomly chosen main belt asteroids. Values with bars too small to be labeled appear to the right.

## 4.1 Issues with Fitting Models

Several general statistical issues occur when NEOWISE model fitting is examined closely. These issues are independent of the models being fit; the physics of thermal modeling is discussed below in section 4.

There is a lack of clarity about the actual model-fitting process used in the NEOWISE studies. Mainzer et al. (Mainzer et al., 2014b) go into the most detail on the process:

> The best-fit values for diameter, visible geometric albedo $p_V$, infrared albedo $p_{IR}$, and beaming parameter $\eta$ were determined using a least-squares optimization that accounted for the measurement uncertainties for bands W1, W2, W3, W4, absolute magnitude $H$, and phase curve slope parameter $G$.

This is insufficient to enable replication of the fitting – it does not distinguish between many different approaches that could have been used. It is nevertheless clear that the NEOWISE papers used a model fitting procedure that is quite outside normal statistical practice. The procedure includes fitting the absolute magnitude $H$—an extremely unconventional choice because $H$ is the product of observation, not a model parameter. As a general rule, observed and measured quantities are not treated as free parameters when constructing models.

NEOWISE authors note that $H$ is measured with some error; the constraint applied to it reflects that uncertainty. In normal statistical practice, this error would contribute to the error estimates for quantities such as $p_v$ that rely on it rather than being used as a fitting parameter. Instead, the NEOWISE papers take the measured value of $H$ as the center value for a constraint on $H$ as a free parameter to be fit (see Table 2).

However, the constraint used is based on the educated guess that the error is 0.3 magnitude for all asteroids, rather than actual observational uncertainty. Moreover, in using such a constraint, the authors seem to be assuming that the statistical distribution of errors in $H$ is uniform within the constraint because they imply it is a free parameter that floats within upper and lower bounds. Observational error is unlikely to produce a uniform distribution, nor hard upper and lower constraints. It is possible that, rather than being a constraint, the NEOWISE group adopted another approach to modeling more realistic error distributions for $H$, but if so it is not described.

As a practical matter, fitting $H$ has the drawback that the analysis must be redone whenever new observational evidence becomes available. This is illustrated by Figure 14 which shows how much the visible band albedo $p_v$ results change when the (November 2015) values of $H$ from MPC are substituted back into the NEOWISE results.

Another problem with fitting $H$ is that fitting results depend sensitively on $H$ due to the exponential relationship in Equation (5). It is unnecessary to involve $H$ in thermal modeling of asteroids—indeed it is numerically unwise, as I show below.

The slope parameter $G$ can be similarly problematic. For some asteroids, $G$ is a parameter measured from lightcurves obtained through visible-band observation, so fitting $G$ may improperly override a measured value in the same manner as discussed above for $H$. For most asteroids, however, $G$ has not been measured and instead is assigned the default value of $G = 0.15$, in which case fitting $G$ might be more defensible.

NEOWISE model fitting is unusual in another regard: fitting has been done for each epoch individually, even when multiple epochs are available. The typical range of the independent variables $r_{as}, \alpha$ is very restricted in typical epochs (see examples in Figures 1 and 7). As a result, one cannot really test model dependence on those variables. No explanation is given in the NEOWISE papers of why this highly unconventional approach is taken.

Instead, the NEOWISE papers attempt to address this choice by comparing the resulting diameter estimates across different epochs for asteroids observed in multiple epochs.

> However, since 87% of the objects with multi-epoch observations have diameters that agree to within 10%, we conclude that multiple observational epochs do not contribute significantly to the differences observed between IRAS and NEOWISE diameters.(Mainzer et al., 2011d)

However that is beside the point; if the separate epochs have similar parameter estimates, then why not combine them into a single least-squares fit?

The likely reason is that it does not work. Mathematically, fits to subsets of the data are *not* the same as an overall fit – any more than tangent lines to a curve are the same as the curve itself. Each individual fit over a single epoch fails to test the dependence of the model on $r_{as}, \alpha$ because of the very limited range those parameters vary across an epoch. Each individual fit will have its only value of the parameter $\eta$ and the other physical parameters.

This choice is inconsistent with the approach taken for asteroids that lack enough bands dominated by thermal emission to fit for $\eta$. In such cases, the NEOWISE authors assign a value of $\eta$ that is taken from the average value found for the same group; $\eta = 1.0$ for main belt asteroids (Masiero et al., 2011), $\eta = 0.77$ for Hildas (Grav et al., 2011a), $\eta = 1.4$ for NEO (Mainzer et al., 2011b). This method allows for a paradoxical situation in which $\eta$ can be forced to have the same value across thousands of different asteroids in a group, but allowed to vary across the observational epochs only days apart for an individual asteroid.

The per-epoch approach by definition rules out pooling data from the 3B and PC mission phases to the FC data. Since those phases lack one or more bands by definition, their observations may not contain enough information in an epoch to perform thermal modeling. Instead, as done in this study, one can pool data from multiple epochs. Indeed, that would be the most natural approach for statistical modeling. The NEOWISE papers do not explain why the per-epoch approach is either valid or desirable.

A separate issue in these papers is that only one kind of model—the NEATM model—is used. While this is typical of previous asteroid thermal modeling studies, most of the machinery of modern statistics is based on comparing alternative models and choosing the one that is best via relative comparison between the models. Many other well-known models are available; several alternatives are examined here in Section 4, along with alternative parameterizations within the NEATM model.

Masiero/MB:NIR, for instance, fits models using a separate IR albedo for the W1, W2 bands $p_{IR1}, p_{IR2}$ rather than using one value for both of them $p_{IR}$, as done in Masiero/MB:Pre. Does the two-albedo model fit the data better than the one-albedo model? If so, is the improvement in fit, as judged by a standard statistical metric, such as the corrected Akaike Information Criterion (AIC$_c$), sufficiently large to justify the extra parameter? These questions are answerable using AIC$_c$, and this potentially can create better fits or reveal new insights.

## 4.2 Issues with Error Analysis

The error analysis presented in the NEOWISE papers included a series of Monte Carlo trials that were performed while varying the parameters. Mainzer et al. (Mainzer et al., 2011d) explain the process in this way:

> Error bars on the model magnitudes and subsolar temperatures were determined for each object by running 25 Monte Carlo (MC) trials that varied the objects' H values by 0.3 mag and the WISE magnitudes by their error bars using Gaussian probability distributions.

The number of Monte Carlo trials listed here is too small to yield an accurate representation of the errors. Other NEOWISE papers cite as many as 100 samples, but even that is an unusually small sample size to use as a basis for reliable error estimates, especially when varying multiple parameters simultaneously.

In order to estimate the errors in the model parameter estimates, such as the diameter *D*, the analysis must consider all sources of error—which includes both sampling error and non-sampling error. Non-sampling errors occur in measurement—i.e., the instrumental errors caused by fundamental WISE observational characteristics. These errors are addressed in the error analysis of the NEOWISE papers. Sampling error also comes into play, however, because the models assume that asteroids are spherical and have uniform surface properties ($p_v, p_{IR}, \eta, \epsilon$). This is an idealized form that only approximates real asteroids. The NEOWISE authors recognize this fact:

> Although we do not expect all, or even most, asteroids to have a spherical shape, our observations covering ~36 hours smooth out rotation effects and allow us to determine the effective diameter of a spherical body with the same physical properties. Long period rotators (P ~days) with large amplitudes, e.g., binary asteroids with mass fraction ~1, will have poor fits resulting in a moderate mismeasurement of albedo and diameter. (Masiero et al., 2011)

It is true that, with sufficiently many observations at many rotational phases, the non-spherical or non-uniform effects will average out, permitting an estimate of the effective spherical equivalent. And it is also true that poorly sampled asteroids are subject to errors in albedo and diameter.

Crucially, however, these factors are relevant not merely for binary asteroids or those having a long rotational period—they affect *all* asteroid observations fit to the model to one degree or another. Unless the asteroid truly is a perfect sphere with uniform albedo, there will be some variation observed at different rotational phases.

This is an expression of the statistical population sampling error, which arises from the fact that observations consist of a finite sample of the asteroids' rotational phases. Although the quoted passage above recognizes the phenomenon qualitatively, no quantitative estimate of this error is made in the NEOWISE error analysis. The consequence is to dramatically underestimate the total error in parameter estimates for the majority of asteroids.

In addition to being problematic in its own right, this underestimate of error has significant consequences for the rest of the analysis. A common method for including observational errors into data analysis is to use weighted least squares (WLS) or minimum $\chi^2$ curve fitting, in which the least-square error for observational data point $i$ is weighted by $1/\sigma_i^2$, where $\sigma_i$ is the estimated error. WLS or minimum $\chi^2$ are only appropriate if the $\sigma_i$ include *all* sources of error. However, as is well known (Feigelson and Babu, 2012; Ivezić et al., 2014; Wall and Jenkins, 2012), this approach fails if the error estimates $\sigma_i$ are too small relative to the actual error because the weights over-constrain possible least-squares solutions.

My own attempts confirm that WLS with the WISE data does fail in this fashion if the $\sigma_i$ from the WISE pipeline are used, particularly for data across multiple epochs. This is understandable because the $\sigma_i$ are only intended to be estimates of the *instrumental measurement error*. This might be the only relevant kind of error in the case of some WISE observations (of stars, for example). For irregularly shaped, rotating asteroids, however, the $\sigma_i$ capture only one aspect of the error. Population sampling error is also present, but it is *not* covered in the $\sigma_i$.

Although I cannot be certain, it seems likely that the unconventional aspects of the NEOWISE analysis may have been motivated by the failure of WLS or minimum $\chi^2$ using the WISE observational $\sigma_i$. That would explain why the $\sigma_i$ were arbitrarily increased (both overall and for saturated data). It may also explain the puzzling 40% rule and the approach of dividing the curve fits into epochs.

Each of these modifications to standard practice would appear have the effect of remedying the failure by either increasing the $\sigma_i$ to approximate the population sampling error, or eliminating inconvenient data that prevented a fit, or by performing the fit over such small time spans that population sampling error was reduced.

These are not appropriate data analysis solutions to the problem: fitting over small intervals is not the same as fitting a model to the overall data set. The problem is conceptual: observational error does not include population sampling error. Note also that normal statistical practice would be to report the failure of conventional modeling rather than simply adopt ad hoc remedies without explaining the motivation.

A better answer to the dilemma posed by the failure of WLS via the WISE instrumental $\sigma_i$ is to adopt a modeling approach that can cope with the population sampling error, as done here. Bootstrap resampling (Feigelson and Babu, 2012; Ivezić et al., 2014; Wall and Jenkins, 2012) is a highly flexible tool for this situation. Bootstrapping allows use of ordinary least-squares (OLS) fitting which avoids the problematic weighting factors that underestimate the error.

## 4.3 Numerical Coincidences in NEOWISE Results

The accuracy of asteroid thermal modeling is best assessed by comparison to asteroids that have their diameter $D$ estimated by other means, such as radar, stellar occultations or spacecraft close approaches (ROS). An early NEOWISE paper on thermal model calibration (Mainzer/TMC in Table 1) references a series of such prior studies (see ROS studies, Table 1).

Surprisingly in 152 cases across 98 asteroids, the diameter $D$ given in the NEOWISE results matches a radar diameter from these references exactly to the quoted accuracy which is the nearest meter, or typically 5-6 digits. This number of exact matches to that precision is extremely unlikely. Since this issue is vitally important to the interpretation and credibility of NEOWISE results it is worth considering in detail.

Mainzer/TMC presents parameters for 50 asteroids, which I initially took to be the results of NEOWISE model fits for objects taken from prior studies, presented for the purpose of comparing the accuracy of diameter estimates with those of radar. The title to Table 1 reads "Spherical NEATM models were created for 50 objects ranging from NEOs to Irregular Satellites in Order to Characterize the Accuracy of Diameter and Albedo Errors Derived from NEOWISE Data". The bottom of Table 1 has the following note: "The diameters and H values used to fit each object from the respective source data (either radar, spacecraft imaging, or occultation) are given." My initial interpretation was that the phrase "diameters and H values used to fit" meant that the diameters and $H$ values in the table were fit, for the objects selected from the respective sources.

However this interpretation appears to be incorrect. Instead, the results presented in Table 1 of Mainzer/TMC appears to have the diameter was set to be equal to the radar diameter by fiat, so the "fit" (presuming that fitting was in fact done) only involved other parameters. Instead of being fit from NEOWISE data, the diameter was apparently set equal to the prior value from the literature. While the wording used in the paper may have intended to convey that, it is at best ambiguous in its as-published form. For example, how could one use Table 1 to "characterize the accuracy" of the NEOWISE diameter estimate if the diameter was simply set equal to the value one would compare it to?

Instead the aim of this paper is apparently to construct spherical asteroids with the known ROS diameters and then compare the flux from them to the observed flux, and then make a rough estimate of the scaling of diameter according to the resulting errors. This is, in effect, the opposite of the technique used to create the ~158,000 NEOWISE model fits, and as such makes little sense as a method to assess its accuracy. This is discussed further below in section 5.5.

Of the 53 results presented for 50 objects in Mainzer/TMC, two are moons (Himalia, Phoebe) rather than asteroids and will not be further considered here. Nine asteroids (128, 164121, 1866, 194, 308, 341843, 522, 7335 and 951) are either not found in the prior references, or they are there with a different diameter than listed in Table 1 of Mainzer/TMC. Asteroid 68216 is referenced to a private communication by L. Benner, and search engines do not reveal subsequent publication of a diameter estimate for that asteroid, so it could not be verified. All of others (40 results for 37 asteroids) exactly match a value in the previous studies.

In light of this it seems reasonable to assume that the correct interpretation of Table 1 in Mainzer/TMC is not as being solely derived from NEOWISE results, but rather a combination of radar/occultation/spacecraft derived diameters being imposed as a constraint upon NEOWISE observations. . Under this interpretation the 10 asteroids that cannot be matched to a prior study referenced in Mainzer/TMC *must* appear in unreferenced prior studies, but I have not been able to find these prior studies.

Adding these 10 to the list of "known" diameters, we find that the diameter matches exactly for 173 cases across 103 asteroids. Of these, 50 of the matches are in Mainzer/TMC. However, there are 117 results for 102 asteroids in Masiero/MB:Pre and 6 results for 3 asteroids in Mainzer/NEO:Pre. Note that three of these asteroids (164121, 341843, 68216) appear in both Masiero/MB:Pre and Mainzer/NEO:Pre despite the fact that one is about main belt asteroids and the other is about NEOs so one would expect the lists to be distinct. So, between the two of Masiero/MB:Pre and Mainzer/NEO:Pre there are 123 exact matches across 102 asteroids. Neither of these papers references prior radar, occultation or spacecraft studies, nor do they mention in any way setting diameters equal to such studies. The asteroids with an exact match are listed in Table 4.

The exact match issue is particularly puzzling since both Masiero/MB:Pre and Mainzer/NEO:Pre specify diameter to the nearest meter but the diameters from ROS sources are typically rounded to the nearest kilometer, which is evident in Table 4. This is quite obvious in the data table excerpt published in Masiero/MB:Pre. Since Masiero/MB:Pre and Mainzer/NEO:Pre both round to the nearest meter, their results almost never have fits which are an integral number of kilometers. Indeed out of 139,372 fits in Masiero/MB:Pre, there are only 5 examples where $D$ is an integral number of kilometers (>9km) which are not one of the exact matches to ROS asteroids. The asteroids in some cases also have other, distinct NEOWISE results published in Masiero/MB:NIR, Grav/Hilda, Mainzer/PP:3 *without* having exactly matching diameters. There are several interesting cases that may shed light on the issue.

Existence of prior diameter data does not guarantee an exact match. There are 21 asteroids (or 17% of those that have estimates in both NEOWISE and prior references) that have diameter estimates in one or more of the prior references yet have a value for $D$ in Masiero/MB:Pre that does not exactly. As an example, asteroid 381 has $D = 129,100$ in (Shevchenko and Tedesco, 2006), a reference used in many other exact matches, but Masiero/MB:Pre has $D = 129,000$, and thus is not an exact match.

In many cases there are multiple diameters within the prior ROS references, but only one is chosen. There does not seem to be any pattern to this – i.e. there do not seem to be preferred references. Nor is there a preference for matches to the ROS source that claims the smallest error.

In the case of asteroid 522, Masiero/MB:Pre has $D = 83,700$ which is an exact match for (Shevchenko and Tedesco, 2006), but Mainzer/TMC has $D = 84,000$. In all other cases the value in Mainzer/TMC matches the one used in Masiero/MB:Pre. Asteroid 128 has $D = 188,000$, but the only reference has $D = 141,000$. If the interpretation of Mainzer/TMC as deliberately setting $D$ to a prior known value is correct then there must be unreferenced sources for asteroids 522 and 128 that were chosen over the other referenced values. If that interpretation is not correct and Mainzer/TMC did not deliberately set diameters equal to reference values, then its exact matches may be due to the same process as Masiero/MB:Pre and Mainzer/NEO:Pre.

One way to ascertain the uniqueness of this situation is through comparable scenarios. Comparing 1728 diameter estimates for asteroids that are in NEOWISE and IRAS (Tedesco et al., 2002b), only one case of an exact match occurs. A second comparable is between IRAS and the known prior studies; 192 asteroids in the IRAS results are also in these prior studies, but there are no exact matches. The exact matches found between NEOWISE and the prior studies are clearly quite different than either IRAS with NEOWISE, or IRAS with the prior studies.

Several examples are plotted in Figure 17, including the NEOWISE curves from Masiero/MB:Pre and Mainzer/NEO:Pre are plotted assuming either the incident solar flux $F_{Sun}$ (*green lines*) or with $F_{Sun}/4$ (dashed magenta lines) and 50 bootstrap fits (*gray lines*.) It is evident that close fitting of the data points is not the cause of the exact matches. For asteroid 114, for example, the NEOWISE curves with $F_{Sun}$ miss two of the four bands. In the case of asteroid 198, the NEOWISE curves do hit every band, but that is mostly because there are some extreme outliers in W1 and W2. In the case of asteroid 488, not only does the diameter match exactly, so does the estimated standard error.

One possibility to examine is whether the exact matches could be due to chance. As a way to estimate the likelihood of an exact match, I constructed a sample data set using the known WISE observations for asteroid 114, replacing the observed WISE magnitudes with calculated magnitudes using NEATM, and assuming that physical parameters of one of the NEOWISE fits with $D = 100,000$ m. When fit with OLS, that data returned $D = 100,000$ m as expected. Monte Carlo trials were run, adding normally distributed errors with zero mean and $\sigma = 0.001$ magnitude added to each WISE magnitude. With 20,000 of these trials one can fit to recover the $D$ estimates, and round them to the nearest meter and count the exact matches. This method estimates that the probability of an exact match to 1 meter rounding is 0.0158 for $\sigma = 0.001$. The likelihood of getting 123 matches is thus approximately 1 chance in $10^{210}$.

Yet even this calculation is overly optimistic. It assumes that all of the data points match the curve exactly, save for the added error. This is not the case in reality as shown in Figure 17, which shows plenty of scatter in the data points. The actual variation between WISE measurements is much larger than 0.001 magnitude. Indeed the NEOWISE group assumes that $\sigma \geq 0.03$ across all observations. With $\sigma = 0.01$ the probability drops approximately another order of magnitude to 0.0015, making the probability of 123 matches even more absurd. In addition, this calculation used all of the 116 data points (across all bands) for asteroid 114, yet NEOWISE fits typically use epochs with fewer points which constrain the fit less, and thus make the probability of an exact match even lower.

Since chance does not appear to be a viable option, that leaves two possibilities. One is that the exact matches are due to an inadvertent error, such as file corruption or a software bug. The other is that the NEOWISE authors intended to force the diameter equal to prior studies, perhaps with a goal of improving the diameter and albedo estimates.

The deliberate scenario seems unlikely, because it would be a serious violation ethical statistical practice (Hogan et al., 2016) Such a dramatic alteration of the NEOWISE study would need to be clearly and explicitly documented in the relevant papers – yet the papers clearly state the opposite. In Masiero/MB:Pre, section 3 is titled "diameter and albedo determination through thermal modeling". It offers no hint that diameters were determined by any other method. Similarly, Mainzer/NEO:Pre contains no discussion of setting the diameter by means other than thermal modeling. The notes to the main results presented in Table 1 of that paper say "This table contains the preliminary thermal fit results based on the First Pass version of the WISE data processing as described in the text." This wording would appear to leave no room for setting diameters by copying values from the ROS literature.

A different ethical dilemma is posed by the conflict between the presumed motive – to get a better diameter estimate – and the main purpose of the paper. Masiero/MB:Pre states clearly that "Using a NEATM thermal model fitting routine we compute diameters for over 100,000 Main Belt asteroids from their IR thermal flux, with errors better than 10%". If in fact the diameters were copied because they were better estimates, then the authors have a duty to warn us that the thermal model estimates for the 102 asteroids with ROS diameters were significantly worse than the ROS diameters and thus were substituted.

Another possible motive was to fix diameter to get better estimates of the other parameters like $\eta$ and $p_{IR}$, but this makes little sense because diameters and visible band albedo (which follows directly from $H$ and diameter) are the main focus of the paper.

Another issue with the intentional case is that there is no unique choice of diameter from radar, occultation or other means. The prior references have a single diameter for only 38% of the 102 asteroids that have exact matches. The remainder have from two to as many as five separate choices across the various referenced studies. If one were to deliberately impose "the" diameter, which one would be chosen? Normally such criteria would also be carefully disclosed. No such discussion exist and no clear pattern emerges from Table 4.

The presumptive goal in using the ROS diameters would be to get a better estimate than with diameters obtained by thermal modeling. However a recent study (Durech et al., 2015) explores the considerable theoretical and observational challenges in doing such "data fusion" across multiple data sources. It is *much* more involved than simply setting diameters equal.

None of these arguments should be considered dispositive, but they show the challenges in assuming that copying the diameters was the intention of the NEOWISE authors.

Regardless of how the diameters came to be set equal, the empirical fact that they were set equal has important consequences. Using ROS diameter rather than thermal modeling determined diameter might produce a better model for the ~150 asteroids where ROS estimates are available, but that leaves ~150,000 where it is not an option. The *only* way to ascertain the accuracy of diameters estimated through thermal modeling is to compare them to the ROS diameters.

The NEOWISE results cannot be validated in this way because the diameter of the test cases have been corrupted with an exact copy of the cases they would be compared to. This undermines the credibility of all of the NEOWISE results.

| Asteroid | NEOWISE Published Results | | | Prior Radar/Occulation/Spacecraft References | | |
|---|---|---|---|---|---|---|
| | $D$ | $\sigma_D$ | Paper | $D$ | $\sigma_D$ | Paper |
| 2 | 544000 | 60714 | Masiero/MB:Pre | 544000 | N/A | Shevchenko et al, 2006 |
| | | | | 522000 | N/A | Shevchenko et al, 2006 |
| | | | | 539000 | 28000 | Durech et al, 2011 |
| 5 | 115000 | 9353 | Masiero/MB:Pre | 115000 | 6000 | Durech et al, 2011 |
| | 115000 | 12000 | Mainzer/TMC | | | |
| | 108290 | 3700 | Masiero/MB:NIR | | | |
| | | | | 120000 | 14000 | Magri et al, 1999 |
| 6 | 185000 | 10688 | Masiero/MB:Pre | 185000 | 10000 | Magri et al, 1999 |
| | 195640 | 5440 | Masiero/MB:NIR | | | |
| | | | | 180000 | 40000 | Durech et al, 2011 |
| 8 | 140000 | 1160 | Masiero/MB:Pre | 140000 | 7000 | Durech et al, 2011 |
| | 140000 | 14000 | Mainzer/TMC | | | |
| | 147490 | 1030 | Masiero/MB:NIR | | | |
| | | | | 138000 | 9000 | Magri et al, 1999 |
| | | | | 141000 | 10000 | Durech et al, 2011 |
| | | | | 160800 | N/A | Shevchenko et al, 2006 |
| 13 | 227000 | 38000 | Mainzer/TMC | 227000 | 30000 | Magri et al, 2007 |
| | 227000 | 25948 | Masiero/MB:Pre | | | |
| 15 | 259000 | 35511 | Masiero/MB:Pre | 259000 | 30000 | Magri et al, 2007 |
| | 231690 | 2230 | Masiero/MB:NIR | | | |
| 19 | 223000 | 43596 | Masiero/MB:Pre | 223000 | 41000 | Magri et al, 1999 |
| | 196370 | 300 | Masiero/MB:NIR | | | |
| | | | | 210000 | 10000 | Timmerson et al, 2010 |
| 36 | 103000 | 11451 | Masiero/MB:Pre | 103000 | 11000 | Magri et al, 2007 |
| | 103000 | 1000 | Mainzer/TMC | | | |
| 38 | 116000 | 15501 | Masiero/MB:Pre | 116000 | 13000 | Magri et al, 2007 |
| | 92250 | 490 | Masiero/MB:NIR | | | |
| 39 | 163000 | 14025 | Masiero/MB:Pre | 163000 | 12000 | Durech et al, 2011 |
| | 163000 | 16000 | Mainzer/TMC | | | |
| | 179480 | 1680 | Masiero/MB:NIR | | | |
| | | | | 177900 | N/A | Shevchenko et al, 2006 |
| 46 | 124000 | 9641 | Masiero/MB:Pre | 124000 | 9000 | Magri et al, 1999 |
| | | | | 124000 | 9000 | Magri et al, 2007 |
| 47 | 138000 | 13000 | Mainzer/TMC | 138000 | N/A | Shevchenko et al, 2006 |
| | 138000 | 11108 | Masiero/MB:Pre | | | |
| | 125140 | 3510 | Masiero/MB:NIR | | | |

| Asteroid | NEOWISE Published Results | | | Prior Radar/Occultation/Spacecraft References | | |
|---|---|---|---|---|---|---|
| | D | $\sigma_D$ | Paper | D | $\sigma_D$ | Paper |
| 50 | 100000 | 7596 | Masiero/MB:Pre | 100000 | 13000 | Magri et al, 2007 |
| | 84070 | 240 | Masiero/MB:NIR | | | |
| 51 | 142600 | 12503 | Masiero/MB:Pre | 142600 | N/A | Shevchenko et al, 2006 |
| | 138160 | 970 | Masiero/MB:NIR | | | |
| 53 | 115000 | 8000 | Mainzer/TMC | 115000 | 14000 | Magri et al, 2007 |
| | 115000 | 10324 | Masiero/MB:Pre | | | |
| 54 | 142000 | 14758 | Masiero/MB:Pre | 142000 | 9000 | Durech et al, 2011 |
| | 142000 | 14000 | Mainzer/TMC | | | |
| | | | | 135000 | 20000 | Durech et al, 2011 |
| | | | | 165000 | 19000 | Magri et al, 2007 |
| 60 | 60000 | 3519 | Masiero/MB:Pre | 60000 | 7000 | Magri et al, 2007 |
| | 43220 | 570 | Masiero/MB:NIR | | | |
| 64 | 50300 | 9389 | Masiero/MB:Pre | 50300 | N/A | Shevchenko et al, 2006 |
| | 58290 | 1080 | Masiero/MB:NIR | | | |
| | | | | 51000 | 10000 | Durech et al, 2011 |
| | | | | 52000 | 10000 | Durech et al, 2011 |
| 80 | 79000 | 9125 | Masiero/MB:Pre | 79000 | 10000 | Magri et al, 1999 |
| | 79000 | 1393 | Masiero/MB:Pre | | | |
| | 68560 | 1030 | Masiero/MB:NIR | | | |
| | 74250 | 3000 | Masiero/MB:NIR | | | |
| | | | | 67000 | 11000 | Durech et al, 2011 |
| 83 | 84000 | 2283 | Masiero/MB:Pre | 84000 | 9000 | Magri et al, 2007 |
| | 84000 | 8000 | Mainzer/TMC | | | |
| | 89640 | 2650 | Masiero/MB:NIR | | | |
| | | | | 81000 | 2000 | Shepard et al, 2010 |
| 84 | 79000 | 4867 | Masiero/MB:Pre | 79000 | 13000 | Magri et al, 1999 |
| 85 | 163000 | 18648 | Masiero/MB:Pre | 163000 | 15000 | Durech et al, 2011 |
| | 163000 | 16000 | Mainzer/TMC | 163000 | 19000 | Magri et al, 2007 |
| | 169460 | 4520 | Masiero/MB:NIR | | | |
| | | | | 163700 | N/A | Shevchenko et al, 2006 |
| | | | | 175900 | N/A | Shevchenko et al, 2006 |
| 94 | 187500 | 7256 | Masiero/MB:Pre | 187500 | N/A | Shevchenko et al, 2006 |
| | 187500 | 27000 | Mainzer/TMC | | | |
| | 173770 | 4190 | Masiero/MB:NIR | | | |
| 97 | 83000 | 6000 | Mainzer/TMC | 83000 | 5000 | Shepard et al, 2010 |
| | 83000 | 5099 | Masiero/MB:Pre | 83000 | 10000 | Magri et al, 1999 |
| | 100720 | 640 | Masiero/MB:NIR | | | |
| 105 | 119000 | 17337 | Masiero/MB:Pre | 119000 | 17000 | Magri et al, 1999 |
| | 119000 | 11000 | Mainzer/TMC | | | |

| Asteroid | NEOWISE Published Results | | | Prior Radar/Occulation/Spacecraft References | | |
|---|---|---|---|---|---|---|
| | D | σ_D | Paper | D | σ_D | Paper |
| | | | | 103700 | N/A | Shevchenko et al, 2006 |
| 109 | 89000 | 6165 | Masiero/MB:Pre | 89000 | 9000 | Magri et al, 2007 |
| | 82590 | 620 | Masiero/MB:NIR | | | |
| | | | | 88200 | N/A | Shevchenko et al, 2006 |
| 110 | 89000 | 6336 | Masiero/MB:Pre | 89000 | 9000 | Shepard et al, 2010 |
| | 88200 | 2710 | Masiero/MB:NIR | | | |
| 111 | 135000 | 18583 | Masiero/MB:Pre | 135000 | 15000 | Magri et al, 2007 |
| | 126340 | 230 | Masiero/MB:NIR | | | |
| 114 | 100000 | 11926 | Masiero/MB:Pre | 100000 | 14000 | Magri et al, 2007 |
| | 100000 | 8874 | Masiero/MB:Pre | | | |
| | 100000 | 9000 | Mainzer/TMC | | | |
| | 100000 | 16000 | Mainzer/TMC | | | |
| | 94180 | 950 | Masiero/MB:NIR | | | |
| | 99150 | 3180 | Masiero/MB:NIR | | | |
| 128 | 188000 | 9002 | Masiero/MB:Pre | 188000 | 29000 | Mainzer/TMC |
| | 188000 | 29000 | Mainzer/TMC | | | |
| | 162510 | 1300 | Masiero/MB:NIR | | | |
| | | | | 141000 | 37000 | Magri et al, 2007 |
| 129 | 129500 | 14772 | Masiero/MB:Pre | 129500 | N/A | Shevchenko et al, 2006 |
| | 128720 | 610 | Masiero/MB:NIR | | | |
| | | | | 113000 | 12000 | Shepard et al, 2010 |
| | | | | 118000 | 19000 | Durech et al, 2011 |
| 134 | 112200 | 10798 | Masiero/MB:Pre | 112200 | N/A | Shevchenko et al, 2006 |
| 135 | 77000 | 8000 | Mainzer/TMC | 77000 | 7000 | Shepard et al, 2010 |
| | 77000 | 7833 | Masiero/MB:Pre | | | |
| | 71040 | 2650 | Masiero/MB:NIR | | | |
| 137 | 144000 | 11272 | Masiero/MB:Pre | 144000 | 16000 | Magri et al, 2007 |
| | 128680 | 530 | Masiero/MB:NIR | | | |
| 139 | 164000 | 19000 | Mainzer/TMC | 164000 | 22000 | Magri et al, 1999 |
| | 164000 | 25212 | Masiero/MB:Pre | | | |
| | 151120 | 1600 | Masiero/MB:NIR | | | |
| | | | | 160200 | N/A | Shevchenko et al, 2006 |
| 141 | 137100 | 14556 | Masiero/MB:Pre | 137100 | N/A | Shevchenko et al, 2006 |
| | 117920 | 1360 | Masiero/MB:NIR | | | |
| 145 | 151000 | 23000 | Mainzer/TMC | 151000 | 18000 | Magri et al, 2007 |
| | 151000 | 11272 | Masiero/MB:Pre | | | |
| | 151000 | 8563 | Masiero/MB:Pre | | | |
| | 124470 | 510 | Masiero/MB:NIR | | | |
| | 127780 | 360 | Masiero/MB:NIR | | | |

| Asteroid | NEOWISE Published Results | | | Prior Radar/Occulation/Spacecraft References | | |
|---|---|---|---|---|---|---|
| | D | $\sigma_D$ | Paper | D | $\sigma_D$ | Paper |
| 182 | 44000 | 15494 | Masiero/MB:Pre | 44000 | 10000 | Magri et al, 2007 |
| | 44000 | 4279 | Masiero/MB:Pre | | | |
| | 39520 | 390 | Masiero/MB:NIR | | | |
| | 44980 | 510 | Masiero/MB:NIR | | | |
| 192 | 93000 | 6795 | Masiero/MB:Pre | 93000 | 9000 | Magri et al, 2007 |
| | 93000 | 8366 | Masiero/MB:Pre | | | |
| | 87120 | 4160 | Masiero/MB:NIR | | | |
| | 98780 | 1240 | Masiero/MB:NIR | | | |
| | | | | 95000 | 13000 | Magri et al, 1999 |
| 198 | 57000 | 7000 | Mainzer/TMC | 57000 | 8000 | Magri et al, 2007 |
| | 57000 | 10064 | Masiero/MB:Pre | | | |
| | 54320 | 340 | Masiero/MB:NIR | | | |
| 208 | 45000 | 5000 | Mainzer/TMC | 45000 | 10000 | Durech et al, 2011 |
| | 45000 | 5000 | Mainzer/TMC | | | |
| | 45000 | 4195 | Masiero/MB:Pre | | | |
| | 45000 | 4591 | Masiero/MB:Pre | | | |
| | 40060 | 590 | Masiero/MB:NIR | | | |
| | 40900 | 600 | Masiero/MB:NIR | | | |
| | | | | 44300 | N/A | Shevchenko et al, 2006 |
| 211 | 143000 | 21629 | Masiero/MB:Pre | 143000 | 16000 | Magri et al, 2007 |
| | 143000 | 13000 | Mainzer/TMC | | | |
| | 141130 | 2490 | Masiero/MB:NIR | | | |
| 216 | 138000 | 19374 | Masiero/MB:Pre | 138000 | 20000 | Magri et al, 2007 |
| | 119190 | 3400 | Masiero/MB:NIR | | | |
| | | | | 104300 | N/A | Shevchenko et al, 2006 |
| | | | | 124000 | 15000 | Shepard et al, 2010 |
| 225 | 128000 | 16129 | Masiero/MB:Pre | 128000 | 16000 | Magri et al, 2007 |
| | 95930 | 1250 | Masiero/MB:NIR | | | |
| 230 | 109000 | 16000 | Mainzer/TMC | 109000 | 14000 | Magri et al, 1999 |
| | 109000 | 13025 | Masiero/MB:Pre | | | |
| | 111330 | 1230 | Masiero/MB:NIR | | | |
| | | | | 101800 | N/A | Shevchenko et al, 2006 |
| 238 | 146500 | 8679 | Masiero/MB:Pre | 146500 | N/A | Shevchenko et al, 2006 |
| | 155660 | 750 | Masiero/MB:NIR | | | |
| | | | | 145300 | N/A | Shevchenko et al, 2006 |
| 247 | 134000 | 13425 | Masiero/MB:Pre | 134000 | 15000 | Magri et al, 2007 |
| | 130940 | 510 | Masiero/MB:NIR | | | |

| Asteroid | NEOWISE Published Results | | | Prior Radar/Occulation/Spacecraft References | | |
|---|---|---|---|---|---|---|
| | D | $\sigma_D$ | Paper | D | $\sigma_D$ | Paper |
| 248 | 54000 | 4913 | Masiero/MB:Pre | 54000 | N/A | Shevchenko et al, 2006 |
| | 50120 | 310 | Masiero/MB:NIR | | | |
| 306 | 51600 | 6333 | Masiero/MB:Pre | 51600 | N/A | Shevchenko et al, 2006 |
| | 47200 | 130 | Masiero/MB:NIR | | | |
| | | | | 49000 | 5000 | Durech et al, 2011 |
| | | | | 53000 | 5000 | Durech et al, 2011 |
| 308 | 144400 | 13864 | Masiero/MB:Pre | 144400 | N/A | Shevchenko et al, 2006 |
| | 144000 | 13000 | Mainzer/TMC | | | |
| | 128580 | 1560 | Masiero/MB:NIR | | | |
| | | | | 117100 | N/A | Shevchenko et al, 2006 |
| 313 | 96000 | 7809 | Masiero/MB:Pre | 96000 | 14000 | Magri et al, 2007 |
| 324 | 229000 | 8145 | Masiero/MB:Pre | 229000 | 12000 | Magri et al, 1999 |
| | | | | 229000 | 12000 | Magri et al, 2007 |
| | 220690 | 1440 | Masiero/MB:NIR | | | |
| | | | | 235500 | N/A | Shevchenko et al, 2006 |
| 334 | 174100 | 12788 | Masiero/MB:Pre | 174100 | N/A | Shevchenko et al, 2006 |
| | 198770 | 5600 | Grav/Hilda | | | |
| 336 | 69000 | 3364 | Masiero/MB:Pre | 69000 | 9000 | Magri et al, 2007 |
| 345 | 99000 | 9000 | Mainzer/TMC | 99000 | N/A | Shevchenko et al, 2006 |
| | 99000 | 11469 | Masiero/MB:Pre | | | |
| 347 | 51000 | 3218 | Masiero/MB:Pre | 51000 | 5000 | Shepard et al, 2010 |
| | 48610 | 120 | Masiero/MB:NIR | | | |
| 350 | 99500 | 10675 | Masiero/MB:Pre | 99500 | N/A | Shevchenko et al, 2006 |
| | 99500 | 6354 | Masiero/MB:Pre | | | |
| | 99500 | 5000 | Mainzer/TMC | | | |
| | 121360 | 2460 | Masiero/MB:NIR | | | |
| | 128730 | 1200 | Masiero/MB:NIR | | | |
| 354 | 165000 | 15613 | Masiero/MB:Pre | 165000 | 18000 | Magri et al, 2007 |
| | 148970 | 420 | Masiero/MB:NIR | | | |
| 356 | 131000 | 9686 | Masiero/MB:Pre | 131000 | 15000 | Magri et al, 1999 |
| | 118480 | 1540 | Masiero/MB:NIR | | | |
| 404 | 98700 | 3450 | Masiero/MB:Pre | 98700 | N/A | Shevchenko et al, 2006 |
| | 94970 | 950 | Masiero/MB:NIR | | | |
| 405 | 125000 | 17427 | Masiero/MB:Pre | 125000 | 16000 | Magri et al, 2007 |
| | 108890 | 310 | Masiero/MB:NIR | | | |
| 420 | 144000 | 5683 | Masiero/MB:Pre | 144000 | N/A | Shevchenko et al, 2006 |
| | 138700 | 3450 | Masiero/MB:NIR | | | |
| 429 | 70000 | 6683 | Masiero/MB:Pre | 70000 | 10000 | Magri et al, 2007 |
| | 72210 | 2190 | Masiero/MB:NIR | | | |

| | NEOWISE Published Results | | | Prior Radar/Occultation/Spacecraft References | | |
|---|---|---|---|---|---|---|
| Asteroid | D | $\sigma_D$ | Paper | D | $\sigma_D$ | Paper |
| 431 | 68600 | 3617 | Masiero/MB:Pre | 68600 | N/A | Shevchenko et al, 2006 |
| | 94580 | 1030 | Masiero/MB:NIR | | | |
| 444 | 163000 | 36000 | Mainzer/TMC | 163000 | 27000 | Magri et al, 2007 |
| | 163000 | 12600 | Masiero/MB:Pre | | | |
| | 163000 | 22144 | Masiero/MB:Pre | | | |
| | 158860 | 2320 | Masiero/MB:NIR | | | |
| | 159330 | 490 | Masiero/MB:NIR | | | |
| | | | | 172400 | N/A | Shevchenko et al, 2006 |
| 488 | 150000 | 21000 | Mainzer/TMC | 150000 | 21000 | Magri et al, 2007 |
| | 150000 | 11326 | Masiero/MB:Pre | | | |
| | 148840 | 3490 | Masiero/MB:NIR | | | |
| 497 | 40000 | 4881 | Masiero/MB:Pre | 40000 | 8000 | Shepard et al, 2010 |
| | 40930 | 320 | Masiero/MB:NIR | | | |
| 522 | 83700 | 4850 | Masiero/MB:Pre | 83700 | N/A | Shevchenko et al, 2006 |
| | 84000 | 9000 | Mainzer/TMC | 84000 | 9000 | Mainzer/TMC |
| 566 | 134000 | 15000 | Mainzer/TMC | 134000 | N/A | Shevchenko et al, 2006 |
| | 134000 | 6627 | Masiero/MB:Pre | | | |
| | 167380 | 3490 | Masiero/MB:NIR | | | |
| 568 | 75800 | 5641 | Masiero/MB:Pre | 75800 | N/A | Shevchenko et al, 2006 |
| | 85190 | 980 | Masiero/MB:NIR | | | |
| 622 | 29000 | 5417 | Masiero/MB:Pre | 29000 | 8000 | Magri et al, 2007 |
| | 21870 | 240 | Masiero/MB:NIR | | | |
| 654 | 127000 | 13000 | Mainzer/TMC | 127000 | 18000 | Magri et al, 2007 |
| | 127000 | 20474 | Masiero/MB:Pre | | | |
| | 116300 | 2380 | Masiero/MB:NIR | | | |
| 678 | 42000 | 2371 | Masiero/MB:Pre | 42000 | 4000 | Shepard et al, 2010 |
| | 39590 | 580 | Masiero/MB:NIR | | | |
| 704 | 312000 | 30000 | Mainzer/TMC | 312000 | 33000 | Magri et al, 2007 |
| | 312000 | 17000 | Mainzer/TMC | | | |
| | 312000 | 19727 | Masiero/MB:Pre | | | |
| | 312000 | 34517 | Masiero/MB:Pre | | | |
| | 306310 | 1030 | Masiero/MB:NIR | | | |
| | 308300 | 1510 | Masiero/MB:NIR | | | |
| | | | | 326100 | N/A | Shevchenko et al, 2006 |
| | | | | 332800 | N/A | Shevchenko et al, 2006 |
| 757 | 36700 | 2272 | Masiero/MB:Pre | 36700 | N/A | Shevchenko et al, 2006 |
| | 32890 | 240 | Masiero/MB:NIR | | | |
| 758 | 85000 | 9365 | Masiero/MB:Pre | 85000 | 7000 | Shepard et al, 2010 |
| | 88980 | 630 | Masiero/MB:NIR | | | |

| Asteroid | NEOWISE Published Results | | | Prior Radar/Occulation/Spacecraft References | | |
|---|---|---|---|---|---|---|
| | $D$ | $\sigma_D$ | Paper | $D$ | $\sigma_D$ | Paper |
| 771 | 29000 | 2544 | Masiero/MB:Pre | 29000 | 2000 | Shepard et al, 2010 |
| | 29000 | 1403 | Masiero/MB:Pre | | | |
| | 28170 | 330 | Masiero/MB:NIR | | | |
| | 29320 | 170 | Masiero/MB:NIR | | | |
| 779 | 77000 | 6578 | Masiero/MB:Pre | 77000 | 2000 | Shepard et al, 2010 |
| | 80570 | 2220 | Masiero/MB:NIR | | | |
| 791 | 82500 | 5957 | Masiero/MB:Pre | 82500 | N/A | Shevchenko et al, 2006 |
| | 99800 | 11030 | Masiero/MB:NIR | | | |
| 796 | 45000 | 5133 | Masiero/MB:Pre | 45000 | 2000 | Shepard et al, 2010 |
| | | | | 45000 | 6000 | Magri et al, 1999 |
| | 43580 | 300 | Masiero/MB:NIR | | | |
| 914 | 77000 | 13126 | Masiero/MB:Pre | 77000 | 10000 | Magri et al, 2007 |
| | 76190 | 490 | Masiero/MB:NIR | | | |
| | | | | 91200 | N/A | Shevchenko et al, 2006 |
| 925 | 58000 | 6000 | Mainzer/TMC | 58000 | 16000 | Durech et al, 2011 |
| | 58000 | 4841 | Masiero/MB:Pre | | | |
| | 57500 | 440 | Masiero/MB:NIR | | | |
| | | | | 59200 | N/A | Shevchenko et al, 2006 |
| 951 | 12000 | 1000 | Mainzer/TMC | 12000 | 1000 | Mainzer/TMC |
| | 12200 | 813 | Masiero/MB:Pre | 12200 | N/A | Thomas et al, 1994 |
| | 13210 | 130 | Masiero/MB:NIR | | | |
| 976 | 65000 | 3598 | Masiero/MB:Pre | 65000 | N/A | Shevchenko et al, 2006 |
| | 83200 | 540 | Masiero/MB:NIR | | | |
| 1512 | 65000 | 7000 | Mainzer/TMC | 65000 | N/A | Shevchenko et al, 2006 |
| | 65000 | 4137 | Masiero/MB:Pre | | | |
| | 79870 | 770 | Grav/Hilda | | | |
| 1627 | 9000 | 817 | Masiero/MB:Pre | 9000 | 900 | Ostro et al, 1990 |
| | 9000 | 1000 | Mainzer/TMC | | | |
| | 8370 | 75 | Mainzer/NEO:Pre | | | |
| | | | | 10200 | N/A | Veeder et al, 1989 |
| 1866 | 8700 | 1000 | Mainzer/TMC | 8700 | 1000 | Mainzer/TMC |
| | 8700 | 590 | Masiero/MB:Pre | | | |
| | 6597 | 189 | Mainzer/NEO:Pre | | | |
| 1963 | 45000 | 7925 | Masiero/MB:Pre | 45000 | 9000 | Magri et al, 2007 |
| | 35540 | 230 | Masiero/MB:NIR | | | |
| 7335 | 900 | 400 | Mainzer/TMC | 900 | 400 | Mainzer/TMC |
| | 932 | 153 | Mainzer/NEO:Pre | | | |
| | 950 | 417 | Masiero/MB:Pre | | | |
| | | | | 1000 | N/A | Mahapatra et al, 2002 |

| Asteroid | NEOWISE Published Results | | | Prior Radar/Occulation/Spacecraft References | | |
|---|---|---|---|---|---|---|
| | $D$ | $\sigma_D$ | Paper | $D$ | $\sigma_D$ | Paper |
| 68216 | 1400 | 200 | Mainzer/TMC | 1400 | 200 | Mainzer/TMC |
| | 1400 | 145 | Mainzer/NEO:Pre | | | |
| | 1400 | 145 | Masiero/MB:Pre | | | |
| | 1400 | 262 | Mainzer/NEO:Pre | | | |
| | 1400 | 262 | Masiero/MB:Pre | | | |
| 164121 | 1100 | 300 | Mainzer/TMC | 1100 | 300 | Mainzer/TMC |
| | 1100 | 186 | Masiero/MB:Pre | | | |
| | 1100 | 88 | Masiero/MB:Pre | | | |
| | 1100 | 186 | Mainzer/NEO:Pre | | | |
| | 1100 | 88 | Mainzer/NEO:Pre | | | |
| | 1717 | 550 | Mainzer/PP:3 | | | |
| | 1738 | 577 | Mainzer/PP:3 | | | |
| 2005 CR37 | 1000 | 123 | Masiero/MB:Pre | 1000 | N/A | Benner et al, 2006 |
| | 1000 | 200 | Mainzer/TMC | | | |
| | 1201 | 236 | Mainzer/NEO:Pre | | | |

**Table 4. NEOWISE fits with diameters that exactly match reference ROS values**. Asteroids where there is an exact match between the NEOWISE results and prior studies referenced by NEOWISE are shown. Cases where there is no match are shaded. The table contains 117 exact matches for 102 asteroids in Masiero/MB:Pre, 6 exact matches for 3 asteriods from Mainzer/NEO:Pre. Cases where Mainzer/TMC appears in the right as a source are presumed to be from an unreferenced source prior to NEOWISE – see text for discussion. Note that three asteroids appear in both Masiero/MB:Pre and Mainzer/NEO:Pre.

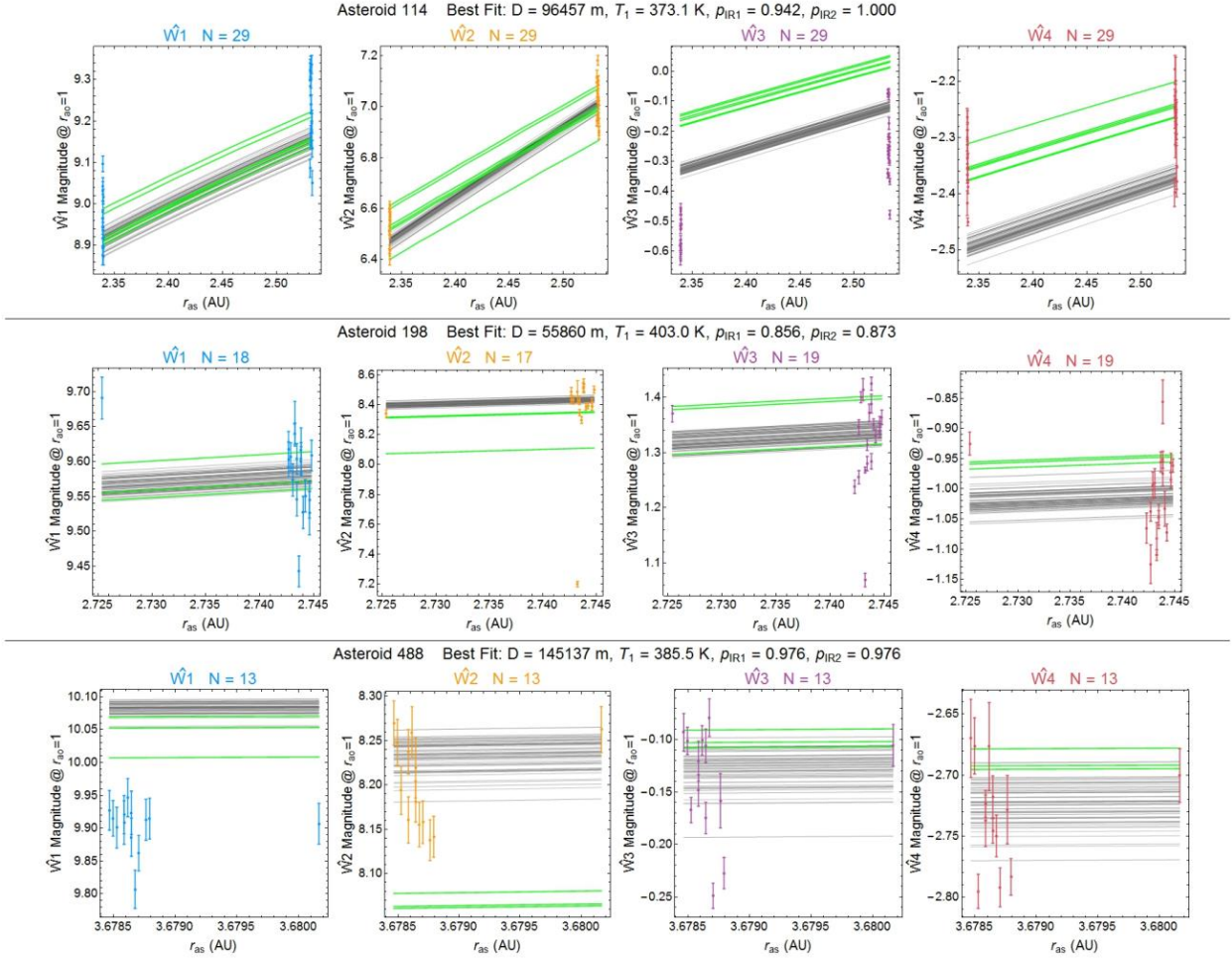

**Figure 17. Plots of example NEOWISE fits with diameters that exactly match reference ROS values.** Each row shows the W1 through W4 data points for an asteroid from the FC data set only, along with the WISE magnitude fluxes for NEOWISE fits, and 50 bootstrap fits to the FC data (*gray lines*).

## 5. Results

In contrast to the best-fit models of the NEOWISE papers, which frequently miss their underlying data bands (Figure 16), a high percentage of the model fits presented here pass through all or most of the data bands (Figure 18).

Only one asteroid, from Mainzer/Tiny (of the 32 having sufficient data), misses three bands; a few asteroids from this and other data sets miss one or two bands. The available data for these asteroids seem to be poorly suited for the thermal models used here. Note that the fits summarized by Figure 18 use fewer free parameters (i.e. $D, T_1, \epsilon_1, \epsilon_2$) than for NEOWISE ($D, \eta, G, H, p_{IR1}, p_{IR2}$.)

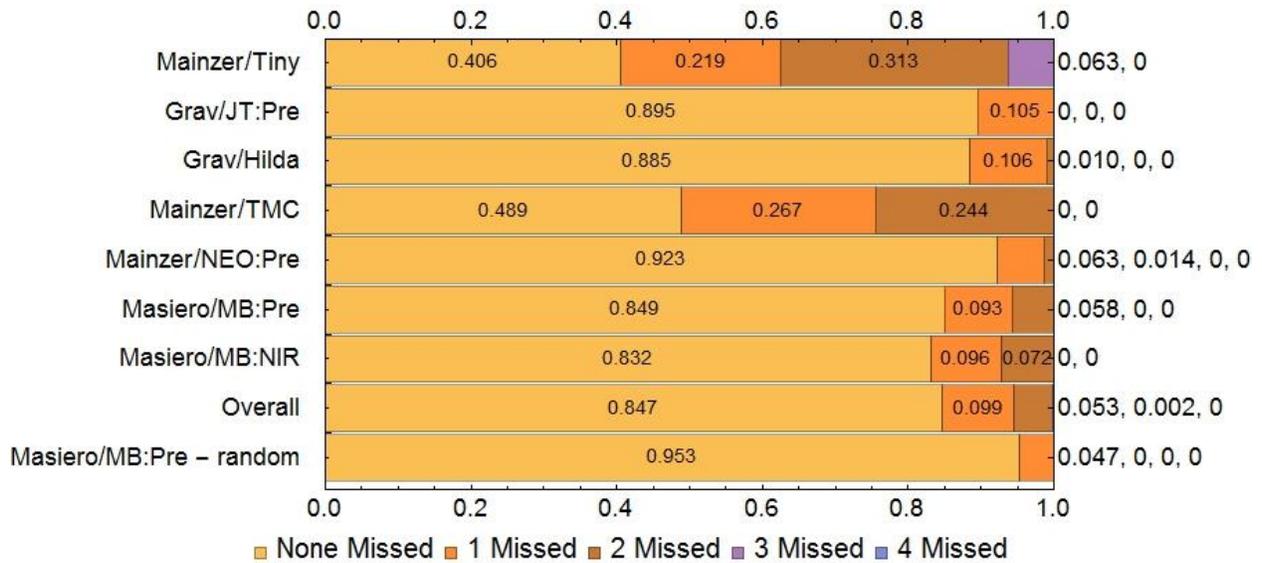

**Figure 18. Verification that model best fits do not miss many WISE bands.** Model fits "miss" a band of data if all of the data points in that band are either below or above the model. Values for bars too small to be labeled appear to the right. In contrast to Figure 16, the vast majority of model fits for this study pass through all of the bands of data (i.e., none missed).

In a departure from prior asteroid modeling, this study uses multiple models and multiple parameterizations, which compete to find the best fit, as evaluated objectively by using $AIC_c$. Figure 19 shows the distribution of fits by model across 500 bootstrap trials for each asteroid. Each of the models is roughly a peer in terms of its contribution to results, as measured by Akaike weights. The blackbody model contributes slightly more than the others for the Mixed data set, whereas NEATM is the largest contributor for the Main Belt Random (MBR) data set.

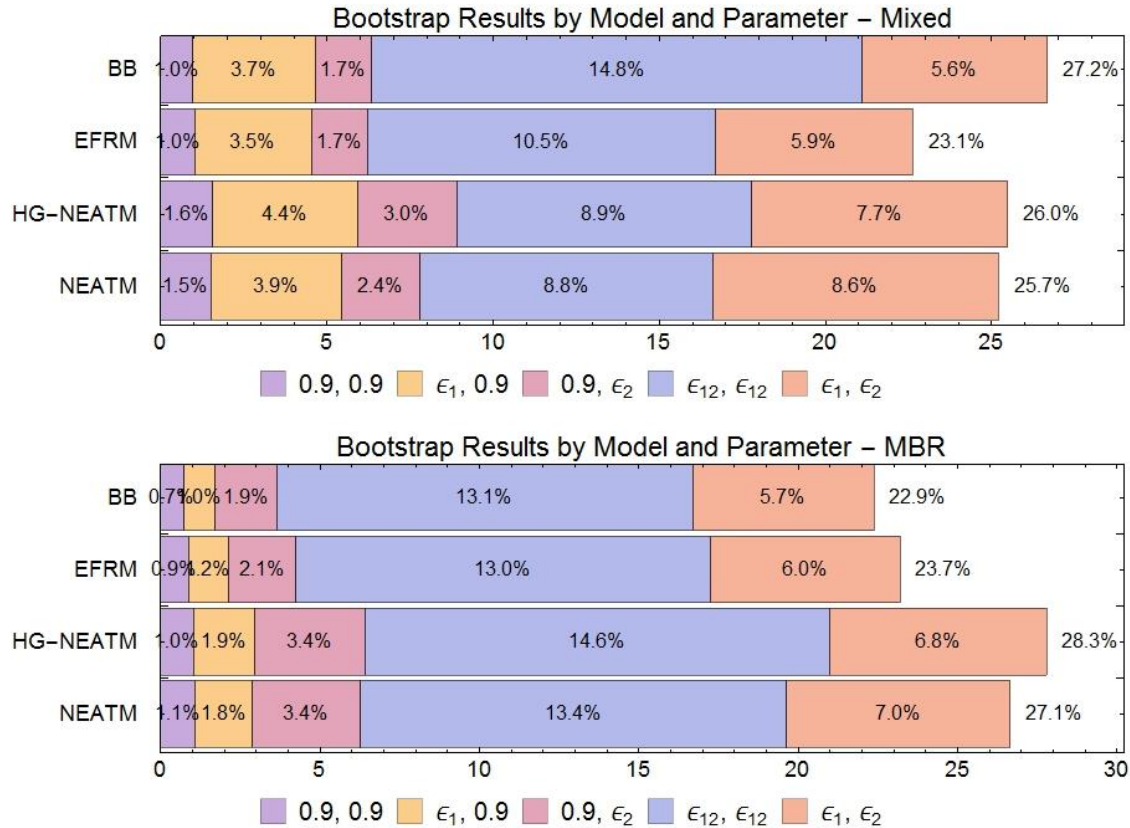

**Figure 19. Distribution of fits by model.** The participation of each model and parameter combination in the bootstrap results, weighted by Akaike weights, is given in the upper chart for the Mixed asteroid set and in the lower chart for the Main Belt Random asteroid set MBR.

The results shown in Figure 19 reveal that the NEATM model does not enjoy a monopoly on being the best fit to asteroid spectra. The isothermal blackbody model (BB) and EFRM models have no phase dependence, and have distinct spectral distributions. The HG-NEATM model has a stronger phase dependence than NEATM does, but the same spectral distribution. Given the success of models other than NEATM in producing best fits (or contributing to the mix of best fits), it seems clear that some diversity of spectral distribution and phase behavior in models is necessary to match the observed NEOWISE asteroid data. Future research could probably identify models that fit even better, at least for some subset of asteroids. This is also a ripe area for thermophysical models that delve deeper into the fundamental causes of variation in the spectra.

The non-randomly sampled Mixed data set shows different characteristics from the randomly selected MBR set. This is not surprising because in comparison with the random set, the Mixed set contains a disproportionate number of both large and small asteroids, and is about half non-main belt asteroids.

The parameterization differences in how $\epsilon_1$ and $\epsilon_2$ are handled are relatively consistent among models. In the mixed asteroid set, the largest contribution to best fits, 9% to 16% for each model and about 43.5% of the total, comes from models in which $\epsilon_1 = \epsilon_2 = \epsilon_{12}$. But the other parameterizations together still capture about half of the contribution from each model. In the MBR data set (lower chart in Figure 19), the contributions are a bit more evenly distributed among the various models and parameterizations.

Figure 20 illustrates the rank-ordered distribution of each model across bootstrap trials. In the case of the NEATM model, there are 4 asteroids in the Mixed data set that have the NEATM model as the only result across 500 bootstrap trials; at the other end of the curve there are about 900 asteroids where NEATM is the most common model and has 25% of the cases. Since there are four models, no asteroid can have the most common model be less than 25%. In between these limits the rank distribution shows the number of asteroids that have each model as the most common one within their bootstrap trial best fits.

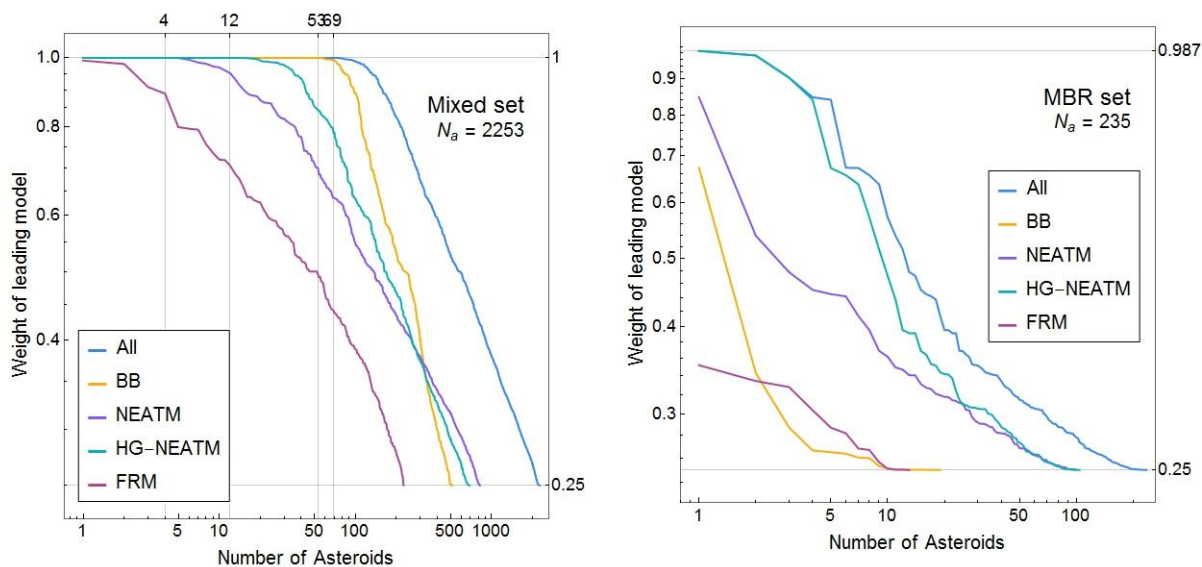

**Figure 20. Asteroid counts by rank distribution of most common model.** The Akaike weighted fraction of bootstrap results is shown on the y-axis against the number of asteroids for the mixed set (*left*) and random main belt set (*right*). When the fraction is 1, all bootstrap samples for that asteroid are from the model; in other cases, the fraction is the $AIC_c$-weighted fraction of bootstrap samples for which the model is the most common by $AIC_c$. In the Mixed asteroid data set (*left*), there are 69 asteroids for which one model is the best fit for all 500 bootstrap trials, for the rest some mixture of models is found. The random main belt data set (*right*) is more diverse—only one asteroid and one model is as high as 98.7% of the bootstrap trials.

The bootstrap trials yield a distribution of each of the key parameters. Histograms for diameter $D$ are shown in Figure 21. For some asteroids, the parameter distribution appears to be approximately a normal distribution having a single maximum. But for many others, the distributions are asymmetric and/or have multiple peaks, as illustrated by the examples.

## 5.1 Diameter Estimates

Unbiased estimates of the parameters and their statistical distributions were calculated from the parameter histograms. Because the NEOWISE results are presented as the mean and standard deviation, the natural error metric is the ratio of the standard deviation of the parameter to its mean. In the case of the bootstrap resampling analysis, it is advantageous to express the error as the ratio of the length of the 95% confidence interval to the median; this metric is better suited to asymmetric and non-normal distributions. In the case of diameter $D$, the two error metrics are given by

$$\frac{\sigma_D}{D} = \frac{\sigma_D}{\text{mean}(D)}, \frac{\Delta D}{4D} = \frac{\text{quantile}(D, 0.975) - \text{quantile}(D, 0.025)}{4\,\text{median}(D)} \quad . \tag{45}$$

In the case of a normally distributed parameter, both metrics are identical. In many real cases, the two produce very similar values, even for non-normally distributed data. But they can differ substantially when distributions are very asymmetric, such as the case of asteroid 19764 shown in Figure 21.

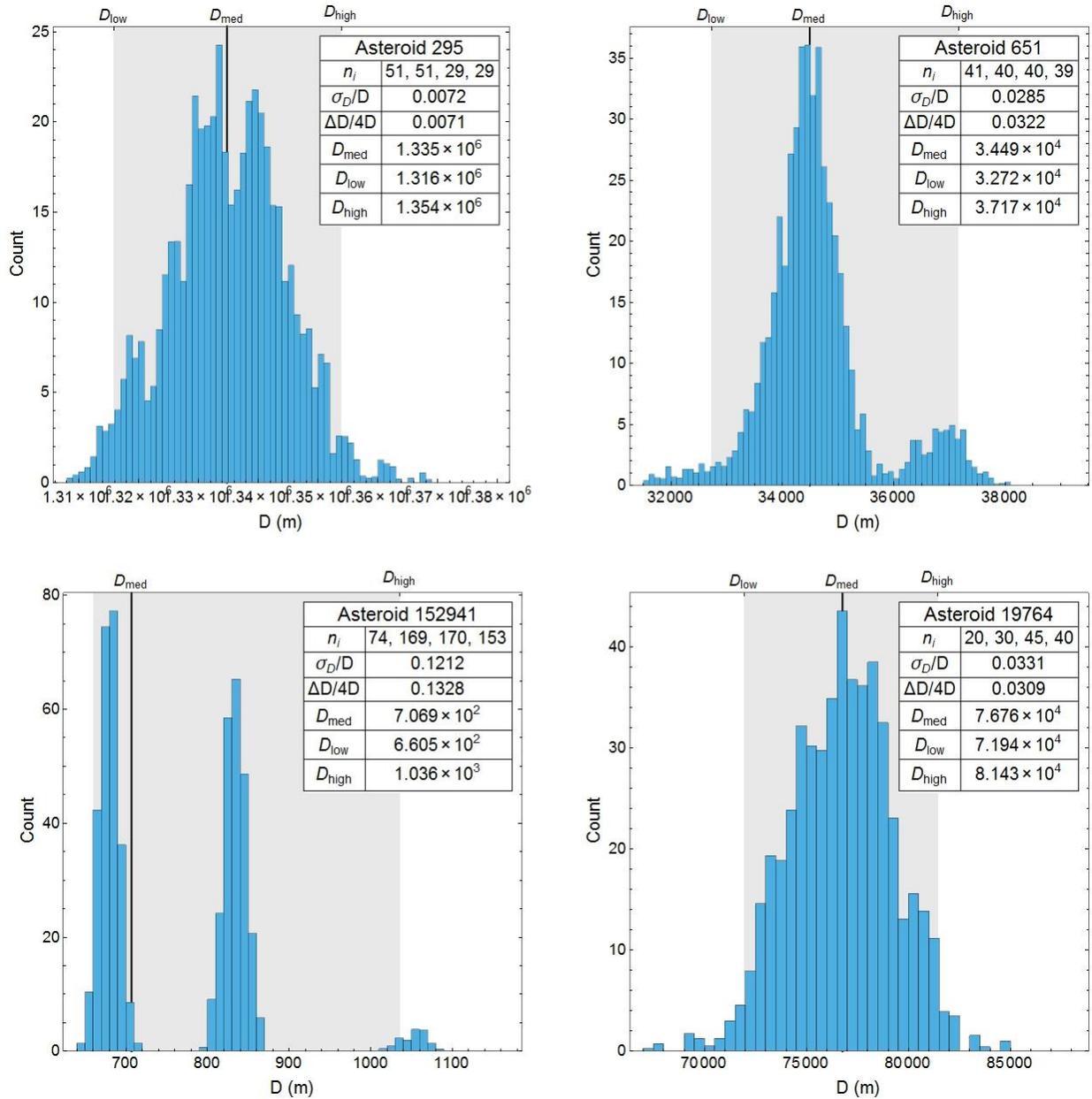

**Figure 21. Histograms of diameter from 500 bootstrap trials for example asteroids.** Although $D$ follows an approximately normal, single-peaked distribution for certain asteroids (*right*), in many other cases the distribution is more complex and shows multiple peaks or asymmetry. The median of the distribution is indicated by a black vertical line; the shaded region between $D_{\text{low}}$ and $D_{\text{high}}$ highlights the 95% confidence interval. Note that these are raw bootstrap trials which may abnormally high values for $D$ (and have $T_1 < 200K$) – see text for discussion.

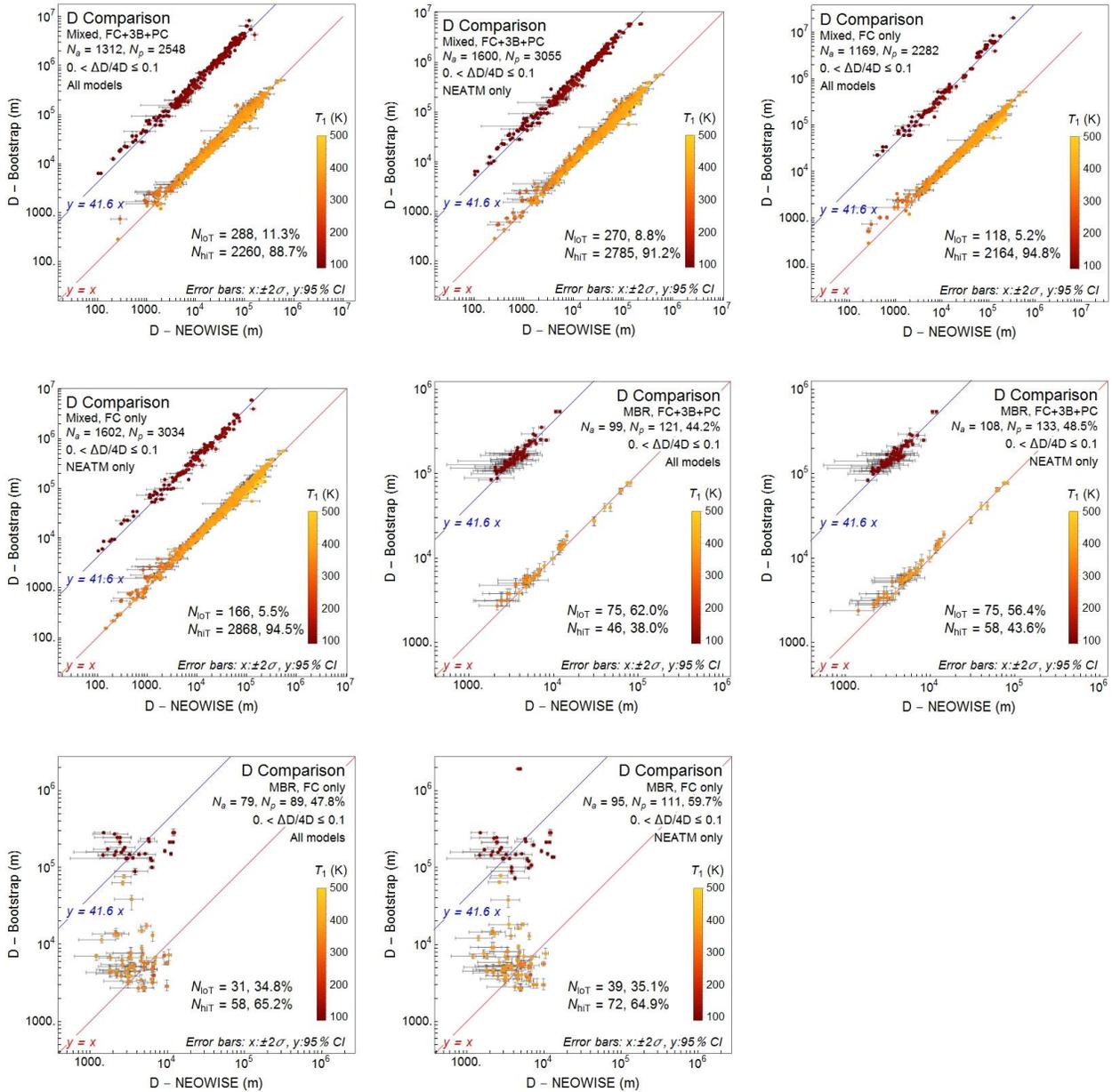

**Figure 22. Comparison of diameter in NEOWISE and bootstrap results.** Each chart plots the diameter $D$ for NEOWISE and the bootstrap results for different data sets and models, filtered to only show points where the error ratio $\Delta D/4D \leq 0.1$ to allow the points to be seen more clearly. For NEOWISE the error bars are $\pm 2\sigma_D$, for bootstrap results the error bars are the 95% confidence interval. The number of asteroids in each plot is $N_a$, the number of data points is $N_p$; they differ because there are multiple NEOWISE fits for some asteroids. The temperature $T_1$ for each bootstrap estimate is color coded using the scale at the right side of each graph. The points cluster either near the line $y = x$ or $y = 41.6\,x$, each of which represent alternative solutions consistent with the examples shown in Figure 15. The number of points with $T_1 < 200$K is $N_{\text{loT}}$, the number with $T_1 \geq 200$K is $N_{\text{hiT}}$.

Figure 22 compares the diameter as determined by the bootstrap analysis to the diameter published in the NEOWISE papers. The figure plots only those asteroids for which $\Delta D/4D \leq 0.1$, since it is in these cases that the error analysis suggests we may have a reasonably accurate estimate. The error bars are 4 times the values in Equation (45); they thus span the 95% confidence interval in the case of the bootstrap results and $\pm 2\sigma$ in the case of the NEOWISE results.

Figure 22 draws on four data sets: the full-data (FC+3B+PC) and the cryo-only (FC) versions of both the Mixed and MBR data sets. It also includes results for all four models (NEATM, BB, EFRM, HG-NEATM) versus results that only use the NEATM model. These variations illustrate that the behavior is robust even as additional data and additional models are incorporated.

Although most of the data points have error bars that overlap with the line $y = x$, indicating that the NEOWISE and bootstrap estimates are similar, however many differ substantially. In nearly all cases shown in Figure 22, a smaller subset of data points cluster near the line $y = 41.6x$, and have $T_1 \leq 200$ K. In one case (MBR data set, FC only, all models), no such points appear on the plot, but they do occur for $\Delta D/4D > 0.1$.

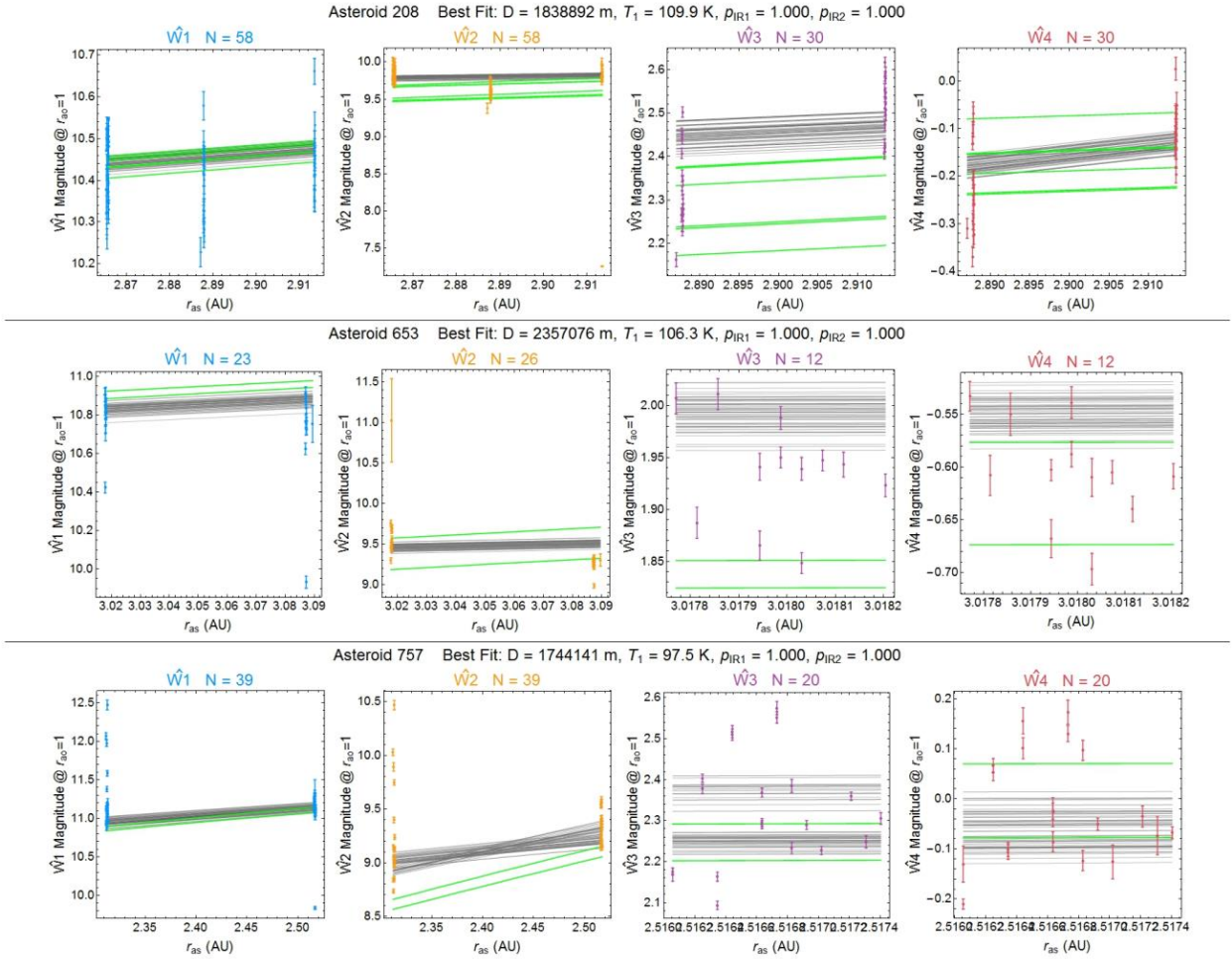

**Figure 23. Bootstrap fits for three example asteroids with low $T_1$.** The plots show three example asteroids from the Mixed (FC+3B+PC) data set fit with 500 bootstrap trials using all four models. In each plot, the best fit to the data points is drawn in black, with the listed parameters. Examples of 50 bootstrap fits are shown (*gray lines*), with shades of gray indicating the AIC$_c$ value (*darker gray* = lower AIC$_c$). Also plotted are NEOWISE fits for epochs of each asteroid (*green lines*)

Figure 23 presents examples of the bootstrap fit for three representative asteroids with low $T_1$. The fits shown in Figure 23 are generally well behaved and draw on sufficient numbers of data points, and a careful examination of the parameter space revealed no evidence that they are trapped in local minima. The error estimates are, in general, no different from those for asteroids that have higher values of $T_1$.

It seems extremely unlikely that the low $T_1$ are physically accurate; I do not propose that ~12% of asteroids are enormous and very cold. However I believe that it is important to attempt to understand why application of NEATM and the other models produce these results. There is something unusual going on with these asteroids which causes straightforward thermal modeling to fail for these cases. The plots in Figure 23 show that they are not obviously pathological in terms of number of data points or error estimates, so I do not believe these cases can be dismissed as being due to observational errors.

In particular, these odd results occurs from a straightforward application of NEATM thermal modeling in the situation where there is indeterminacy in the thermally dominated bands. Conversely any attempts to "fix" the problem are therefore *not* NEATM modeling, but rather some new and modified model. This phenomenon has not previously been noted in the asteroid thermal modeling literature, but it could occur in contexts other than WISE/NEOWISE as studied here.

The low $T_1$ fits imply high values of IR albedo (at least one of $p_{IR1} \approx 1, p_{IR2} \approx 1$)—or, equivalently, low values of emissivity ($\epsilon_1, \epsilon_2$). Although one could impose an upper limit on $p_{IR}$, doing so only makes the fits worse; the points shift slightly from being near the line $y = 41.6x$ to a lower multiple, but still far from $y = x$. Moreover, the fits shown do match the spectrum, and there is no justification for arbitrarily imposing a constraint.

The proximate reason for these problematic fits is the indeterminacy in $\widehat{W3} - \widehat{W4}$ that is explored in Figure 7. Two different pairs of parameters $(T_1, D)$ fit any given $\widehat{W3}, \widehat{W4}$ flux values because the same value of $\widehat{W3} - \widehat{W4}$ can occur at two different values of $T_1$ (see Figure 7), and the absolute $\widehat{W3}, \widehat{W4}$ depends on $D$. The "tie breaker" between the two different solutions must therefore come from the W1 and W2 bands.

In many cases the data from the W1, W2 bands may not be very temperature sensitive, as can be seen in Figure 6. In the cases where the bootstrap results yield $T_1 < 200$ K, the W1 and W2 band data have broken the tie in favor of the lower $T_1$ solution, which consequently implies a much large diameter $D$. Within the framework of the NEATM (and other models in this study), this is a direct consequence of the multiple solutions for $T_1$ shown in Figure 7. If WISE data included more purely thermal bands (e.g., if the W3 band had been split into two narrower bands), then this indeterminacy could likely be reduced.

As shown by the examples in Figure 23, the W1, W2 band data for these cases is not obviously problematic or pathological – they have relatively low scatter and are well fit by the bootstrap fits. The fact that they are well fit shows their role as "tie breakers".   The cases in Figure 23 have larger number of W1, W2 data points because they include data from FC+3B+PC, but low $T_1$ fits also occur for FC only cases in which the number of data points in the bands is more comparable.

The number and percentage of fits with  $T_1 < 200$K is also shown in Figure 22.  Cases with all four models have significantly fewer fits with $T_1 < 200$K than cases which use the NEATM model only, for both the Mixed and MBR data . Note that the number of models used also affects the count of fits with $\Delta D/4D > 0.1$ in each graph, so the absolute counts $N_{loT}$ may be a better indicator than the percentage.   This trend suggests that one reason for low $T_1$ solutions being chosen in the curve fitting is the overall fit quality.  With more or better models we may find fewer low $T_1$ cases.

The NEOWISE papers impose constraints to escape the problem. In particular they impose a limit $\eta \leq \pi$, which in effect prohibits the low $T_1$ solution at the expense of having poor fits, as shown in Figures 14, 15 and the green lines in Figure 23. However, there is no guarantee that this constraint means that the NEOWISE analysis finds the alternative, high $T_1$ solution (i.e. the second solution in Figure 7).  Indeed the trend that more models reduces the number of low $T_1$ fits suggests that this phenomenon requires more and better curve fitting rather than imposing constraints that create worse fits.

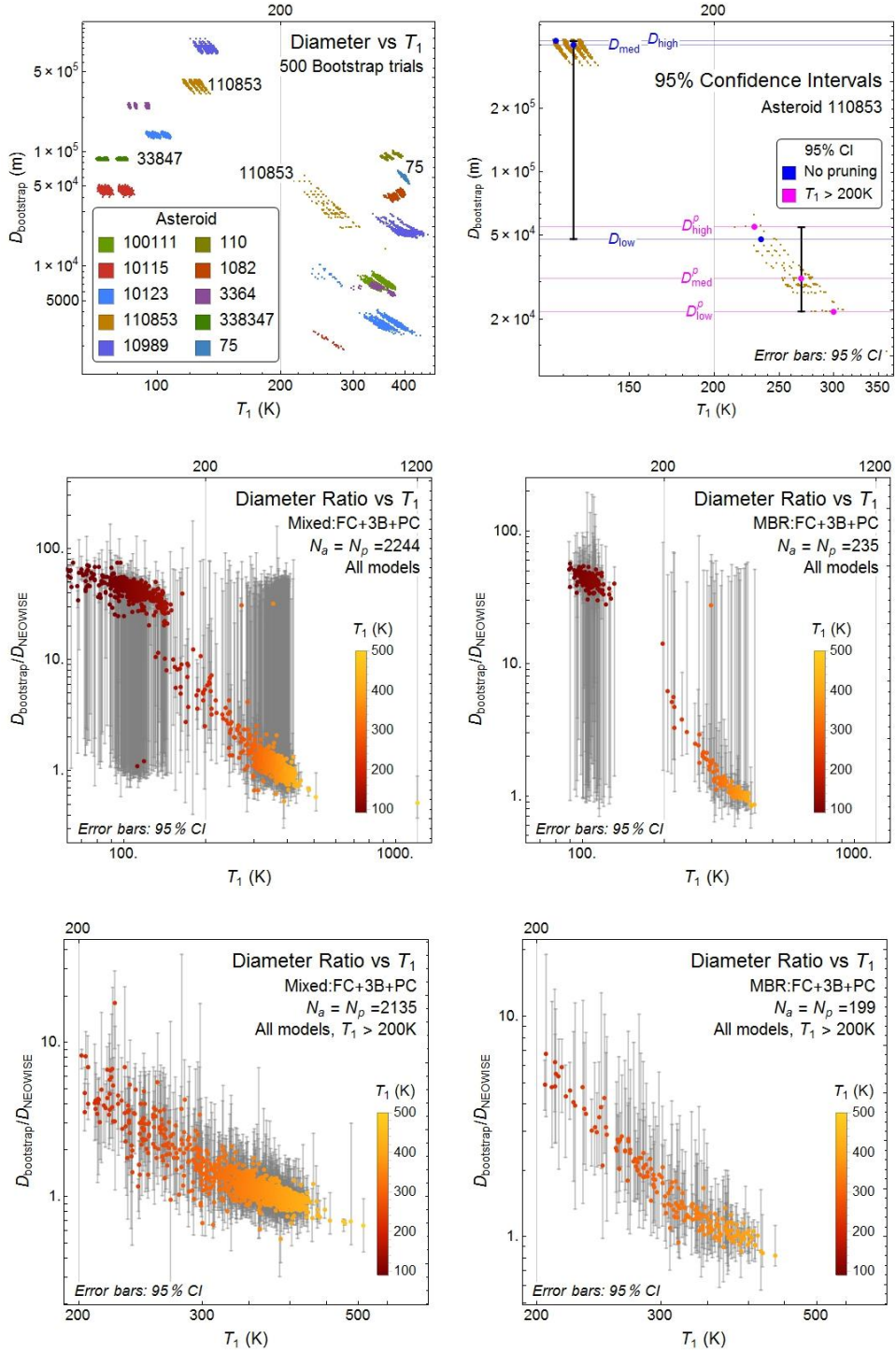

**Figure 24. Role of low $T_1$ fits in bootstrap results.** The $T_1$ versus $D_{\text{bootstrap}}$ point for 500 bootstrap trials are shown for ten example asteroids (*top left*). Some asteroids, like 33847 have bootstrap trials that all have $T_1 \leq 200\text{K}$, others like asteroid 75 exclusively have $T_1 > 200\text{K}$. Still others like 110853 have some bootstrap trials on each side of the $T_1 = 200\text{K}$ line. Pruning out fits with $T_1 \leq 200\text{K}$ changes the 95% confidence interval bounds $D_{\text{low}}, D_{\text{med}}, D_{\text{high}}$, and their pruned equivalents $D^p_{\text{low}}, D^p_{\text{med}}, D^p_{\text{high}}$ are shown for asteroid 110853 (*top right*). A plot of the ratio diameters NEOWISE and the bootstrap results from Monte Carlo evaluation is shown without pruning (*middle row*) and with pruning out fits with $T_1 \leq 200\text{K}$, (*bottom row*).

Nor does this offer an explanation of why the W1, W2 bands are biased in favor of the low $T_1$ solution. The answer is likely to lie in the effect of the $\epsilon_3, \epsilon_4$ assumptions shown in Figure 7, but further research will be needed to determine this. Something is going on with the low $T_1$ fits, and it behooves us to note that rather than just adopting arbitrary means to force them to conform to prior expectations.

The low $T_1$ fits shown in Figures 20 and 21 occur for the median diameter and temperature from a sample of 500 bootstrap trials. However they also occur for individual bootstrap trials as shown in Figure 24. Ten example asteroids are plotted. Some, like asteroids 75, 110, 1082 and 100111 only have bootstrap trials that have $T_1 > 200K$. Others like asteroid 33847 has $T_1 < 200K$ for all of its bootstrap trials. A third group, including 3364, 10115, 10123, 10989 and 110853 have bootstrap trials that cluster on both sides of the $T_1 = 200K$ line.

The presence of the low $T_1$ bootstrap trials changes the median value, and 95% confidence interval error bars as shown at the top right for asteroid 110853. If we artificially prune the fits with $T_1 \leq 200K$, it affects the medians and error bars.

We can use the method of Figure 11 to estimate the ratio of diameters. The center row of Figure 24 shows the ratio for all bootstrap trials. The large error bars are due to asteroids that have trials that fall on both sides of $T_1 = 200K$. Many of the fits which have a median value that falls above 200K actually have some fits that are low $T_1$, like asteroid 110853. The low error fits plotted in Figure 22 have all of their bootstrap trials on the same side of $T_1 = 200K$, but the higher error fits typically have trials on both sides.

The bottom row of Figure 24 shows the effect of artificially pruning out bootstrap trials with $T_1 \leq 200K$. Asteroids which have all of their trials below this threshold are removed from the analysis, (34 from the Mixed set with FC+3B+PC and one from MBR.) The remainder of the asteroids have greatly reduced estimated errors and their medians also shift. It is clearly evident that the bootstrap diameters are not comparable to the NEOWISE diameters. While some are near the line $D_{\text{bootstrap}}/D_{\text{NEOWISE}} \approx 1$ both median values (temperature color-coded points) and the error bars (gray) range way above and below.

Table 5 gives the estimated ratio $D_{\text{bootstrap}}/D_{\text{NEOWISE}}$ and its 95% confidence interval for each of the different combinations of data set and models studied here, with and without artificial pruning of trials with $T_1 \leq 200K$.

These figures tell us that the bootstrap methods do not in general give us the same results – the median is only slightly bigger but the scatter is enormous. Note that this numbers are with the artificial pruning; without pruning the lower bound and median are roughly similar but the upper bound becomes more like ~30.

| Data Set, Models, and Pruning | | | | $D_{\text{bootstrap}}/D_{\text{NEOWISE}}$ | | | |
|---|---|---|---|---|---|---|---|
| Pruning | Data Set | Bands | Models | 95% low | Median | 95% high | $E_\rho$ |
| None | Mixed | FC+3B+PC | All | 0.786 | 1.068 | 54.988 | 28.833 |
| | | | NEATM | 0.845 | 1.107 | 53.564 | 25.184 |
| | | FC only | All | 0.778 | 1.046 | 53.906 | 22.477 |
| | | | NEATM | 0.844 | 1.079 | 51.225 | 19.907 |
| | MBR | FC+3B+PC | All | 0.799 | 34.509 | 73.071 | 47.606 |
| | | | NEATM | 0.827 | 31.874 | 72.219 | 46.430 |
| | | FC only | All | 0.332 | 2.604 | 100.000 | 47.567 |
| | | | NEATM | 0.346 | 2.341 | 100.000 | 47.421 |
| $T_1 \leq 210K$ | Mixed | FC+3B+PC | All | 0.775 | 1.036 | 3.737 | 1.206 |
| | | | NEATM | 0.834 | 1.085 | 3.521 | 1.147 |
| | | FC only | All | 0.769 | 1.027 | 2.752 | 0.978 |
| | | | NEATM | 0.835 | 1.066 | 2.570 | 0.766 |
| | MBR | FC+3B+PC | All | 0.760 | 1.310 | 7.633 | 2.375 |
| | | | NEATM | 0.792 | 1.333 | 7.988 | 2.552 |
| | | FC only | All | 0.762 | 1.169 | 4.864 | 1.273 |
| | | | NEATM | 0.801 | 1.204 | 5.655 | 1.462 |

**Table 5. Ratio of diameter between NEOWISE and bootstrap results.** The table shows the aggregate ratio of diameters for different cases obtained by Monte Carlo sampling the bootstrap trials.

The large ranges in the ratios is primarily a consequence of scatter in the data points which is evident in most plots of WISE/NEOWISE data including examples in Figures 8, 14 and 21. There is a large variation in the WISE magnitudes, and this includes all of the bands. This is likely due to irregular asteroid shape or non-uniform surface properties, although in some cases it may also be due to the asteroid being a binary or higher order system.

## 5.2 Diameter Distribution

The distribution of asteroid sizes is important to many aspects of solar system science. Given the results summarized in Table 5 it is worth asking how much the distribution is affected by the analysis techniques used here versus those of NEOWISE.

Table 5 shows that any individual asteroid is likely to have a diameter that is quite different under NEOWISE and bootstrap analysis. Across a large number of asteroids we would expect the Table 5 results to create a net shift to larger diameters both because the median is greater than one, but also because the tails of the distribution are asymmetric.

The effect on the overall distribution is shown in Figure 25, which plots histograms of the probability distribution and cumulative probability distribution of diameters for the combination of the Mixed and MBR data sets; 2488 asteroids in total. There is a shift to larger diameters. In the case of asteroids with $D < 83,387$ m, the cumulative distribution is always lower for the bootstrap results than for NEOWISE. At $D = 200$ m, there are a factor of 2 fewer asteroids in the bootstrap distribution, dropping to equal by $D = 131,334$ m.

Note that this finding is conditioned on WISE/NEOWISE observation and may reflect some observational biases. For example this does not necessarily mean that there are, in actuality, fewer asteroids at small diameters. Instead the interpretation is that the asteroids that WISE/NEOWISE observes are larger than the NEOWISE results would imply, which in turn suggests that it may be less sensitive to small asteroids than previously believed.

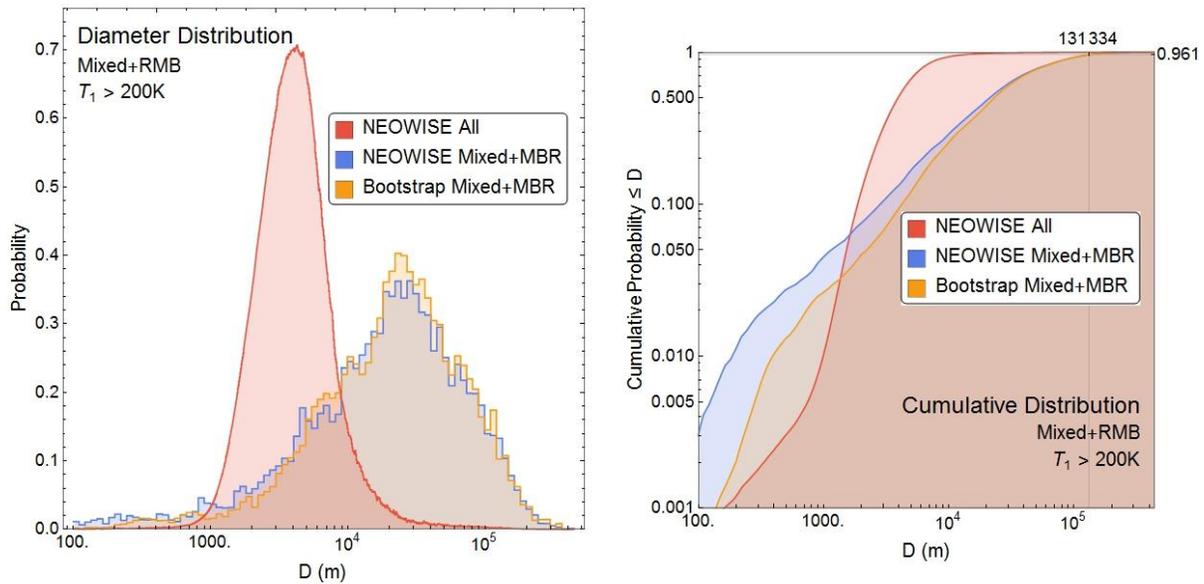

**Figure 25. Comparison of estimated diameter distributions.** The histogram of diameters of all 2488 asteroids in the Mixed and MBR data sets are shown for both the probability $P(x = D)$ (*left*) and cumulative probability $P(x \leq D)$ (*right*) based on a Monte Carlo sampling of 10,000 trials for each asteroid. They are compared to histograms for NEOWISE results for the same asteroids, as well as all NEOWISE asteroids. The bootstrap and NEOWISE distributions for the Mixed+MBR asteroids are similar to each other but quite different than the overall NEOWISE histogram. The bootstrap distribution has fewer asteroids with $D < 131,334$ m than NEOWISE for the same asteroids.

In order to reconstruct and estimate the actual distribution of asteroids one must take account of potential observational biases. Doing this comprehensively would require a larger set of asteroids than studied here.

It is interesting to see that the overall NEOWISE distribution of diameters is markedly different from the Mixed and MBR subsets asteroids studied here; they clearly are not typical, at least in the sense of the histogram, to NEOWISE asteroids overall. This is a cautionary note in extrapolating conclusions found for these asteroids to the larger set.

## 5.3 Diameter Error Estimates

The error estimation by NEOWISE and the bootstrap method can be compared in a number of ways (Figure 26). Plots in the leftmost column show how the error for each asteroid is changed by the analysis – if the error is unchanged it lies on the red $y = x$ line, but if it differs it moves away from that line as shown by the blue and magenta arrows. In some cases, the bootstrap analysis reduces the error estimate versus NEOWISE, but in the majority of cases the error increases. This is consistent with the observation above that the NEOWISE error analysis does not include population sampling errors.

The NEOWISE papers use $\Delta W$, the maximum peak-to-trough value for the $\widehat{Wi}$, as a measure of potential non-spherical behavior.

$$\Delta W = \max_{i=1,..4} \left( \max_n \widehat{Wi}_n - \min_n \widehat{Wi}_n \right) \quad . \tag{46}$$

One might suspect that it could be correlated to the error estimate but as shown in Figure 26, the correlation is relative weak; a linear regression has a value of $R^2 = 0.39$ for Mixed and $R^2 = 0.34$ for the MBR data set. One might similarly expect that the number of WISE data points for the asteroid influences the error, but that relationship, too, appears to be weaker, with $R^2 = 0.22$ for the Mixed data set and $R^2 = 0.19$ for the MBR data set. If we include both $\Delta W$ and the data count there is a stronger correlation, with $R^2 = 0.62$, which makes sense because error comes not simply from the number of the data points, or from the variation between data points (i.e. $\Delta W$), but rather from a combination of the two. With sufficiently large number of data points even high $\Delta W$ can be tamed and a low error estimate achieved. Interactions between the model and the number of data points and degree of variation in each band is also important.

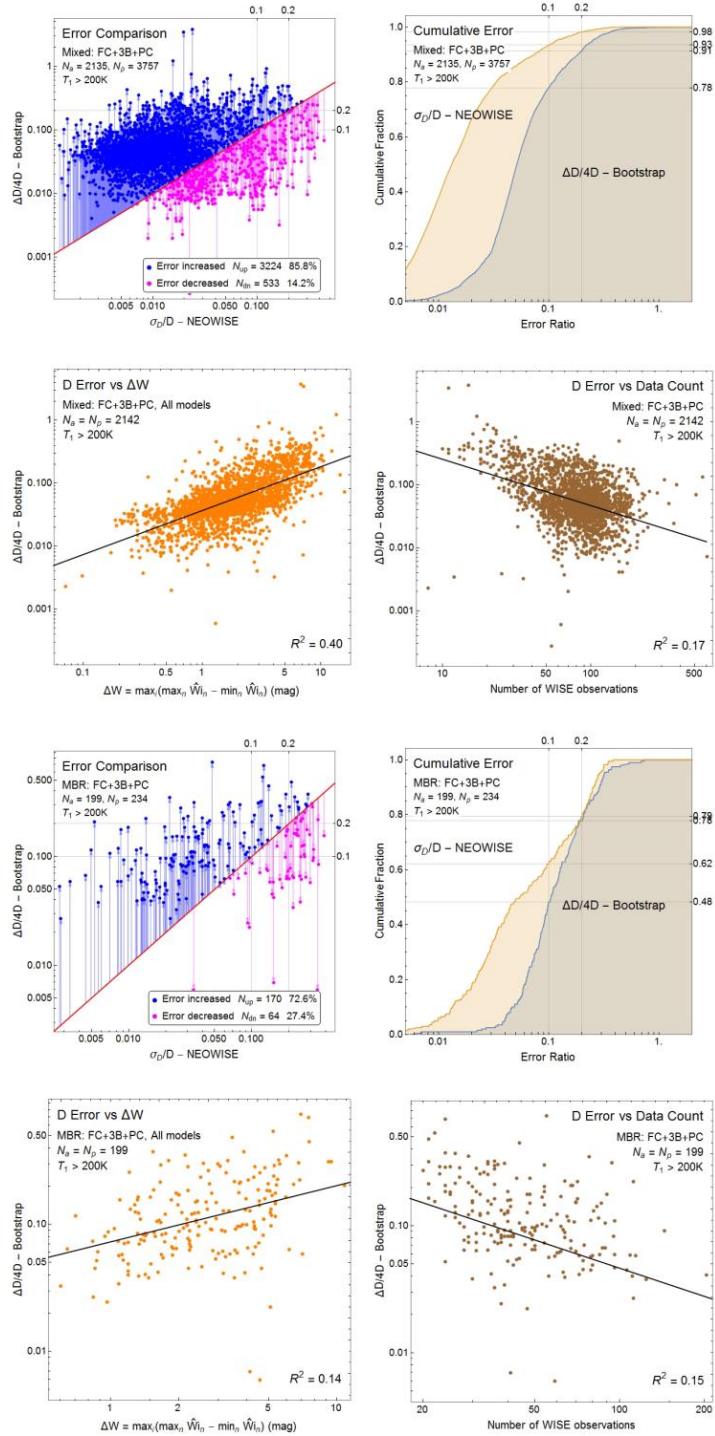

**Figure 26. Comparison of estimated diameter errors.** Errors are compared in four different ways for diameters estimated from the Mixed asteroid data set (*top two rows*) and from the MBR data set (*bottom two rows*). In the leftmost column, the ratio of the diameter to its standard deviation $\sigma_D/D$ is shown for NEOWISE paper results; $\Delta D/4D$ is shown for the bootstrap analysis. Points on the red line $y = x$ have the same error estimate for both NEOWISE and bootstrap models. In the majority of cases, the bootstrap error estimate is larger (*blue points*) than are the NEOWISE estimates for the same asteroids. In other cases (*magenta points*) the NEOWISE errors are smaller. The second column shows the cumulative distribution of errors. The scatter plots show the error $\sigma_D/D$ versus the maximum peak-to-trough ratio (*left column*) and number of data points (*right column*).

The cumulative distribution plots in Figure 26 show that for the Mixed asteroid set, the NEOWISE error analysis estimates that for 93% of the asteroids, *D* is estimated to better than $\pm10\%$, and to better than $\pm20\%$ for 98% of the asteroids. This error profile is consistent with the result, stated in multiple NEOWISE papers, that the diameter estimates can be made to within 10%. However it is important to note that error analysis is about estimating the *precision* internal to the model. It does not tell us what the *accuracy* is – i.e. how well does the model predict the diameters of real asteroids. This is in an important distinction, as will be discussed in a section below. It would appear that the reported NEOWISE results may have blurred this distinction. Also as discussed above the error analysis does not include population sampling error.

Even within the NEOWISE results, significantly fewer random main belt asteroids (60%) have an error estimate that is $\pm10\%$ or better, and 79% of which have $\pm20\%$ or less error. The NEOWISE error estimates are overly optimistic because they ignore population sampling errors. For comparison, among the bootstrap estimates only 32% of random main belt asteroids and 54% of the Mixed set have associated error estimates of $\pm10\%$ or better.

Note that estimates shown in the cumulative distribution in Figure 26 are within the bootstrap sample, which is limited to asteroids which have at least three data points in each WISE band. However, not all asteroids have sufficient data to be eligible for bootstrap modeling, as done in this study, because they do not meet the minimum criterion of having at least two data points in each WISE band.

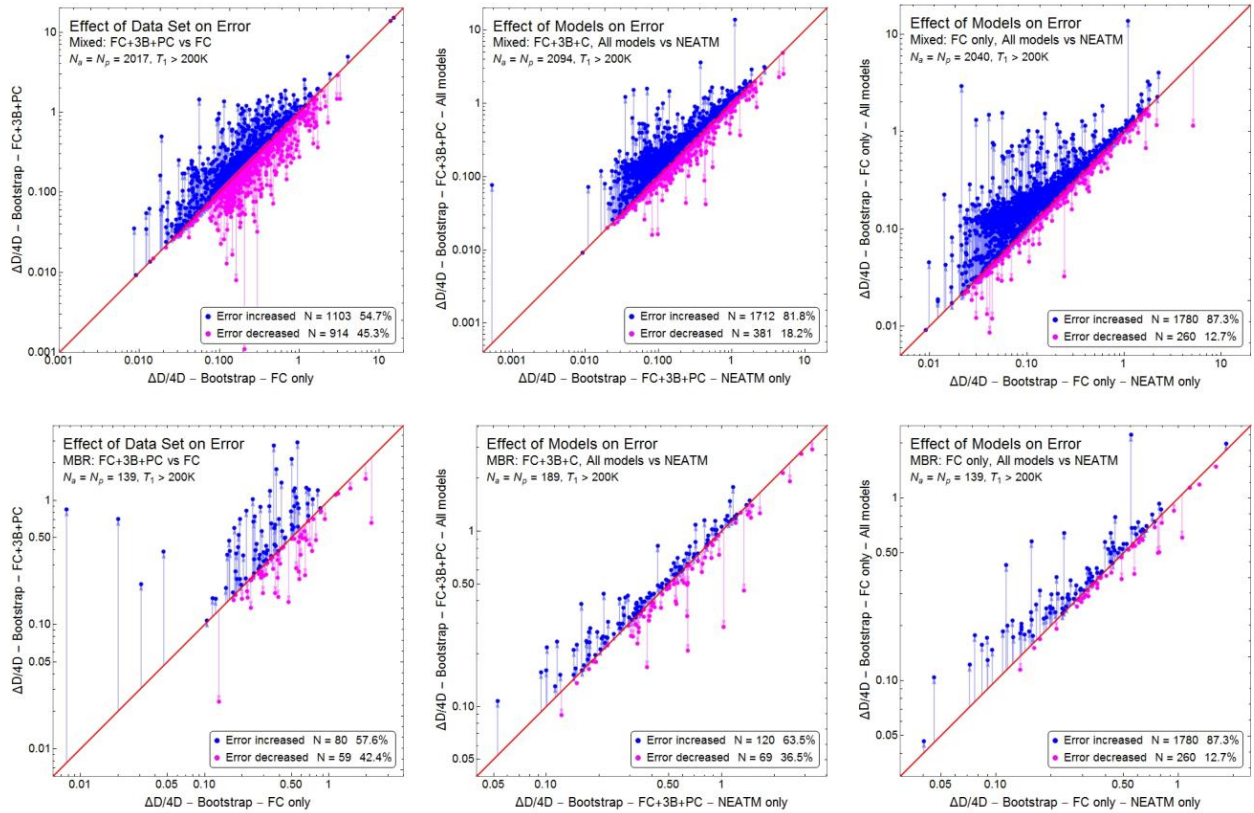

**Figure 27. Impact of additional data and models on error estimates.** $\Delta D/4D$ error values are compared for pairs of models and data sets to explore how errors are affected by adding more data or models. Top row: Mixed asteroid set (FC+3B+PC or FC only). Bottom row: MBR (FC+3B+PC or FC only). Comparison of panels in the far left column reveals the effect of adding more data to the analysis for the same asteroids. The center and left columns show the impact on error estimates of bootstrap trials run with all models versus those run with the NEATM model only.

Figure 27 explores the question of what role the inclusion of 3-band and post-cryo data—or the inclusion of additional models—plays in the error analysis. The effect of adding 3B+PC to the FC data is slightly negative. Some error estimates increase dramatically, whereas some decrease dramatically. Overall, the two effects are out of balance but not overwhelmingly so: error estimates increase for about 55% of the asteroids but decrease for 45% when using the Mixed set with FC+3B+PC data, and closer to 58% - 42% for the MBR set. Since the scientific goal is to approach the truth, adding 3B+PC data to the FC data set is clearly helpful; for 55% of the asteroids it reveals more variation that the FC data set alone, for the remaining 45% the added data sharpens the estimates.

This has interesting implications for the NEOWISE reactivation mission (Mainzer et al., 2014a) which continues to collect data. Like the PC data set, it only contains W1, W2 observations. Figure 27 suggests that it will be valuable to add this data to the present analysis.

In contrast, adding models has a more dramatic effect on estimated errors, increasing them for 82% of asteroid and decreasing errors for just 18%. The percentage of points with errors increased, or decreased by bootstrap analysis is similar for the random main belt cases and for FC-only data.

Counterintuitively, adding data points increases estimated error when the additional data reveals variation in the observed flux. Asteroids may have irregular surfaces and shapes, so observations from previously unobserved rotational phases can inform the analysis that the asteroid is more irregular than inferred from the smaller data set. It is only when an asteroid has been fully observed from all rotational phases (and likely, multiple solar phase angles $\alpha$ as well) that additional observations improve error estimates monotonically by damping errors.

Likewise, adding models also may increase estimated errors because additional models ironically allow better curve fitting (compare Figure 16 to Figure 18). In the bootstrap trials, the larger the number of models, the more of them that flexibly fit the various bootstrap samples, resulting in greater variance in the parameters. Another factor is that NEOWISE model fits use six free parameters, while the models here have only four. Ironically the four-parameter curves fit the data better (Figures 14, 15, 16) but at the cost of more variation within those parameters.

The origin of the error is easy to see in plots of WISE/NEOWISE data points, such as Figures 8, 14 and 21. Although it is not the main purpose of these figures, it is obvious that there is considerable scatter in the data points in each of the bands. This behavior is typical of plots for the asteroids studied here. Cases exist where there is more or less scatter, but generically the scatter in the data points in at least or more band is much greater than the instrumental error bars. This scatter creates the potential for error. In cases where there is sufficient population sampling relative to the scatter one can get very low estimated error – for example asteroid 464 has an estimated diameter error of 0.4% from 210 data points (FC+3B+PC). However there are many other asteroids where the combination of scatter and data point count leads to larger errors.

## 5.4 Estimates of other parameters

The diameter estimates reported by NEOWISE can differ significantly from those produced by the bootstrap method, so it is not surprising that other parameters differ as well. The temperature parameter $T_1$ must vary with the diameter (as seen in Figure 22). The IR albedo parameters $p_{IR1}, p_{IR2}$ are altered significantly by the application of Kirchhoff's law; within the NEOWISE assumptions, they are essentially free parameters that can vary the amount of reflected sunlight without having a concomitant impact on thermal emission. The visible band albedo $p_v$ is a function of diameter and, in addition, is calculated by using the current (November 2015) $H$ values from the MPC. As shown in Figure 6, changing $H$ impacts $p_v$ even within the NEOWISE framework.

Figure 28 shows the distribution of the absolute subsolar temperature $T_1$. The histogram for all NEOWISE fits shows a very unusual distribution, with a sharp peak that cuts off at $T_1 = 400$K; its origin is unclear. The subset of NEOWISE fits studied here (combining both the main belt random and Mixed data sets) has a median that is almost the same (399.7 versus 404.3K) without the same extreme peak in the histogram. The bootstrap histogram is colder, with median 376.2K. Note that this is the result after pruning the low $T_1$ fits – with them included the distribution is shifted even farther toward low temperature.

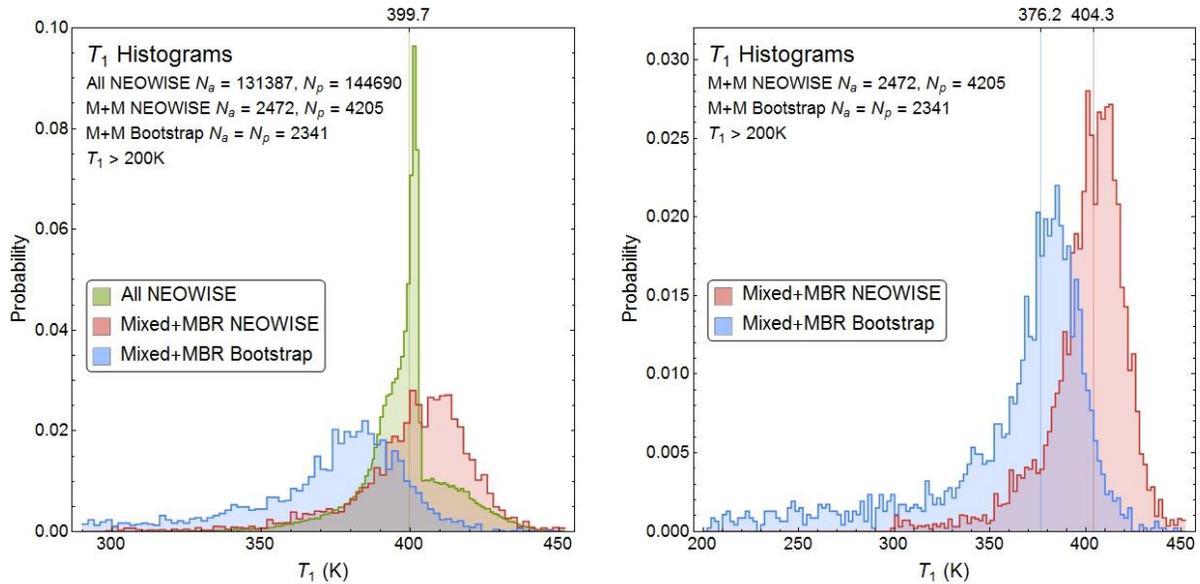

**Figure 28. Comparison of $T_1$.** The left plot shows histograms for the $T_1$ parameter for all NEOWISE fits (*green*) along with the median of the distribution (*vertical line*), NEOWISE fits for the Mixed + MBR asteroids in this study (*red*) and $T_1$ from bootstrap (*blue*). The plot on the right shows only the latter two, along with their median temperatures (*vertical lines*).

Figure 28 also shows that the overall histogram of $T_1$ for all NEOWISE asteroids (the green histogram) is markedly different than the histograms for the Mixed+MBR subset studied here. This is analogous, and likely related, to the difference in diameter distribution histograms seen in Figure 25. The selected asteroids, even including the random main belt, are *not* a representative sample of the NEOWISE data overall.

Figure 29 shows the distribution of the visible band albedo $p_v$. As with Figure 28 and 25, the histogram of all NEOWISE fits differs noticeably from the histogram for the subset studied here in the Mixed+MBR data sets. In the case of $p_v$, the Mixed+MBR subset is shifted toward lower $p_v$ (i.e. darker) asteroids than the NEOWISE data set as a whole. As above, the Mixed+MBR subset is *not* representative of the overall behavior, at least as measured by the histogram.

The bootstrap distribution is in general similar to the NEOWISE values for the Mixed+MBR subset, but it is also shifted toward lower $p_v$. This is shown clearly in the cumulative distribution; the bootstrap distribution of $p_v$ has much higher count for low $p_v$ than NEOWISE for the same asteroids, which is also darker than NEOWISE overall.

Even though the distributions for $p_v$ are not very different, the $p_v$ value for any individual asteroid varies a lot, as can be seen in the scatter plot in the lower right of Figure 29. This is due to both the differences between the NEOWISE and bootstrap methodology for fitting, but also because the bootstrap version uses current values of $H$ from MPC.

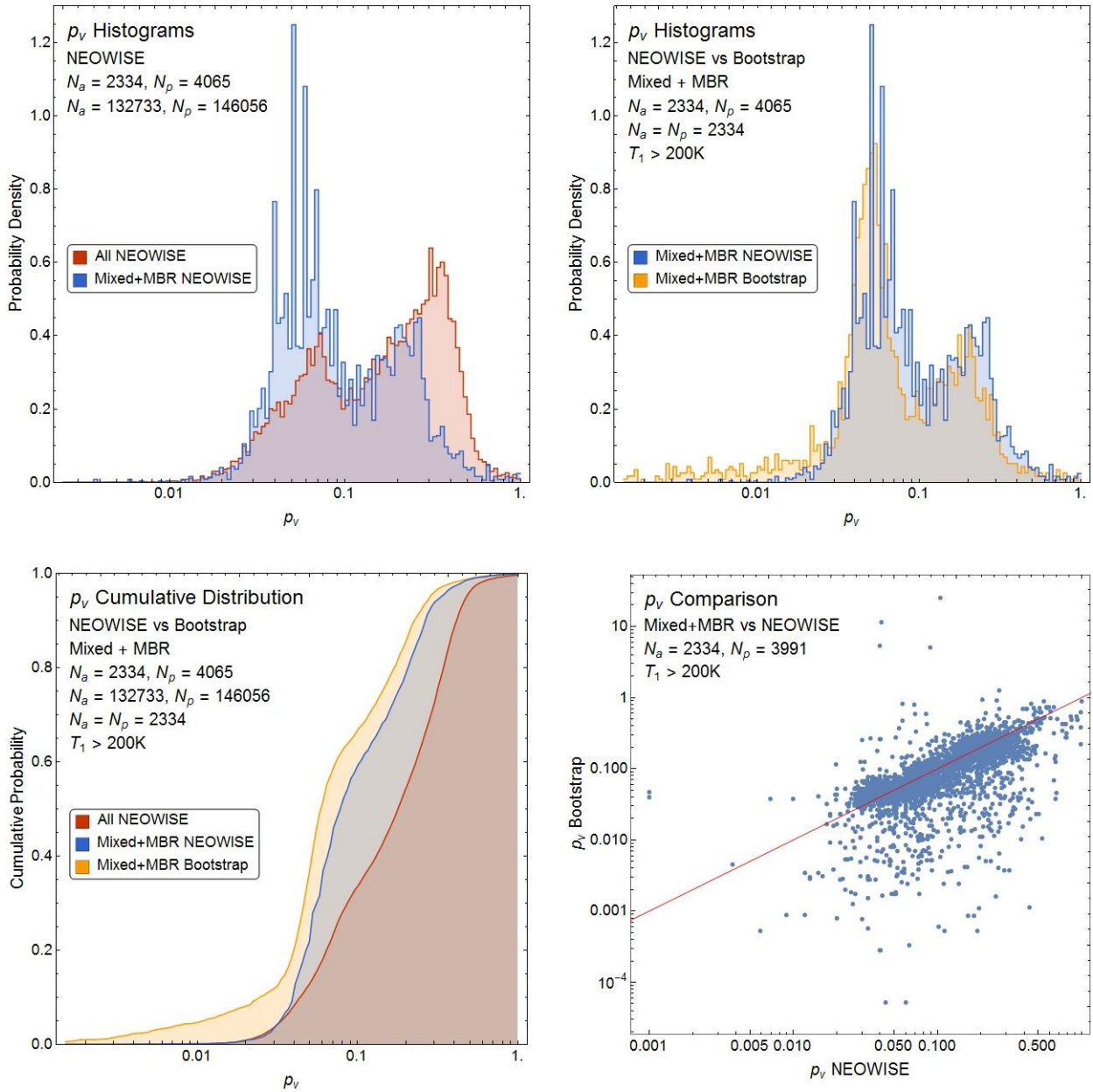

**Figure 29. Comparison of albedo $p_v$.** Histograms for the distribution of $p_v$ for all NEOWISE fits (*red*) is shown in comparison to the histogram for the NEOWISE fits for the Mixed +MBR asteroids (*blue*) *upper left*. The comparison to $p_v$ from the bootstrap analysis is upper right. The cumulative distribution for these distributions is shown (*lower left*) along with a scatter plot that shows how the NEOWISE and bootstrap values for individual asteroids vary (*lower right*).

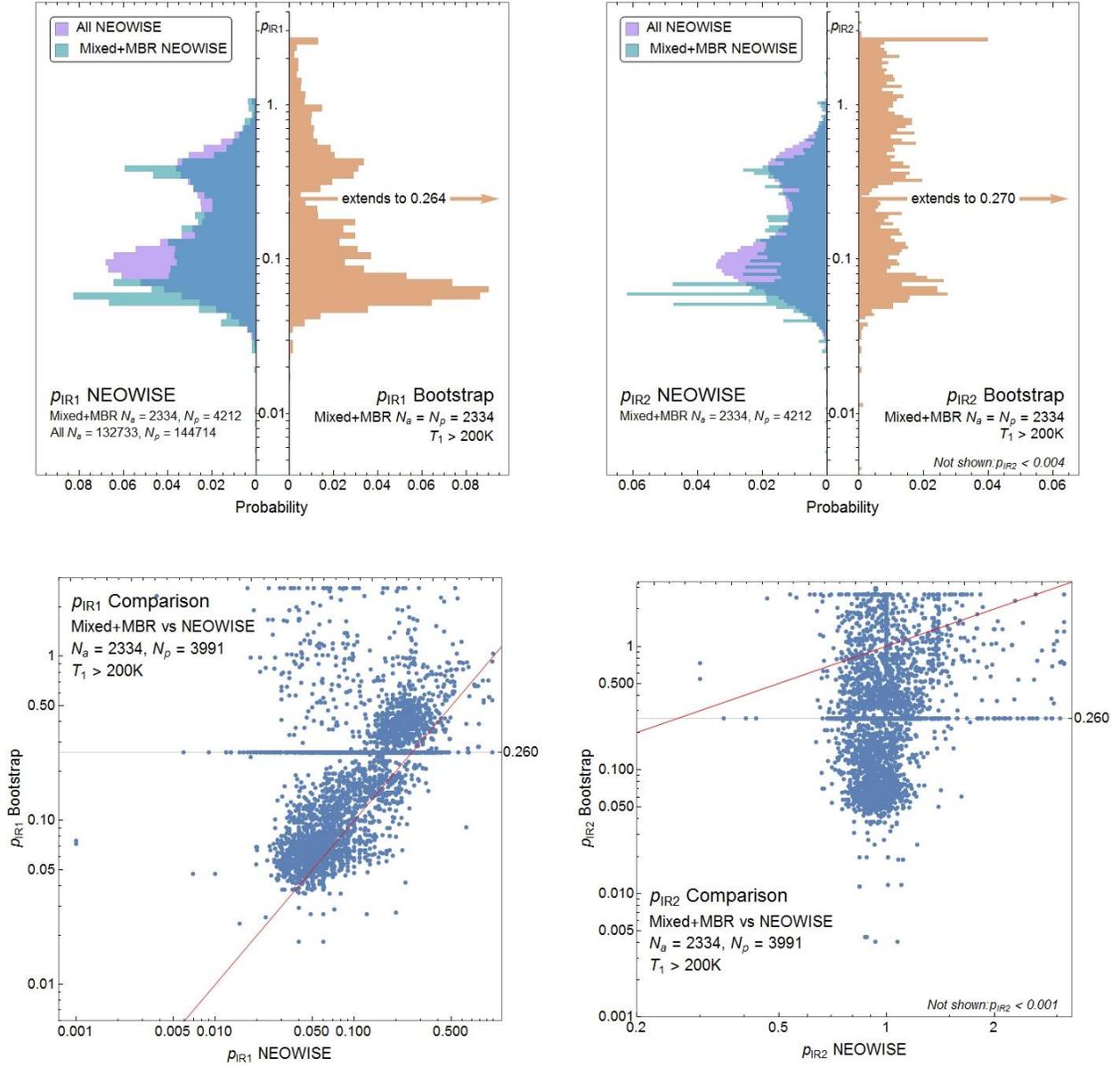

**Figure 30. Comparison of IR albedos $p_{IR1}, p_{IR2}$**. The top row compares histograms from the NEOWISE values for $p_{IR1}$ and $p_{IR2}$ with the corresponding values from the bootstrap distribution. Note that most NEOWISE fits have $p_{IR1} = p_{IR1}$, and as a result the NEOWISE histograms for each are quite similar. The bottom row shows how $p_{IR1}, p_{IR2}$ vary for individual asteroids. Note that $p_{IR1} = p_{IR1} = 0.26$ is over-represented in these plots, as explained in the text.

Plots for the IR albedos $p_{IR1}$ and $p_{IR2}$ are shown in Figure 30. The histograms for $p_{IR1}$ and $p_{IR2}$ both show large spikes at $p_{IR1} = 0.26$ and $p_{IR2} = 0.26$, which are due to the presence of $\epsilon_1 = 0.9$ and $\epsilon_2 = 0.9$ in the parameterizations in Equation (34). When $\epsilon_1 = 0.9$ and $G = 0.15$, $p_{IR1} = 0.26$, and likewise for $\epsilon_2, p_{IR2}$. These parameterizations occur with relatively low frequency in the bootstrap trials, but since all those that do occur wind up in the same histogram bin, they cause a large spike in that bin. Similarly they create a line of points in scatter plots on the bottom row of Figure 30.

There are two ways this spike could be eliminated. One would be to offer other parameterizations with other values than 0.9. The reason for 0.9 in Equation (34) is simply that it is the choice used for $\epsilon_3, \epsilon_4$. Another way to reduce the spike is to have more realistic values for $G$. While $G = 0.15$ is the accepted default value for the HG phase system, it is unlikely that all asteroids have exactly the same value. The spike is caused by having many asteroids with identical values of both $\epsilon$ and $G$.

The plots of the histograms of $p_{IR1}$ and $p_{IR2}$ in Figure 30 show considerable differences between the NEOWISE values and those obtained from the bootstrap approach. This is not surprising because as shown above, there is an important difference between NEOWISE and the present study: Kirchhoff's law. Kirchhoff's law has a direct and obvious impact on albedo in the W1, W2 bands..

More generally, Figure 30 shows that the NEOWISE results for parameters $p_{IR1}$, $p_{IR2}$ or $p_v$ are radically different than results from the methodology studied here. This suggests that studies that use the NEOWISE results for these parameters to classify asteroids or for other applications may need to be revisited (Alí-Lagoa et al., 2014, 2013; Bauer et al., 2013; Durech et al., 2015; Emery et al., 2015; Fernández et al., 2009; Grav et al., 2012; Hanuš et al., 2015; Mainzer et al., 2012c, 2011e, 2012a, 2012b; Masiero et al., 2015a, 2015b, 2014, 2013, 2012a, 2012b; Nesvorný et al., 2015; Pravec et al., 2012).

## 5.5 Comparison with Diameters Estimated by Other Methods

Error estimates reveal the precision with which thermal modeling estimates, but they do not tell us the accuracy of the modeling framework. The accuracy question is this: how well do the thermal-model estimates of diameter and other physical parameters match reality?

The NEOWISE papers, and reviews which summarize them advertise high accuracy:

> NEATM-derived diameters generally reproduce measurements from radar, stellar occultations, and in situ spacecraft visits to within ±10%, given multiple thermally-dominated IR measurements that adequately sample an asteroid's rotational light curve with good signal-to-noise ratio (SNR) and an accurate determination of distance from knowledge of its orbit (Mainzer et al. 2011c). It is worth noting that the accuracy of the diameters of objects used to confirm the performance of radiometric thermal models (such as radar or stellar occultations) is typically ~10%. (Mainzer et al., 2015b)

This implies that the correspondence between NEATM model estimates and those from radar or other means has been calculated yielding a quantified numerical answer $\leq \pm 10\%$. In the passage above there is a caveat that the error estimate applies only with certain pre-conditions, but those caveats are frequently dropped from other references (as in the quote from (Masiero et al., 2011) in section 1 above.)

Careful examination of the NEOWISE paper referred to above as "Mainzer et al. 2011c" – referenced as (Mainzer et al., 2011c) , or Mainzer/TMC in this study – and other papers reveals no such calculation. Because this point is vitally important to assessing the accuracy and credibility of the NEOWISE results it is worth understanding it in detail

Mainzer/TMC presents a set of 50 objects which apparently have had their diameter set equal to the diameter from the ROS literature (see section 4.3 above). The WISE magnitudes were then calculated for these hypothetical objects, and compared to the actual fluxes for a limited set of just 46-52 data point in each band for 50 test objects – which apparently means only about 1 observation per object. The mean and standard deviation of the residuals across the 50 asteroids was taken as Table 2 in that paper (Mainzer et al., 2011c). Here is the relevant quote in which the accuracy of the estimates is mentioned:

> The offsets and errors given in Table 2 can be regarded as the minimum systematic errors in magnitude for minor planets observed by WISE/NEOWISE. Since diameter is proportional to the square root of the thermal flux (Equation (1)), the minimum systematic diameter error due to uncertainties in the color correction is proportional to one-half the error in flux. These magnitude errors result in a minimum systematic error of ~5%–10% for diameters derived from WISE data; they are of similar magnitude to the diameter uncertainties of most of the underlying radar and spacecraft measurements, which are ~10% (references are given in Table 1). (Mainzer et al., 2011c)

It is very important to recognize the limitations inherent in this approach. It is not a calculation – it is a very simplistic scaling argument with a factor of two (5% to 10%) based on a tiny subset of the data (~50 data points for ~50 objects.) It is both conceptually and numerically simple to calculate the actual accuracy – as done in section 3.4 above, but that was not done.

In addition the approach taken, which could be called the "reverse" method – comparing fluxes from hypothetical objects to observed flux, is emphatically not the same as the "forward method" – fitting a NEATM model to observed fluxes. The process of least squares model fitting is explicitly nonlinear, so there is every expectation that the reverse and forward methods would give different answers.

The NEOWISE team applied the forward method – fitting models to observations for ~158,000 asteroids that do not have ROS diameters. The clear and obvious thing to do is apply the same method to asteroids with ROS diameters then compute the difference (as in section 3.4 above). This is the only way to get a handle on NEOWISE diameter estimation, but this was not done in the Mainzer/TMC (Mainzer et al., 2011c), nor does it appear to have been done in any other NEOWISE paper that I have found.

Instead another NEOWISE paper, which also has "thermal model calibration" in its title, (Mainzer et al., 2011d) compares modeled diameters from IRAS (Tedesco et al., 2002b)and RW (Ryan and Woodward, 2010) to the ROS set. The ROS asteroids are chosen on the basis that:

> These objects have been selected to have W3 peak-to-peak amplitudes <0.5 mag to avoid the worst complications caused by applying spherical models to highly elongated objects. (Mainzer et al., 2011d)

This selection process removes some of the value of the comparison. If the goal is to assess the quality level of more than 157,000 asteroids, as stated, it is not appropriate to select only the best of them. Instead one would like test cases that are typical of the larger population. Indeed, the criterion applied to W3 magnitudes would eliminate much of the population.

Somehow the NEOWISE authors reach the following:

> We conclude that the diameters for minor planets derived from NEOWISE are generally in good agreement with those found by IRAS and are likely more free of systematic biases than the diameters provided in either Tedesco et al. (2002) or Ryan & Woodward (2010). Together with Mainzer et al. (2011b), this demonstrates that the NEOWISE data set will produce good quality physical parameters for the >157,000 minor planets it contains. (Mainzer et al., 2011d)

Such an indirect and non-quantitative assessment does not support such a sweeping and important conclusion, particularly because a direct quantitative assessment can easily be made (Figure 11 and equations (41) and (43)).

Direct comparison of the diameters from thermal modeling to their ROS counterparts is the *only* way to judge the accuracy of NEOWISE diameter estimates. Diameter is closely bound with albedo $p_v$ via Equation (44), with IR albedos $p_{IR1}$ and $p_{IR2}$ via equations (26) and (28) and with $T_1$ and $\eta$ via Equation (20). As a consequence, diameter comparison is vital to all physical parameter estimation. In addition, diameter is the only physical parameter where we have independent assessments of via radar, stellar occultations for more than a handful of asteroids. It is thus all the more puzzling that the NEOWISE papers do not assess diameter accuracy. It is hard to see how any conclusion about "good quality" can be made without knowing how accurate the diameter estimates are.

As shown in section 4.3 it appears that the vast majority of asteroids with ROS diameter results have had their NEOWISE diameter set exactly equal to the ROS diameter (whether by intention or accidentally is unclear.) This prevents third party studies, including the present one, from making a an assessment of the accuracy. So not only is there insufficient basis for the NEOWISE accuracy claims the fact that the diameters were copied (whether accidentally or deliberately is unclear) actively prevents assessment of their accuracy.

Figure 31 plots comparisons of asteroids from the ROS sources in Table 1 with thermal modeling results from four sources: bootstrap results of this study, NEOWISE fits, IRAS and the RW-NEATM results. When there are multiple ROS sources for a single asteroid, all examples which have error estimates are used. In the case of NEOWISE there are also multiple results for each asteroid, so in general the number of points can be greater than the number of asteroids (i.e. $N_p \geq N_a$.) This is one reason that the NEOWISE results show scatter away from the line $y = x$ despite having been set exactly equal to the ROS diameters – some NEOWISE results were set equal to some, but not all ROS diameters.

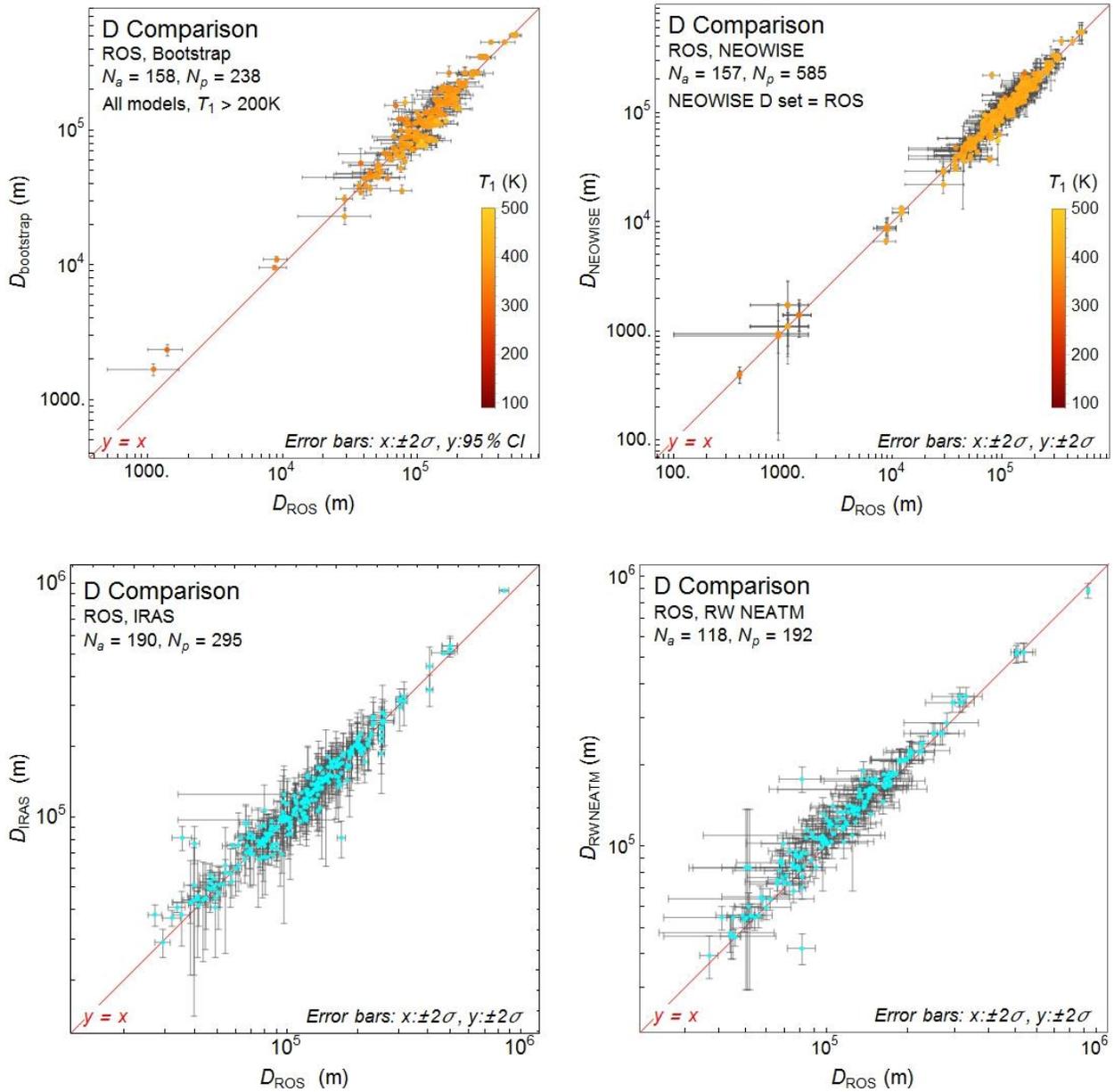

**Figure 31. Comparison of diameter estimates derived from thermal modeling to estimates made by radar, occultations and spacecraft.** The diameters $D$ (with error estimates) for asteroids from the ROS literature are compared with are compared to estimates of $D$ from thermal modeling. *Top left*: bootstrap results for the Mixed data set (FC+3B+PC) with all models with $T_1 < 200K$. *Top right:* NEOWISE results (note issue in Section 4.3 and Table 4 that the diameters were set equal). *Bottom left*: results from IRAS. *Bottom left*: results from NEATM modeling from RW (Ryan and Woodward, 2010). Temperature parameter $T_1$ is not available for IRAS or RW results.

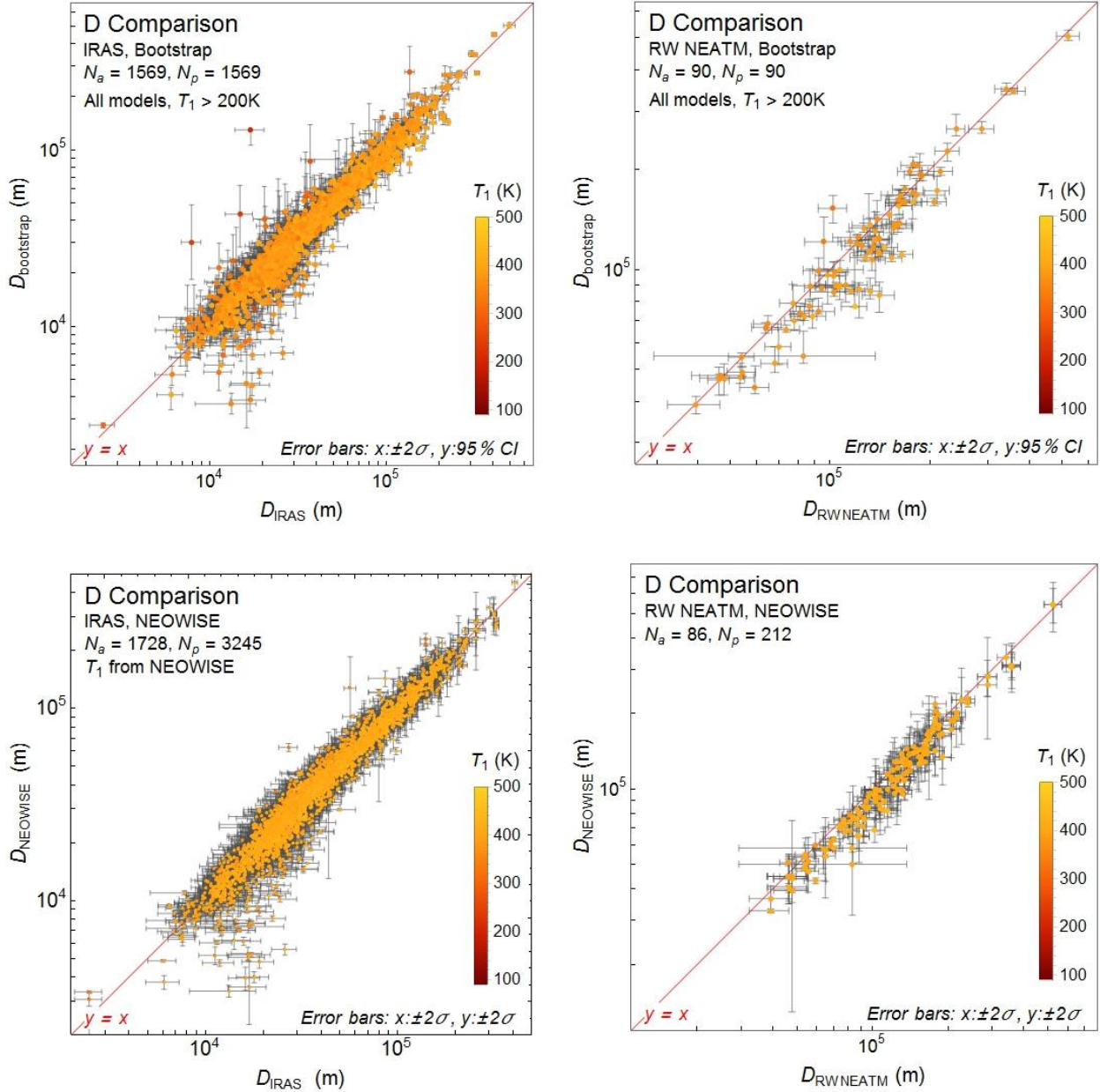

**Figure 32. Comparison of *D* estimated by thermal modeling for WISE/NEOWISE data with other thermal modeling.** The plots show diameter data points (with error estimates) obtained from IRAS and RW-NEATM (see Table 1) with those for the bootstrap method of this study (Mixed data set, all models, with $T_1 < 200K$), and NEOWISE.

A different comparison is to match the results of this study, and NEOWISE with other thermal modeling efforts, particularly IRAS and RW-NEATM. This is shown in Figure 32.

Table 6 presents the statistical evaluation of comparisons between thermal modeling and ROS estimates, as well as between the bootstrap and NEOWISE estimates and thermal modeling from IRAS and RW-STM and RW-NEATM.

Despite having the diameters copied from ROS sources, NEOWISE still only has accuracy $\leq \pm 27.9\%$ compared to ROS at the 1-$\sigma$ level. This is due to several factors. First, not all ROS sources, nor all NEOWISE estimates agree with one another for the same asteroid. In addition both the ROS and NEOWISE estimates have estimated errors which compound. The claim that NEOWISE is accurate to $\pm 10\%$ is directly refuted by the NEOWISE figures and the ROS sources referenced by Mainzer/TMC – even with the data tampered with by coping from a ROS source. Presumably this means that true NEOWISE results for these asteroids must be even further off.

| Estimate | Compare | Data Set, Models, and Pruning ||||  $D_{estimate}/D_{compare}$ ||| $E_\rho$ |
|---|---|---|---|---|---|---|---|---|---|
| | | Pruning | Data Set | Bands | Models | 95% low | Median | 95% high | |
| Bootstrap | ROS | None | Mixed | FC+3B+PC | All | 0.610 | 1.019 | 61.645 | 21.107 |
| | | | | | NEATM | 0.643 | 1.101 | 49.818 | 13.821 |
| | | | | FC only | All | 0.582 | 0.970 | 65.108 | 21.819 |
| | | | | | NEATM | 0.619 | 1.038 | 46.731 | 10.353 |
| | | $T_1 \leq 210K$ | Mixed | FC+3B+PC | All | 0.581 | 0.988 | 2.096 | **0.315** |
| | | | | | NEATM | 0.645 | 1.084 | 1.991 | 0.418 |
| | | | | FC only | All | 0.560 | 0.944 | 1.965 | 0.234 |
| | | | | | NEATM | 0.619 | 1.031 | 1.744 | 0.309 |
| | IRAS | $T_1 \leq 210K$ | Mixed | FC+3B+PC | All | 0.602 | 1.013 | 1.537 | **0.319** |
| | RW-STM | None | Mixed | FC+3B+PC | All | 0.678 | 1.062 | 69.203 | 20.569 |
| | | $T_1 \leq 210K$ | | | | 0.674 | 1.046 | 1.638 | 0.275 |
| | RW-NEATM | None | Mixed | FC+3B+PC | All | 0.607 | 0.936 | 51.231 | 16.238 |
| | | $T_1 \leq 210K$ | | | | 0.601 | 0.920 | 1.365 | **0.085** |
| NEOWISE | ROS | | | | | 0.555 | 0.986 | 1.866 | **0.279** |
| | IRAS | | | | | 0.576 | 1.001 | 1.409 | 0.193 |
| | RW-STM | | | | | 0.677 | 1.028 | 1.482 | 0.223 |
| | RW-NEATM | | | | | 0.609 | 0.911 | 1.273 | 0.049 |
| IRAS | ROS | | | | | 0.614 | 0.986 | 1.605 | 0.212 |
| RW-STM | ROS | | | | | 0.573 | 0.954 | 1.838 | 0.202 |
| | IRAS | | | | | 0.217 | 1.017 | 6.995 | 2.022 |
| RW-NEATM | ROS | | | | | 0.684 | 1.085 | 2.202 | 0.352 |
| | IRAS | | | | | 0.275 | 1.159 | 7.602 | 2.313 |

**Table 6. Comparison of thermal model derived diameters to those from other sources.** The Estimate column contains the source of diameter estimates, while the Compare column contains the source of estimates that they are being compared to. The Data Set, Models and Pruning columns are relevant only for the bootstrap estimates. Source references are in Table 1. The ratio of diameters statistics are calculated as with Figure 11. Bold cases are explained in the text.

The bootstrap method with pruning of fits with $T_1 < 200K$ matches ROS diameters to within $\pm 31.5\%$, and IRAS diameter to within $\pm 31.9\%$, all at the $\pm 1\sigma$ level. It very closely matches the RW-NEATM results to within $\pm 8.5\%$.

# 6. Discussion

The methods presented here return to first principles to re-derive simple asteroid thermal models in the presence of reflected sunlight. Dependence on the absolute magnitude $H$ is avoided, and several new thermal models are developed as simple extensions of NEATM, FRM and simple blackbodies. Kirchhoff's law is important in situations with reflected sunlight. Using multiple competing models improves the curve fitting and allows relative assessment of goodness-of-fit. Each of these modifications leads to higher fidelity modeling with a sound statistical basis.

Higher fidelity "simple" thermal models may seem to be a contradiction, since thermophysical modeling can offer much more sophisticated physics. Why not leave simple models as-is? While thermophysical models are definitively to be preferred to the simple models described here, there is still a place for simple models. The advent of large scale observational data sets like WISE/NEOWISE means that large scale screening of thousands of asteroids can be important. As an example, the HG-NEATM model described here is a better fit for some asteroids – likely due to its stronger phase angle dependence – these asteroids are thus ripe for further investigation. Screening directly with more complex models may be impractical. The application of better "simple" models can inspire complex models grounded in physics.

A related question is why bother with Kirchhoff's law? Since simple thermal models are so simple they already lack fidelity in several other ways. There is a misperception that NEATM and related thermals already violate physical law, and even violate the conservation of energy. This is emphatically not the case – NEATM is explicitly about imposing conservation of energy on a system where we know there are unobserved sources and sinks of energy – "beaming", errors in the approximations for $A$, $\epsilon$, and thermal interia. Harris introduced the $\eta$ parameter to quantitatively measure the net unobserved energy and thereby restore conservation of energy.

Another reason for Kirchhoff is that some instruments, like WISE, have a significant fraction of their observations in bands that contain reflected sunlight. To correctly use that data we must use Kirchhoff. A third reason is that the IR albedo ($p_{IR1}$ and $p_{IR2}$) is of interest and we can hardly solve correctly for IR albedo without Kirchhoff's law, which governs both albedo and emissivity. NEOWISE does this incorrectly and as a result the results they get for "IR albedo" are not that at all, and likely have little if any scientific value.

The models in this study are fit to all observational data for each asteroid. As a consequence, data from the 3-band and post-cryo mission phases is included with the full-cryo mission data. No ad hoc rules are used to modify error estimates or to eliminate data. The corrected Akaike information criterion (AIC$_c$) is used both to select among models and to select among plausible parameter assumptions for each model. Bootstrap resampling provides error estimates. In short, the analysis is based entirely on a straightforward application of standard statistical methods.

The data analysis method employed in the NEOWISE papers is quite different. At a physics level they differ in treatment of Kirchhoff's law. At the statistical data analysis level the NEOWISE approach involves numerous ad hoc rules for modifying error estimates, excluding data from analysis and restricting analysis to artificial epochs. Error estimates were based only on limited Monte Carlo simulation of non-sampling errors. Measured parameters, such as $H$ and $G$ were treated as free variables, with observational evidence input only as constraints. Only one model was fit, and no goodness-of-fit metrics were used. Errors appear to have been made in whatever led to the evident internal inconsistencies (Figure 12). Most NEOWISE "best fits" do not fit very well, missing entire bands of the data (Figures 14, 15, 16) despite having more free parameters. Other non-standard methods may have been used as well; NEOWISE data-analysis procedures have not been described in sufficient detail to enable either transparency or replication.

A disturbing number of numerical coincidences are found in the NEOWISE diameter estimates that exactly match diameters in the reference ROS data set in 123 cases (Table 4). This is quite bizarre. While no firm conclusion can be drawn without further investigation, the odds that these occurred by chance appear to be vanishingly small. Unfortunately this situation makes it impossible to check the accuracy of NEOWISE diameter estimates – the prime test cases have had their diameters copied into the NEOWISE results. Until this is cleared up I believe this is reason by itself to doubt the utility of all NEOWISE results.

Ideally, different modeling approaches should nevertheless give consistent results. That is not the case here. This study finds asteroid diameters that sometimes match those found by the NEOWISE group but differ substantially in many cases (Figures 21, 22, Table 5.) Parameters such as IR or visible-band albedo are much more divergent (Figures 28, 29 and 30.)

Substantive differences also appear in the error estimates produced by the two methods (Figures 23, 24). The results of this study reveal that population sampling errors can be large and can dramatically lower the precision of the estimates made from the WISE/NEOWISE data. This is even true if one artificially prunes the low $T_1$ fits from bootstrap trials (Table 5).

The accuracy of NEOWISE fits when compared to asteroids of known diameter, which was claimed to be ±10%, is actually much worse. In part that is because the ±10% claim appears not to be based on a quantitative calculation at all; and in addition there is the issue of the copied diameters which makes a true comparison moot.

The bootstrap methods here show the $1\sigma$ accuracy of ±31.5% on the ROS data sets. Note however that the internal error estimates for these asteroids are much smaller than the typical case. In addition it appears clear that statistical distributions across all NEOWISE results are quite different than those taken over the subset of asteroids studied here (Figures 25, 28, 29 and 30) so great caution should be used in extrapolating the results broadly to the entire WISE/NEOWISE data set.

Considering that the results of the NEOWISE method appear to be better—error estimates are lower, and diameter estimates correlate more closely to those from other methods —one might well ask why not simply use the approach that seems to work best?

The statistician's answer is that one must adopt neutral and unbiased analysis procedures. If that gives you a less accurate or even paradoxical answer then so be it. That is what the data is telling you, and there may be a lesson hidden in the failure. Just this sort of situation exists with the low $T_1$ fits found here. The fact that unbiased application of the NEATM model produces these physically unlikely results is explained in part by the indeterminacy in the purely thermal W3, W4 bands shown in Figure 7, but likely is also due to something about the asteroid physical properties.

The NEOWISE approach, in contrast, might be said to suffer from the "calibration fallacy," the notion that one can legitimately "calibrate" an analytical procedure by adopting ad hoc rules that as a side effect of eliminating outliers and awkward cases and forcing results to conform to prior expectations. It appears that a conceptual problem in assuming instrumental measurement error would include population sampling error created a fitting failure, which then lead to ad-hoc procedures trying to remedy the original mistake.

To be clear, some kinds of calibration can be appropriate. Data from a particular instrument may require special analysis due to conditions of the instrument or its observing conditions, for example. It is entirely valid to measure calibration parameters and to use an analysis suited to an instrument. However, any such procedure poses a question for the investigator: is the special method required to cope with a specific property of the instrument or observation in a neutral or unbiased way, or could the rule or procedure cause an undue influence on the final results?

As one example, eliminating the indeterminacy in W3, W4 by artificially constraining the parameters may mask an interesting physical effect. It also is no longer the NEATM model, but instead is an ad-hoc procedure that may, or may not be valid. The same is true for eliminating data with the 40% rule, or model fitting only on small subsets of the data (epochs), or arbitrarily increasing measurement errors. Such extraordinary intervention in the data requires commensurately strong evidence than it is unbiased and not simply there to conform to expectations.

Several aspects of the NEOWISE analysis that I have characterized as "unconventional" are only so when viewed through the larger context of scientific statistical analysis; within the scope of previous asteroid thermal modeling studies, some of these methods have been used prior to the NEOWISE studies, and may even be commonplace. Violating Kirchhoff's law appears to be common to a number of studies (Emery et al., 2006; Hargrove et al., 2012; Lim et al., 2011; Mueller et al., 2011; Trilling et al., 2010). So, too, are studies that use only a single model (usually NEATM) and a single parameterization and that neglect to calculate goodness-of-fit measures. I have focused here on NEOWISE primarily because it is represents by far the largest and most recent study of its kind. As a consequence, understanding the limitations and accuracy of the NEOWISE studies is critical to our understanding of asteroids.

Some of the limitations of NEOWISE results are due to the specific data analysis methods used in their preparation, but there are also more general issues that apply to asteroid thermal modeling more generally.

The large scatter in WISE/NEOWISE data (examples in Figures 8, 14, 16, 21) appears to be real and intrinsic and is the proximate cause of the higher error estimates found here. It has long been known that asteroids, especially at smaller sizes, are irregular and non-spherical. It is also known that some asteroids are not a single body but instead are binary or trinary systems or have even more constituents. These factors may contribute to the scatter, but regardless of the origin, asteroid thermal modeling must confront this variability with sufficient observation of different rotational phases, and data analysis methods like bootstrap resampling that can cope with the resulting errors.

The goal of precise diameter estimation by thermal modeling may prove illusive for the majority of asteroids until we have sufficiently dense observations sampling enough asteroid rotational phases that the scatter is tamed. Some improvement may come from the restarted NEOWISE project which continues to collect data (Mainzer et al., 2014a), although it shares the limitation of the post-cryo mission in being limited to the W1 and W2 bands. Using the analysis framework developed in this study that data can potentially improve estimates for asteroids that have already been observed. Ideally solar system astronomy will have new IR space telescopes, with new data sets. Until then the relatively coarse and imprecise diameter estimates revealed here (Table 5, 6) are likely the best that can be done with the WISE/NEOWISE data in the majority of cases.

Better sampling is required for precise estimates, but the accuracy with which thermal modeling estimates the actual diameter is also an issue. Accuracy depends in part on precision, but also a connection to orthogonal means to estimate diameter. The argument that IR observations and thermal modeling accurately estimate asteroid diameters is based on comparison to a distressingly small number of asteroids (~150), which may not be representative of the broader asteroid population with respect to emissivity or other factors.

Thermal modeling has at least two important areas of uncertainty and indeterminacy. The widespread assumption that $\epsilon = 0.9$ across thermal IR wavelengths is not true for many asteroid constituent proxies (Figure 4) and can have a large impact on results (Figure 7). This raises the possibility that we must estimate $\epsilon$ by other means, and suggests that assigning the value of the beaming parameter $\eta$ on the basis of coarse asteroid groups (i.e. main belt, NEO) is likely invalid because $\eta$ will vary by material and surface properties. Understanding the impact these factors have on thermal modeling is a challenge for further research.

The other indeterminacy is the surprising finding (Figure 7) that in the presence of reflected sunlight the thermally dominated W3, W4 bands can admit multiple solutions at different temperatures. This leads to the problematic low $T_1$ fit and bootstrap trials (Figures 20, 21, 22, 27 and 28.)

Despite the challenges of high scatter, the WISE/NEOWISE observations are still the largest and most important data set of asteroid thermal IR observations available to astronomers, and will likely retain that position for quite some time. There is still a role for the WISE/NEOWISE raw data for asteroid modeling, even in cases where the error estimates are regrettably high. In cases where the error is important, a small percentage of the asteroids are sampled adequately and do have good estimates even as it stands. A small percentage can still be thousands of asteroids in a survey of this size and scope. Further understanding of the open issues in thermal modeling discussed here may be able to increase that further, at least to a point. That is why it is important to understand both the data and its weaknesses.

Some of the conclusions found here may apply to other infrared asteroid studies. Caution must be exercised because the specific features found here may not apply to all studies. In particular, observations that include more spectral bands, tighter spectral resolution, or bands that are less contaminated by reflected solar radiation would may avoid some of the problems inherent in the WISE/NEOWISE observational data. Older studies also typically had lower sensitivity and may have been unable to observe smaller, and possibly more irregular asteroids.

The results here have strong implications for future IR telescopes for asteroid observation, such as the proposed NEOCam mission. NEOCam uses only two IR bands; band 1 ($4 - 5.2 \mu$m), and band 2 ($6 - 10 \mu$m) (Mainzer et al., 2015a; McMurtry et al., 2013). The NEOCam band 1 is roughly similar to WISE band W2, while band 2 overlaps with W3, it is much narrower. So instead of having four bands, two of which are typically thermally dominated (W3, W4) and two of which have some degree of reflected sunlight (W1,W2) the NEOCam mission will have one thermally dominated and one sunlight dominated band. It is unclear what level of thermal modeling, if any, will be possible with this passband structure.

# 7. Acknowledgements

This publication makes use of data products from the Widefield Infrared Survey Explorer, which is a joint project of the University of California, Los Angeles, and the Jet Propulsion Laboratory/California Institute of Technology, funded by the National Aeronautics and Space Administration. This publication also makes use of data products from NEOWISE, which is a project of the Jet Propulsion Laboratory/California Institute of Technology, funded by the Planetary Science Division of the National Aeronautics and Space Administration. Wayt Gibbs, Dhileep Sivam, and Daniel McCoy assisted in the preparation of the manuscript, and Andrei Modoran with data downloading.

I also wish to thank helpful discussions, correspondence, and data from Victor Alí Lagoa, Francesca DeMeo, José Galache, Bruce Hapke, Alan Harris, Alan Harris (DLR), Željko Ivezić, Seth Koren, Tony Pan, Thomas Statler, and Edward Wright.

# 8. Appendix—Result Tables

The parameter estimates for asteroids studied here are listed in the following tables.  Full versions will be available online.

| Asteroid | Parameter | Mean | σ | Median | 95% confidence interval | |
|---|---|---|---|---|---|---|
| 10 | H | 5.43 | | | | |
| | G | 0.15 | | | | |
| | D (m) | 451032.5 | 5879.213 | 450727.7 | 438654.2 | 462495.4 |
| | eps1 | 0.974478 | 0.00066 | 0.974472 | 0.973156 | 0.975799 |
| | eps2 | 1 | 0 | 1 | 1 | 1 |
| | pv | 0.058459 | 0.00152 | 0.058502 | 0.055569 | 0.061774 |
| | T1 (K) | 377.5577 | 0.879213 | 377.6076 | 375.6789 | 379.1764 |
| 100111 | H | 14.9 | | | | |
| | G | 0.15 | | | | |
| | D (m) | 7069.968 | 431.5667 | 7040.549 | 6354.797 | 7959.829 |
| | eps1 | 0.880747 | 0.030065 | 0.9 | 0.804738 | 0.9 |
| | eps2 | 0.535486 | 0.092724 | 0.547643 | 0.330965 | 0.691145 |
| | pv | 0.039172 | 0.004701 | 0.039069 | 0.030566 | 0.047956 |
| | T1 (K) | 350.7929 | 13.43086 | 352.2549 | 323.8841 | 373.0514 |
| 100926 | H | 16.7 | | | | |
| | G | 0.15 | | | | |
| | D (m) | 39482.84 | 13332.57 | 42622.26 | 4427.585 | 52501.02 |
| | eps1 | 0.997873 | 0.005355 | 0.999824 | 0.983057 | 0.999874 |
| | eps2 | 0.986755 | 0.033047 | 0.998849 | 0.9 | 0.999175 |
| | pv | 0.002227 | 0.00563 | 0.000203 | 0.000134 | 0.018824 |
| | T1 (K) | 111.6188 | 39.33066 | 105.0681 | 20.87383 | 208.0778 |
| 10115 | H | 17 | | | | |
| | G | 0.15 | | | | |
| | D (m) | 45581.73 | 3964.773 | 45857.89 | 42036.84 | 49212.05 |
| | eps1 | 0.998776 | 0.008692 | 0.999606 | 0.998708 | 0.999812 |
| | eps2 | 0.997137 | 0.025387 | 0.999205 | 0.998703 | 0.999492 |
| | pv | 0.000531 | 0.004948 | 0.000133 | 0.000116 | 0.000158 |
| | T1 (K) | 80.718 | 15.83287 | 80.9404 | 72.72334 | 86.79721 |
| 10123 | H | 14.2 | | | | |
| | G | 0.15 | | | | |
| | D (m) | 89325.62 | 65817.38 | 136873.4 | 2819.071 | 145577.7 |
| | eps1 | 0.708213 | 0.401851 | 0.99966 | 3.22E-13 | 0.999804 |
| | eps2 | 0.708137 | 0.401712 | 0.999414 | 3.22E-13 | 0.99961 |
| | pv | 0.138255 | 0.185553 | 0.000197 | 0.000174 | 0.464342 |
| | T1 (K) | 199.0953 | 128.2261 | 105.5811 | 95.67905 | 394.0723 |

**Table A1. Bootstrap estimates of asteroid parameters – Mixed data set.** The table (and online data) list the parameters derived from WISE/NEOWISE data using the Mixed: FC+3B+PC data set and all four models described in the text. Each parameter is the Akaike weighted averaged of the best fit models across 500 bootstrap trials with four models and five parameterizations of each model. Note that $H$, $G$ are from MPC November 2015, $D$ is in meters, $T_1$ in degrees K. This is a sample, the entire data set is available online.

| Asteroid | Parameter | Mean | σ | Median | 95% confidence interval | |
|---|---|---|---|---|---|---|
| 10 | H | 5.43 | | | | |
| | G | 0.15 | | | | |
| | D (m) | 451032.5 | 5879.213 | 450727.7 | 438654.2 | 462495.4 |
| | eps1 | 0.974478 | 0.00066 | 0.974472 | 0.973156 | 0.975799 |
| | eps2 | 1 | 0 | 1 | 1 | 1 |
| | pv | 0.058459 | 0.00152 | 0.058502 | 0.055569 | 0.061774 |
| | T1 (K) | 377.5577 | 0.879213 | 377.6076 | 375.6789 | 379.1764 |
| 100111 | H | 14.9 | | | | |
| | G | 0.15 | | | | |
| | D (m) | 7069.968 | 431.5667 | 7040.549 | 6354.797 | 7959.829 |
| | eps1 | 0.880747 | 0.030065 | 0.9 | 0.804738 | 0.9 |
| | eps2 | 0.535486 | 0.092724 | 0.547643 | 0.330965 | 0.691145 |
| | pv | 0.039172 | 0.004701 | 0.039069 | 0.030566 | 0.047956 |
| | T1 (K) | 350.7929 | 13.43086 | 352.2549 | 323.8841 | 373.0514 |
| 100926 | H | 16.7 | | | | |
| | G | 0.15 | | | | |
| | D (m) | 4718.303 | 369.326 | 4661.057 | 4214.85 | 5574.094 |
| | eps1 | 0.9838 | 0.00312 | 0.984007 | 0.981288 | 0.985943 |
| | eps2 | 0.899424 | 0.016693 | 0.9 | 0.9 | 0.9 |
| | pv | 0.016913 | 0.003805 | 0.016986 | 0.011877 | 0.020772 |
| | T1 (K) | 206.1704 | 2.986142 | 206.3217 | 200.8229 | 210.6606 |
| 10115 | H | 17 | | | | |
| | G | 0.15 | | | | |
| | D (m) | 2199.26 | 233.8238 | 2108.719 | 1853.096 | 2664.261 |
| | eps1 | 0.9 | 0 | 0.9 | 0.9 | 0.9 |
| | eps2 | 0.692677 | 0.057877 | 0.700193 | 0.579832 | 0.77345 |
| | pv | 0.05971 | 0.012159 | 0.062952 | 0.039436 | 0.081518 |
| | T1 (K) | 264.4225 | 12.45553 | 267.9381 | 240.9117 | 285.2781 |
| 10123 | H | 14.2 | | | | |
| | G | 0.15 | | | | |
| | D (m) | 3204.821 | 576.8242 | 3123.089 | 2723.174 | 3861.113 |
| | eps1 | 0.208126 | 0.206265 | 0.175041 | 5.07E-14 | 0.617468 |
| | eps2 | 0.208314 | 0.206545 | 0.175041 | 5.07E-14 | 0.618943 |
| | pv | 0.37512 | 0.068264 | 0.378339 | 0.247528 | 0.497622 |
| | T1 (K) | 366.0622 | 21.92538 | 366.3971 | 328.7042 | 403.3306 |

**Table A2. Bootstrap estimates of asteroid parameters – Mixed data set – with pruning.** The table (and online data) list the parameters derived from WISE/NEOWISE data using the Mixed: FC+3B+PC data set and all four models described in the text. Each parameter is the Akaike weighted averaged of the best fit models across 500 bootstrap trials with four models and five parameterizations of each model. Trials with $T_1 \leq 200K$ pruned. Note that $H, G$ are from MPC November 2015, $D$ is in meters, $T_1$ in degrees K. This is a sample, the entire data set is available online.

| Asteroid | Parameter | Mean | σ | Median | 95% confidence interval | |
|---|---|---|---|---|---|---|
| 10088 | H | 13.5 | | | | |
| | G | 0.15 | | | | |
| | D (m) | 6174.043 | 1582.048 | 5851.911 | 3890.804 | 9099.526 |
| | eps1 | 0.781358 | 0.125185 | 0.809626 | 0.518792 | 0.9 |
| | eps2 | 0.3703 | 0.341259 | 0.465466 | 0 | 0.811938 |
| | pv | 0.224288 | 0.112841 | 0.205331 | 0.08492 | 0.464485 |
| | T1 (K) | 309.1995 | 28.78408 | 309.0183 | 255.4772 | 363.9725 |
| 102587 | H | 13.7 | | | | |
| | G | 0.15 | | | | |
| | D (m) | 60559.22 | 72685.25 | 2787.911 | 2345.98 | 153197.9 |
| | eps1 | 0.873679 | 0.10823 | 0.837212 | 0.697198 | 0.999939 |
| | eps2 | 0.832774 | 0.196463 | 0.9 | 0.297109 | 0.999635 |
| | pv | 0.502376 | 0.409517 | 0.752475 | 0.000249 | 1.062677 |
| | T1 (K) | 286.8796 | 156.3663 | 405.395 | 86.50139 | 426.5318 |
| 103187 | H | 14.4 | | | | |
| | G | 0.15 | | | | |
| | D (m) | 173514 | 3510.864 | 173669.6 | 165900 | 179971.5 |
| | eps1 | 0.999714 | 0.000116 | 0.999745 | 0.99938 | 0.999854 |
| | eps2 | 0.998869 | 0.007011 | 0.999374 | 0.999183 | 0.999501 |
| | pv | 0.000102 | 4.17E-06 | 0.000102 | 9.48E-05 | 0.000112 |
| | T1 (K) | 101.4336 | 3.789823 | 102.4059 | 95.17476 | 106.5952 |
| 104195 | H | 15.5 | | | | |
| | G | 0.15 | | | | |
| | D (m) | 168008.9 | 11540.88 | 168550.7 | 162504.8 | 175044.3 |
| | eps1 | 0.999366 | 0.006845 | 0.999838 | 0.999808 | 0.999868 |
| | eps2 | 0.993055 | 0.064732 | 0.998901 | 0.984996 | 0.999588 |
| | pv | 0.000169 | 0.001889 | 3.92E-05 | 3.64E-05 | 4.22E-05 |
| | T1 (K) | 99.89358 | 14.16245 | 100.7973 | 93.1012 | 103.7168 |
| 104738 | H | 15 | | | | |
| | G | 0.15 | | | | |
| | D (m) | 252961.1 | 159747.5 | 356461.7 | 8890.728 | 365030.4 |
| | eps1 | 0.928327 | 0.130794 | 0.999696 | 0.589934 | 0.999861 |
| | eps2 | 0.964333 | 0.050579 | 0.999674 | 0.9 | 0.999845 |
| | pv | 0.004578 | 0.007381 | 1.39E-05 | 1.33E-05 | 0.022345 |
| | T1 (K) | 190.8703 | 112.8318 | 119.9236 | 111.0683 | 389.8424 |

**Table A3. Bootstrap estimates of asteroid parameters – random main belt data set.** The table (and online data) list the parameters derived from WISE/NEOWISE data using the MBR: FC+3B+PC data set and all four models described in the text. Each parameter is the Akaike weighted averaged of the best fit models across 500 bootstrap trials with four models and five parameterizations of each model. Note that $H$, $G$ are from MPC November 2015, $D$ is in meters, $T_1$ in degrees K. This is a sample, the entire data set is available online.

| Asteroid | Parameter | Mean | σ | Median | 95% confidence interval | |
|---|---|---|---|---|---|---|
| | H | 13.5 | | | | |
| | G | 0.15 | | | | |
| | D (m) | 6174.043 | 1582.048 | 5851.911 | 3890.804 | 9099.526 |
| | eps1 | 0.781358 | 0.125185 | 0.809626 | 0.518792 | 0.9 |
| | eps2 | 0.3703 | 0.341259 | 0.465466 | 0 | 0.811938 |
| | pv | 0.224288 | 0.112841 | 0.205331 | 0.08492 | 0.464485 |
| 10088 | $T_1$ (K) | 309.1995 | 28.78408 | 309.0183 | 255.4772 | 363.9725 |
| | H | 13.7 | | | | |
| | G | 0.15 | | | | |
| | D (m) | 2687.763 | 186.2749 | 2689.496 | 2305.838 | 3122.314 |
| | eps1 | 0.7936 | 0.051125 | 0.800046 | 0.681629 | 0.9 |
| | eps2 | 0.726948 | 0.184992 | 0.785949 | 0.236862 | 0.9 |
| | pv | 0.820882 | 0.112104 | 0.808552 | 0.599925 | 1.099999 |
| 102587 | $T_1$ (K) | 411.1854 | 10.86292 | 412.0551 | 387.3414 | 429.3719 |
| | H | 15.5 | | | | |
| | G | 0.15 | | | | |
| | D (m) | 6368.85 | 195.8119 | 6316.857 | 6130.722 | 6694.126 |
| | eps1 | 0.9 | 0 | 0.9 | 0.9 | 0.9 |
| | eps2 | 0.062723 | 0.094648 | 1.37E-10 | 1.63E-14 | 0.2782 |
| | pv | 0.027548 | 0.001658 | 0.027929 | 0.024869 | 0.02965 |
| 104195 | $T_1$ (K) | 297.8889 | 11.98477 | 302.0745 | 269.0971 | 310.5245 |
| | H | 15 | | | | |
| | G | 0.15 | | | | |
| | D (m) | 11278.55 | 1958.501 | 10849.93 | 8154.152 | 16179.03 |
| | eps1 | 0.765387 | 0.134598 | 0.773099 | 0.435687 | 0.9 |
| | eps2 | 0.895869 | 0.022178 | 0.9 | 0.835085 | 0.9 |
| | pv | 0.015036 | 0.00471 | 0.015004 | 0.006748 | 0.026564 |
| 104738 | $T_1$ (K) | 360.2345 | 25.31853 | 367.5668 | 303.0081 | 407.5218 |
| | H | 15.4 | | | | |
| | G | 0.15 | | | | |
| | D (m) | 14364.2 | 1521.393 | 14281.27 | 11567.34 | 17729.94 |
| | eps1 | 0.900186 | 0.002168 | 0.9 | 0.9 | 0.9 |
| | eps2 | 0.805472 | 0.062393 | 0.804788 | 0.675161 | 0.903774 |
| | pv | 0.006121 | 0.001288 | 0.005991 | 0.003887 | 0.009132 |
| 105821 | $T_1$ (K) | 331.5558 | 16.39175 | 332.0549 | 300.1422 | 359.5854 |

**Table A4. Bootstrap estimates of asteroid parameters – random main belt data set – with pruning.** The table (and online data) list the parameters derived from WISE/NEOWISE data using the MBR: FC+3B+PC data set and all four models described in the text. Each parameter is the Akaike weighted averaged of the best fit models across 500 bootstrap trials with four models and five parameterizations of each model. Trials with $T_1 \leq 200K$ pruned. Note that H and G are from the This is a sample, the entire data set is available online. Note that $H, G$ are from MPC November 2015, $D$ is in meters, $T_1$ in degrees K.